\documentclass[a4paper,11pt]{article}
\pdfoutput=1
\usepackage{jcappub}

\usepackage{amsmath, amssymb, amsthm, graphicx, epsfig, fancyhdr,epsfig, slashed, bm}

\usepackage{tikzsymbols}
\usepackage{natbib}
\usepackage{float}

\usepackage{tikz,xcolor,hyperref}

\usepackage{bm}
\usepackage{url}
\usepackage{booktabs}
\usepackage{multirow}
\usepackage{changepage}

\usepackage[normalem]{ulem} 

\usepackage{natbib}
\usepackage{subfigure}
\usepackage{lmodern}

\usepackage{orcidlink}

\newcommand{\be}{\begin{equation}}
\newcommand{\ee}{\end{equation}}
\newcommand{\bea}{\begin{eqnarray}}
\newcommand{\eea}{\end{eqnarray}}

\newcommand{\GeV}{\mathinner{\mathrm{GeV}}}

\definecolor{deeppink}{RGB}{255,20,147}

\definecolor{lime}{HTML}{A6CE39}
\DeclareRobustCommand{\orcidicon}{
	\begin{tikzpicture}
	\draw[lime, fill=lime] (0,0) 
	circle [radius=0.2] 
	node[white] {{\fontfamily{qag}\selectfont \tiny ID}};
	\draw[white, fill=white] (-0.0625,0.095) 
	circle [radius=0.007];
	\end{tikzpicture}
	\hspace{-2mm}
}

\foreach \x in {A, ..., Z}{\expandafter\xdef\csname orcid\x\endcsname{\noexpand\href{https://orcid.org/\csname orcidauthor\x\endcsname}
			{\noexpand\orcidicon}}
}
\begin{document}

\title{Primordial Magnetogenesis and Gravitational Waves from ALP-assisted Phase Transition}

\author[a]{Pankaj Borah\orcidlink{0000-0003-2715-271X},}
\author[b]{P. S. Bhupal Dev\orcidlink{0000-0003-4655-2866},}
\author[c]{Anish Ghoshal\orcidlink{0000-0001-7045-302X}}

\affiliation[a]{Department of Physics, Indian Institute of Technology Delhi, \\
Hauz Khas, New Delhi 110016, India}
\affiliation[b]{Department of Physics and McDonnell Center for the Space Sciences, \\
Washington University, St. Louis, MO 63130, USA}
\affiliation[c]{Department of Physics and Astronomy, University of Sussex, Brighton BN1 9QH, United Kingdom}
\emailAdd{pankaj.borah@physics.iitd.ac.in}
\emailAdd{bdev@wustl.edu}
\emailAdd{A.Ghoshal@sussex.ac.uk}

\abstract{Sufficiently strong first-order phase transitions (FOPTs) in the early Universe can simultaneously produce an observable stochastic gravitational wave background (SGWB) and a large-scale primordial magnetic field (PMF). The recent $3.8\sigma$ evidence for a non-zero intergalactic magnetic field from anisotropic pair-halo searches using ${\it Fermi}$-LAT data further motivates a cosmological origin. We investigate an FOPT-origin of both cosmic signatures, namely, PMF and SGWB, and their correlation, within a minimal axion-like particle (ALP) framework in which a global $U(1)$ symmetry is spontaneously broken through radiative corrections, with the ALP sector coupled to the Standard Model (SM) via a Higgs-portal. We compute the present-day PMF amplitude and coherence length for both maximally helical and non-helical configurations, accounting for inverse cascade effects. For maximally helical configurations, we find peak field strengths up to $B_0 \sim 10^{-9}$ G at coherence length $\lambda_0 \sim 10^{-3}-10^{-1}$ Mpc, consistent with lower bounds on the IGMF inferred from blazar observations by MAGIC, H.E.S.S. and ${\it Fermi}$-LAT. We show that the ALP parameter region consistent with $\gamma$-ray blazar data (assuming maximal helicity) simultaneously produces stochastic GW signals detectable at future space-based interferometers, such as LISA, over the ALP decay constant range $10^3~{\rm GeV} \lesssim f_a \lesssim 10^5$ GeV. We demonstrate that these correlated constraints can be directly mapped onto effective ALP couplings to SM particles, e.g., photons, gluons, and fermions. This establishes a multi-messenger complementarity between cosmological observables and laboratory/astrophysical ALP searches, with combined constraints preferring relatively heavy ALPs, $m_a \gtrsim 0.1~\text{GeV}$, accessible to next-generation intensity and energy-frontier experiments.}

\maketitle

\section{Introduction}

The direct detection of gravitational waves (GWs) by the LIGO, VIRGO, and KAGRA (LVK) collaborations have opened a new observational window into our Universe (see, e.g., Refs.~\cite{Chatziioannou:2024hju,Keitel:2025npi} and references therein). Compact binary coalescences, including mergers of neutron stars and stellar-mass black holes, are now routinely observed with high precision.  Continued improvements in detector sensitivity, together with forthcoming facilities, are expected to significantly enhance measurement accuracy and broaden the spectrum of observable GW sources. Among these, stochastic gravitational-wave backgrounds (SGWBs) constitute a particularly important class, arising either from the superposition of unresolved astrophysical populations or from genuinely non-localised cosmological processes.

A detection of a primordial SGWB would provide a powerful probe of the early Universe. Unlike photons or neutrinos, primordial GWs propagate essentially freely from their production epoch to today, with only Planck-suppressed interactions. As a result, they retain direct information about the physical conditions at their origin. To date, searches by the LVK collaborations have not yielded evidence for a SGWB, leading instead to upper bounds at frequencies in the kHz range~\cite{LIGOScientific:2016jlg, LIGOScientific:2019vic, KAGRA:2021kbb}. On the other hand, the Pulsar Timing Array (PTA) collaborations~\cite{NANOGrav:2023gor, EPTA:2023fyk, Reardon:2023gzh, Xu:2023wog,NANOGrav:2023hvm} have reported compelling evidence for a SGWB in the nHz regime, which may admit both astrophysical and cosmological interpretations~\cite{NANOGrav:2023hfp, NANOGrav:2023hvm, EPTA:2023xxk}. Forthcoming and future observational programs~\cite{Corbin:2005ny, 
Punturo:2010zz,  TianQin:2015yph, LISA:2017pwj, 
Ruan:2018tsw, Weltman:2018zrl, 
Reitze:2019iox, 
AEDGE:2019nxb, Sesana:2019vho, Badurina:2019hst,
Kawamura:2020pcg, 
MAGIS-100:2021etm, 
Garcia-Bellido:2021zgu, 
Aggarwal:2020olq, Berlin:2021txa, Herman:2022fau, Bringmann:2023gba,Valero:2024ncz} are expected to substantially improve sensitivities across a wide frequency range, spanning from nHz to kHz and potentially extending beyond.

From a theoretical perspective, one of the most well-motivated cosmological sources of SGWB is a strong first-order phase transition (FOPT), which can arise from spontaneous symmetry breaking occurring in the early Universe~\cite{Mazumdar:2018dfl}. During an FOPT, bubbles of the broken phase nucleate within the symmetric background, subsequently expanding, colliding, and eventually percolating. The dynamics of bubble walls, together with the induced bulk motion of the surrounding plasma, constitute highly energetic processes that source GWs~\cite{Witten:1984rs,Hogan:1986dsh}. The resulting signal forms a stochastic background with distinct spectral shape, with the frequency at the peak of the
signal scaling as $f_p \propto T$ (see e.g., Refs.~\cite{LISACosmologyWorkingGroup:2022jok,Athron:2023xlk} for reviews).  In particular, the SGWB from an FOPT at the electroweak (EW) scale of $T \sim 100~\text{GeV}$ peaks around mHz frequencies and falls within the sensitivity band of space-based GW observatories like  LISA, whereas transitions occurring at much higher scales, e.g., $T\sim \mathcal{O}(\text{EeV})$, yield signals in the kHz regime, accessible to ground-based GW detectors like LIGO. Importantly, within the Standard Model (SM), both the EW~\cite{Kajantie:1996mn, Laine:2015kra} and QCD~\cite{Bhattacharya:2014ara}  symmetry-breaking transitions are crossovers rather than first order. Therefore, the observation of an SGWB consistent with an FOPT would constitute compelling evidence for beyond-the-SM (BSM) physics.

Complementary to SGWB, primordial magnetogenesis provides an additional window into the early Universe dynamics. The presence of coherent magnetic fields in the intergalactic medium are indirectly suggested by blazar observations~\cite{Neronov:2010gir, MAGIC:2022piy, HESS:2023zwb}. Recent multi-source analysis in Ref.~\cite{Zhang:2026lte} has further strengthened these indications. Specifically, a search for anisotropic pair halos using 14 years of \textit{Fermi}-LAT data from 21 high-synchrotron-peaked BL Lac (HBL) sources excluded the null intergalactic magnetic field (IGMF) hypothesis at $3.8\sigma$. Assuming a non-helical magnetic field with coherence length of 1~Mpc, the analysis yields a best-fit field strength of $B_{0} = 2.8 \times 10^{-16} \text{ G}$ with a 99\% confidence interval of $[0.9,\,8.9]\times10^{-16}$ G, providing a concrete observational benchmark for primordial magnetogenesis scenarios. The origin of these IGMFs is a longstanding puzzle, with two main possibilities: astrophysical and cosmological. For sufficiently long blazar activity timescales, plasma instabilities might develop and cool the photons before they up-scatter CMB photons~\cite{Broderick:2011av, Schlickeiser:2013eca}. However, subsequent studies indicate that energy losses of such pairs through plasma instabilities are negligible~\cite{Sironi:2013qfa,Rafighi:2017ise,Perry:2021rgv}, with recent experimental results even suggesting their absence~\cite{Arrowsmith:2025apl}. Consequently, sources of primordial magnetogenesis particularly associated with cosmic inflation or FOPT are strongly favoured. In the inflationary case, however, correlated baryon isocurvature can pose a challenge, rendering these scenarios less compelling unless inflation occurs after the EWPT to avoid this~\cite{Kamada:2020bmb}.
Alternatively, primordial magnetogenesis can be nicely achieved during FOPTs, involving either the EW ~\cite{Vachaspati:1991nm, Zhang:2019vsb, Ellis:2019tjf, Di:2020kbw, Olea-Romacho:2023rhh, Vachaspati:2024vbw}, QCD~\cite{Sigl:1996dm, Tevzadze:2012kk} or some dark sector~\cite{Balaji:2024rvo, Balaji:2025tun,ArteagaTupia:2025awh}  phase transition (PT). Notably, the idea of magnetogenesis driven by an EW FOPT, first introduced in Ref.~\cite{Vachaspati:1991nm}, suggests that magnetic fields can be generated through EW sphaleron-induced processes~\cite{Vachaspati:2001nb, Copi:2008he}. During an FOPT, the nucleation, expansion, and collision of bubbles induce highly turbulent plasma flows with large Reynolds numbers, driving the system into a magnetohydrodynamic (MHD) turbulent regime~\cite{Witten:1984rs, Hogan:1986dsh, Kamionkowski:1993fg, Brandenburg:1996fc, Christensson:2000sp, Kahniashvili:2010gp, Brandenburg:2017neh}. In addition, baryon number violating processes can generate magnetic fields with non-vanishing helicity through changes in the Chern–Simons number~\cite{Vachaspati:2001nb, Copi:2008he, Chu:2011tx}. This helicity plays a central role in the subsequent evolution, triggering an inverse cascade that transfers magnetic energy from small to larger length scales~\cite{Kahniashvili:2012uj, Copi:2008he}, thereby enhancing the coherence length of the field. Such helical structures are naturally linked to $\mathcal{C}$ and $\mathcal{CP}$ violation~\cite{Forbes:2000gr, Brandenburg:2017neh}, as required by baryogenesis~\cite{Sakharov:1967dj}, and may therefore encode information about fundamental symmetry breaking in the early Universe. Moreover, helicity in the velocity field can further sustain inverse cascades even when the initial magnetic configuration is non-helical, reinforcing the growth of large-scale correlations. On the other hand, astrophysical origins, such as the Biermann battery~\cite{PhysRev.82.863} followed by dynamo amplification~\cite{AlvesBatista:2021sln}, can generate magnetic fields in localised environments; however, they face challenges in explaining the presence of coherent fields in cosmic voids~\cite{Dolag:2010ni}. Thus, cosmological scenarios involving FOPTs that naturally accommodate magnetic fields with large coherence lengths remain an attractive explanation for the origin of the observed IGMFs.

Achieving a sufficiently strong FOPT is, therefore, essential for generating an observable GW signal and PMF. A particularly well-motivated realisation arises in theories that are approximately conformal, or scale-invariant~\cite{Meissner:2006zh}. In such setups, symmetry breaking can proceed radiatively via the Coleman–Weinberg  mechanism~\cite{Coleman:1973jx}, with a small explicit breaking of scale invariance controlling the shape of the effective potential. 
More broadly, classically scale-invariant frameworks provide a unified setting for several phenomenological and cosmological applications. The flatness of the scalar potential can support slow-roll inflation~\cite{Ghoshal:2022hyc,Ghoshal:2022qxk,Ghoshal:2024hfk}, while the constrained structure of couplings allows for viable WIMP or FIMP dark matter candidates~\cite{Hambye:2018qjv,Barman:2021lot,Barman:2022njh}. Furthermore, the absence of explicit mass scales implies that quadratic divergences that one otherwise encounters in the SM are avoided~\cite{Bardeen:1995kv}, and all physical scales arise via dimensional transmutation~\cite{Coleman:1973jx}. Although this mechanism fails to reproduce the observed EW scale within the SM alone, it can be successfully realised in extensions with additional scalar degrees of freedom ({\it d.o.f.})~\cite{Hempfling:1996ht,Espinosa:2007qk,Chang:2007ki,Foot:2007as,Foot:2007ay,Foot:2007iy,Meissner:2006zh,Meissner:2008gj,Iso:2009ss,Iso:2009nw,Barman:2021lot}. 
A characteristic feature of these scenarios is the suppression of the tunnelling rate, which generically leads to substantial supercooling~\cite{Witten:1980ez}. As a consequence, the transition is delayed and proceeds in a vacuum-dominated background. Bubble nucleation occurs at relatively low temperatures, and the subsequent expansion and collision of bubble walls take place with minimal plasma friction. This prolongs the duration of the transition and enhances the fraction of vacuum energy released into bulk motion, thereby amplifying the resulting GW signal~\cite{Jinno:2016knw,DelleRose:2019pgi,VonHarling:2019rgb,Ellis:2020nnr,Ghoshal:2020vud}.

In this context, axion-like particle (ALP) scenarios are particularly compelling. They naturally appear as pseudo Nambu-Goldstone bosons in many BSM extensions with a spontaneously broken global $U(1)$ symmetry. Notably, ALPs can address several open questions in the SM, such as the strong $\mathcal{CP}$ problem via the Peccei–Quinn (PQ) mechanism~\cite{Peccei:1977hh, Peccei:1977ur, Weinberg:1977ma,Wilczek:1977pj}, the hierarchy problem through the relaxion framework~\cite{Graham:2015cka}, and may also play important cosmological roles in inflation~\cite{Freese:1990rb, Adams:1992bn, Daido:2017wwb}, dark matter~\cite{Preskill:1982cy, Abbott:1982af,Dine:1982ah}, dark energy~\cite{Jain:2004gi,Kim:2009cp,Kim:2013jka,Choi:2019jck,Brandenberger:2020gaz,Yao:2023qve}, and baryogenesis~\cite{Daido:2015gqa,DeSimone:2016bok,Jeong:2018jqe,Co:2019wyp,Im:2021xoy, Foster:2022ajl}. Importantly, a defining feature of these scenarios is that their phenomenology is primarily controlled by the $U(1)$ symmetry-breaking scale $f_a$, which determines not only the masses and couplings of the ALP sector but also its finite-temperature dynamics. Consequently, the finite-temperature dynamics associated with spontaneous $U(1)$ breaking constitute an intrinsic prediction of ALP models, making the study of the corresponding phase transition an important aspect of hidden-sector cosmology. If the transition is first order, the accompanying out-of-equilibrium dynamics can leave a variety of observable cosmological relics. In particular, radiatively induced symmetry breaking gives rise to an approximately scale-invariant effective potential, naturally leading to strong supercooling and a vacuum-dominated phase transition. Such transitions simultaneously produce a SGWB signal and seed primordial magnetic fields through the turbulent plasma dynamics accompanying bubble expansion and collision. The two observables therefore constitute complementary manifestations of the same underlying symmetry-breaking dynamics, providing a powerful probe of the thermal history of the ALP sector.

In this work, we investigate a minimal ALP setup with spontaneous radiative breaking of a global $U(1)$ symmetry. The resulting FOPT is controlled by a small set of parameters, making the thermal history of the model highly predictive. We study the correlated cosmological signatures arising from this transition by simultaneously analysing the primordial magnetic fields and stochastic GW backgrounds. We demonstrate that the generated magnetic fields can satisfy observational constraints on the strength of the IGMFs from the {\it Fermi}-LAT,  MAGIC and H.E.S.S.  collaborations~\cite{MAGIC:2022piy, HESS:2023zwb}, after accounting for their subsequent evolution, including inverse cascade effects. In parallel, the associated SGWB can reach amplitudes accessible to current and future GW detectors. A central result of our analysis is the emergence of a correlated parameter space where both the observables— magnetic fields and SGWB signals—are simultaneously viable. Imposing the requirement of generating magnetic fields consistent with blazar observations, together with the prospects for detecting a SGWB signal, significantly enhances the complementarity between cosmological probes and laboratory as well as astrophysical searches for ALPs. This multi-messenger interplay enables a more precise delineation of the viable parameter space of the ALP framework in terms of the ALP decay constant $f_a$. In particular, the parameter regions yielding observable GW signals are tightly correlated with those producing magnetic fields of sufficient strength and coherence length, allowing these combined requirements to be directly mapped onto the underlying ALP parameter space. Conversely, existing bounds on ALP couplings to SM particles translate into restrictions on the FOPT dynamics, and hence on the resulting GW spectrum and magnetic field properties. This establishes radiatively induced ALP-assisted phase transitions as a predictive framework in which the thermal history of a hidden sector can be tested through correlated cosmological observables, while remaining complementary to laboratory and astrophysical searches for ALPs.

The remainder of this paper is organised as follows. In Sec.~\ref{sec:FOPT-ALP}, we describe the dynamics of the FOPT in the ALP framework, including finite-temperature effects and analytic approximations relevant to the supercooled regime. In Sec.~\ref{sec:the-model}, we present an explicit model realisation and discuss the parameter space of interest. Sec.~\ref{sec:pmf-from-fopt} reviews the generation and evolution of PMFs, their coherence length, and the bounds on the IGMF from blazar observations. In Sec.~\ref{sec:GWs}, we provide some details on the computation of the resulting GW spectra from a strong FOPT. In Sec.~\ref{sec:results}, we present our numerical results and discuss the interplay between GWs, magnetic fields, and complementary constraints from laboratory ALP searches. Finally, we summarise our findings and conclude in Sec.~\ref{sec:summary}.

\section{ALP-assisted First-order Phase Transition}
\label{sec:FOPT-ALP}

In this study, we consider a scenario where a global $U(1)$ symmetry is dynamically broken through radiative corrections~\cite{Gildener:1976ih}. The scalar sector in this framework consists of a collection of classically massless fields, with some of them carrying $U(1)$ charge. The tree-level scalar potential is purely quartic and can be written as
\bea
V(\phi) = \frac{\lambda_{ijkl}}{4} \phi_i \phi_j \phi_k \phi_l \,.
\eea 
It was shown in Ref.~\cite{Gildener:1976ih} that renormalisation group (RG) evolution of the quartic couplings generically leads to the existence of a scale $\Lambda$ at which a particular linear combination of couplings vanishes. At this scale, the potential develops a flat direction, which can be specified by a unit vector $\vec{n}$ in field space. parametrising the fields as $\vec{\phi} = \vec{n} \sigma$, the tree-level potential identically vanishes along the $\sigma$ direction. Consequently, the dynamics along this flat direction are governed entirely by radiative corrections. The one-loop effective potential can then be expressed as
\bea
\label{eq:effpot0}
V_\textrm{eff}(\sigma) = \frac{\beta_{\lambda_\textrm{eff}}}{4} \sigma^4 \left( \log \frac{\sigma}{\hat{f}}  - \frac{1}{4}\right) \, ,
\eea
where $\lambda_\textrm{eff}(\mu) = \lambda_{ijkl}(\mu) n_i n_j n_k n_l$ denotes the effective quartic coupling along the flat direction, satisfying $\lambda_\textrm{eff}(\Lambda) = 0$, and $\beta_{\lambda_\textrm{eff}}$ is its corresponding beta function. For $\beta_{\lambda_\textrm{eff}} > 0$, the potential develops a non-trivial minimum at $\sigma = \hat{f}$, leading to spontaneous symmetry breaking and the $\sigma$ field acquiring a mass of $m_\sigma^2 = \beta_{\lambda_\textrm{eff}} \hat{f}^2$. For recent discussion on radiative symmetry breaking and the associated PT dynamics, see e.g., Refs.~\cite{Levi:2022bzt,Salvio:2023qgb,Salvio:2026bco}.

\subsection{Finite-temperature effects} \label{sec:finite-T-effects}

In the vicinity of the origin, $\sigma \simeq 0$, the effective potential $V_\textrm{eff}$ exhibits a nearly flat profile, as its second derivative vanishes. In this regime, thermal corrections play a crucial role~\cite{Witten:1980ez}. In particular, finite-temperature effects generate a positive quadratic contribution for any $T>0$, thereby lifting the origin and converting it from a local maximum into a metastable minimum. These effects are captured by the finite-temperature contribution to the effective potential~\cite{Dolan:1973qd}:
\bea
V_T(\sigma, T) = \frac{T^4}{2 \pi^2} \sum_{\rm b} J_B\left( \frac{m_{\rm b}^2(\sigma)}{T^2} \right) + \frac{T^4}{2 \pi^2} \sum_{\rm f} J_F\left( \frac{m_{\rm f}^2(\sigma)}{T^2} \right) \,,
\eea
where $m_{\rm b,f}(\sigma)$ denote the field-dependent masses of the bosonic and fermionic {\it d.o.f.}, respectively. The thermal functions $J_{B,F}$ are defined as~\cite{Anderson:1991zb}
\begin{align}
    J_{B/F}(y^2) = \int_0^\infty dt~t^2 \log{\left[1\mp \exp{\left(-\sqrt{t^2+y^2}\right)}\right]} \, .
    \label{eq:JBF}
\end{align}
Near $\sigma \simeq 0$, the vanishing curvature ensures that the high-temperature expansion, $y^2\ll 1$, remains applicable. In this regime, the free energy along the flat direction can be approximated as
\bea
\label{eq:free-energy-flat-direction}
F(\sigma, T) \simeq - \frac{\pi^2}{90} g_* T^4 +  \tilde{a} \frac{T^2}{24} \sigma^2  + \frac{\beta_{\lambda_\textrm{eff}}}{4} \sigma^4 \left( \log \frac{\sigma}{\hat{f}}  - \frac{1}{4}\right) + V_0  \, ,
\eea
where $g_* = N_{\rm b} + \frac{7}{8} N_{\rm f}$ represents the total number of relativistic {\it d.o.f.} in the metastable phase. The coefficient $\tilde{a}$ is defined as $\tilde{a} \sigma^2 \equiv \left[ \sum_{\rm b} m_{\rm b}^2(\sigma)  +  \frac{1}{2} \sum_{\rm f} m_{\rm f}^2(\sigma) \right]$ and the term $V_0 = - V_\textrm{eff}(\sigma = \hat{f}) = \beta_{\lambda_\textrm{eff}} \hat{f}^4/16$ is introduced such that the vacuum energy vanishes at the true minimum.

Moving away from the origin, the validity of the high-temperature expansion persists as long as the field value satisfies $\sigma \lesssim T/{\hat g}$, where $\hat g$ characterises the typical coupling entering the field-dependent masses, $m(\sigma) \sim \hat g \sigma$. In the supercooled regime, the structure of the logarithmic term in Eq.~(\ref{eq:free-energy-flat-direction}) simplifies considerably. One can rewrite
\bea
\log \frac{\sigma}{\hat{f}} = \log \frac{\hat g \sigma}{T} + \log \frac{T}{\hat{g} \hat{f}}  \simeq \log \frac{T}{\hat{g} \hat{f}} \equiv  \log \frac{T}{M} \,,
\eea
where $M \equiv \hat{g} \hat{f}$ denotes the characteristic heavy mass scale at the true vacuum. This approximation is justified during supercooling ($T \ll M$), since the dominant contribution to the bounce action arises from field values near the barrier, for which $\hat g \sigma/T \sim 1$. Under these conditions, the free energy can be cast into a simple polynomial form:
\bea
\label{eq:approxpot}
F(\sigma, T) \simeq \frac{m^2(T)}{2} \sigma^2 - \frac{\lambda(T)}{4} \sigma^4 \, ,
\eea
with temperature-dependent coefficients $m^2(T) = \tilde{a} T^2/12$ and $\lambda(T) = \beta_{\lambda_{\textrm{eff}}} \log (M/T)$. Field-independent contributions have been omitted, as they do not affect the transition rate or tunnelling dynamics.

The PT associated with the potential in Eq.~(\ref{eq:approxpot}) is governed by the three-dimensional Euclidean bounce action $S_3/T \approx 18.897 \, m(T)/(\lambda(T) T) \equiv A_3/\log(M/T)$. Here, the mild (logarithmic) temperature dependence reflects the approximate scale invariance of the system. The corresponding nucleation rate per unit volume takes the form~\cite{Linde:1977mm,Coleman:1977py,Callan:1977pt,Affleck:1980ac,Linde:1980tt,Linde:1981zj}
\bea
\Gamma \simeq T^4 \left( \frac{S_3}{2\pi T} \right)^\frac{3}{2} \exp(-S_3/T) \, .
\eea
Because $S_3/T$ varies only slowly with temperature, the transition typically proceeds after a significant period of supercooling. In addition to thermally induced tunnelling, vacuum decay via four-dimensional $O(4)$ symmetric bounces may also contribute. When the corresponding action $S_4$ becomes smaller than $S_3/T$, quantum tunnelling can enhance the nucleation probability. However, within the parameter space of interest, we find that the thermal nucleation consistently provides the dominant contribution.

The onset of the transition is determined by the condition $\Gamma/H^4 = 1$~\cite{Enqvist:1991xw}. The Hubble expansion rate is given by
\bea
H^2 = \frac{1}{3M_\textrm{Pl}^2} \left[ \frac{\pi^2}{30} g_* T^4  +  V_0 \right] \simeq \frac{V_0}{3M_\textrm{Pl}^2} \equiv H_I^2 \,,
\label{eq:2p9}
\eea
where $M_{\rm Pl}=2.4\times 10^{18}$ GeV is the reduced Planck mass. The approximation on the right-hand side of Eq.~\eqref{eq:2p9} applies in the supercooled regime, where the vacuum energy dominates over the radiation component, leading to a quasi-de Sitter (inflationary) expansion. Substituting this into the nucleation criterion, one obtains~\cite{Conaci:2024tlc}
\bea
T_n \simeq \sqrt{M H_I} \exp \left( \frac{1}{2} \sqrt{-A_3 + \log^2(M/H_I)} \right) \,,
\eea
together with a lower bound on the nucleation temperature, given by
\begin{equation}
T_n^{\rm min} = \sqrt{M H_I}
\simeq
0.1\hat{f} \left(\frac{\hat{f}}{M_{\rm Pl}}\right)^{1/2} .
\end{equation}

The strength of the PT is quantified by the parameter $\alpha$, defined as the released vacuum energy relative to the radiation energy density. It can be expressed in terms of the free-energy difference between the false and true vacua as
\bea
\alpha = \frac{1}{\rho_\textrm{rad}} \left( 1 - T \frac{\partial}{\partial T} \right) \bigg[ F(0, T) - F(\sigma_n, T)  \bigg]\bigg|_{T_n} \,.
\eea
In the supercooled regime, where vacuum energy dominates, this simplifies to
\begin{equation}
\alpha \simeq \frac{V_0}{\rho_{\rm rad}(T_n)}
= \left( \frac{T_{\rm inf}}{T_n} \right)^4 ,
\end{equation}
with $T_\textrm{inf} = (30 V_0/(g_* \pi^2))^{1/4}$ denoting the temperature at which vacuum domination sets in\footnote{Here, we assume that $g_*$ remains approximately constant between $T_\textrm{inf}$ and $T_n$.}.

The inverse duration of the PT is characterised by the parameter $\beta$~\cite{Grojean:2006bp}:
\bea
\frac{\beta}{H} = - \frac{d \log \Gamma}{d \log T} \bigg|_{T=T_n} \simeq - 4 + T \frac{d (S_3/T)}{d T}\bigg|_{T=T_n} = - 4  + \frac{A_3}{\log^2(M/T_n)} \,,
\eea
This expression highlights that, in the presence of strong supercooling, $\beta/H$ can approach $\mathcal{O}(1)$, thereby enhancing the power spectrum of the GW signal.

Once the transition completes, the Universe reheats to a temperature
\bea
T_{\rm reh} = T_\textrm{inf} \min\left(1, \frac{\Gamma_d}{H_I} \right)^{1/2} \,,
\eea
where $\Gamma_d$ denotes the decay rate into the SM plasma. In what follows, for the sake of simplicity, we will assume efficient reheating, such that $T_{\rm reh} \simeq T_\textrm{inf}$.

Finally, we close this section by noting that, within the approximations discussed above, the PT dynamics of the system is effectively determined by three quantities: the symmetry-breaking scale $\hat{f}$, the beta function $\beta_{\lambda_\textrm{eff}}$, and the thermal coefficient $\tilde{a}$ (see also Refs.~\cite{Levi:2022bzt,Salvio:2023qgb, Salvio:2026bco} for more details). These parameters are fixed once a concrete ultraviolet (UV) realisation of the model is specified.

\section{The Model}
\label{sec:the-model}
In this section, we present a concise overview of the model framework, considering a scenario where the global $U(1)$ symmetry is radiatively broken, similar to that discussed in Refs.~\cite{DelleRose:2019pgi, VonHarling:2019rgb, Conaci:2024tlc} (see also Refs.~\cite{Hambye:2013dgv, Iso:2017uuu, Hambye:2018qjv, Arteaga:2024vde} for similar models in the context of EWPT), inspired by the KSVZ-type UV completion~\cite{Kim:1979if,Shifman:1978by}\footnote{In the original KSVZ model~\cite{Kim:1979if,Shifman:1978by}, the PT is second-order.}. In this framework, we have a pair of complex scalar fields $\phi_1$ and $\phi_2$, a pair of vector-like fermions $\psi$ and $\psi^c$, 
and a $U(1)_S$ (with $S$ denoting a ``secluded'' gauge sector) gauge field $A^\mu$ with gauge coupling $g$. All new particle states are neutral under the SM gauge group\footnote{One can, in principle, also consider $\psi\, (\psi^c)$ in the fundamental\, (anti-fundamental) representation of colour, and assign non-zero SM $U(1)_Y$ hypercharges similar to Ref.~\cite{DelleRose:2019pgi}. In this scenario, the domain wall number is equal to one, and they can decay by mixing with the right-handed quarks.}. Under the new $U(1)_S$ gauge symmetry, only $\phi_1$ is charged, while being neutral under the global $U(1)$. The $\phi_2$, instead, transforms under the global $U(1)$ and its phase corresponds to the ALP. We extend the framework of Ref.~\cite{DelleRose:2019pgi} by coupling the ALP sector to the SM Higgs doublet $H$ through a Higgs-portal quartic coupling.
The Lagrangian of the model is given by
\bea
\label{eq:elemLag}
\allowdisplaybreaks
\mathcal L &&= - \frac{F^{\prime 2}}{4 g^2} + |D^{\prime}_\mu \phi_1|^2 + |D_\mu H|^2 + |\partial_\mu \phi_2|^2 + \bar{\psi} i \gamma^\mu \partial_{\mu} \psi + \bar{\psi}^{c} i \gamma^\mu \partial_{\mu} \psi^{c} \\ \nonumber 
&&- \lambda_H|H|^4 - \lambda_1 |\phi_1|^4 - \lambda_2 |\phi_2|^4
-  \lambda_{12} |\phi_1|^2 |\phi_2|^2
-\lambda_{h2}|H|^2|\phi_2|^2 + (y \phi_2 \psi \psi^c + \textrm{H.c.} ) ,
\eea
where $F^{\prime \mu\nu}$ and $D^{\prime \mu} = \partial^\mu - i g A^{\prime \mu}$ are the field strength and the covariant derivative of the $U(1)_S$ gauge field, respectively. Note that our choice of $U(1)_S$ realisation following Ref.~\cite{DelleRose:2019pgi}, rather than a non-Abelian $SU(2)$ extension as in Ref.~\cite{Conaci:2024tlc}, is motivated by the structure of the radiatively generated effective potential (see below). Although Abelian gauge theories generically exhibit a Landau pole in the ultraviolet, this does not impact our analysis, as we restrict to parameter regions where the gauge coupling remains perturbative over the relevant scales\footnote{We have verified this explicitly by evolving the couplings using both one- and two-loop $\beta$-functions, which were obtained using \texttt{PyR@TE 3}~\cite{Sartore:2020gou}. For the parameter space relevant to the PT and magnetogenesis, the gauge coupling remains perturbative up to scales $f_a \gtrsim 10^{12}$ GeV, without encountering a Landau pole.}. 

In Eq.~\eqref{eq:elemLag}, the coupling $\lambda_{h1} |\phi_1|^2 |H|^2$, while allowed by the symmetry, is omitted for simplicity as a tree-level boundary condition at the conformal scale. This is natural, since $\phi_1$ is charged under $U(1)_S$ with no connection to the ALP {\it d.o.f.}, and the only physically motivated portal to the SM is through $\phi_2$, the field whose global $U(1)$ phase is the ALP. Furthermore, in the gauge-dominance regime  $\lambda_1 \simeq \lambda_2 \ll g^2$ (see below), the radiatively induced value  $\lambda_{h1}^{\text{loop}} \sim \lambda_{12} \lambda_{h2} / (16\pi^2)$ is further suppressed and does not affect any of the PT observables.

In this work, we consider the usual Gildener-Weinberg approach~\cite{Gildener:1976ih} and, for simplicity, assume the flat direction along $H=0$ at the tree-level. 
A simple criterion for the original flat direction to remain valid is $\lambda_{h2}(\mu)>0$, where $\mu$ is the Gildener–Weinberg scale. We keep the two-field flat direction in $(\phi_1,\phi_2)$,
\bea
\label{eq:GW-flat-condition}
\lambda_{12}(\mu)=-2\sqrt{\lambda_1(\mu)\lambda_2(\mu)},\qquad
\tan^2\theta=\sqrt{\frac{\lambda_1}{\lambda_2}},
\eea
and set $|H|=0$ on the ray. We parameterise the background fields as
\bea
\label{eq:defmodel}
r_1\equiv|\phi_1|=\frac{\sigma}{\sqrt2}\cos\theta,\qquad
r_2\equiv|\phi_2|=\frac{\sigma}{\sqrt2}\sin\theta.
\eea
Along the Gildener-Weinberg flat direction, parametrised by $\sigma$, the corresponding field-dependent masses of the gauge field $A^\mu$, the fermion $\psi$, the heavy scalar mode $\tau$ in the radial direction orthogonal to $\sigma$, and the SM Higgs field $H$ are respectively given by
\bea
M_A = g \, \cos \theta \, \sigma \,, \quad M_\psi = \frac{y}{\sqrt{2}} \sin \theta \, \sigma \,, \quad M_\tau = (4 \lambda_1 \lambda_2)^{1/4} \sigma \,, \quad M_H = \sqrt{\frac{\lambda_{h2}}{2}} \sigma \sin\theta \, .
\label{eq:masses}
\eea
The phase of $\phi_2$, corresponding to the ALP field $a$, remains exactly massless at this level, while $\sigma$ acquires a loop-suppressed mass through the Coleman-Weinberg mechanism. With the rotation angle $\theta$ defined in Eqs.~(\ref{eq:GW-flat-condition}) and (\ref{eq:defmodel}), the axion decay constant $f_a$ is identified as $f_a = \hat{f} \sin \theta$, where $\hat{f} = \langle \sigma \rangle$ (see Sec.~\ref{sec:FOPT-ALP}). The ALP mass $m_a$ is usually generated by some higher-dimensional operators or non-perturbative effects, which are independent of $f_a$ (see e.g., Refs.~\cite{Dias:2014osa, DiLuzio:2020wdo, Bauer:2020jbp}).

Here, the Higgs portal coupling, $\lambda_{h2}>0$, is fixed consistently with the measured Higgs mass by matching the Higgs quartic coupling at the EW scale, $\lambda_H(v)=m_h^2/(2v^2)$, where $m_h\simeq 125$ GeV is the observed SM Higgs mass and $v\simeq 246.2$ GeV is the EW vacuum expectation value, to its UV value obtained through RG evolution up to the conformal scale $\mu\simeq f_a$.\footnote{
We employ \texttt{PyR@TE 3}~\cite{Sartore:2020gou} to compute the one- and two-loop $\beta$-functions for the scalar quartic, gauge, and Yukawa couplings. For comparison, similar one-loop results can be found in Ref.~\cite{DelleRose:2019pgi}.}  At this threshold, the heavy scalar modes of the ALP sector and the portal interaction modify the running and induce finite matching corrections, ensuring continuity between the SM and the full theory.  The couplings are evolved from the EW scale to $\mu=f_a$ with the flat-direction conditions imposed there, and $\lambda_{h2}$ is adjusted such that {$\lambda_{h2}(f_a)=2m_h^2/f_a^2$}, corresponding to $m_h\simeq125~\mathrm{GeV}$ in the RG-improved sense.  This RG-consistent determination of $\lambda_{h2}$ preserves perturbativity and it simultaneously yields the Higgs effective mass $M_H^2(f_a)=\tfrac12\lambda_{h2}f_a^2\sin^2\theta$, which, for positive $\lambda_{h2}$, keeps the Higgs at the origin, maintaining the radiative dynamics of the original model~\cite{Conaci:2024tlc}.

Along the Gildener–Weinberg flat direction defined in Eqs.~(\ref{eq:GW-flat-condition}) and (\ref{eq:defmodel}), the tree-level scalar potential vanishes identically, and the dynamics of the radial field $\sigma$ is entirely governed by radiative corrections. At one-loop, the zero-temperature effective potential takes the Coleman–Weinberg form given in Eq.~(\ref{eq:effpot0}). For the present model, the $\beta$ function of the effective quartic coupling, $\beta_{\lambda_{\textrm{eff}}}$, receives contributions from the hidden gauge boson, the vector-like fermions, the heavy scalar orthogonal to the flat direction, and the Higgs portal interaction. Explicitly, we obtain\footnote{In the $SU(2)$ case, the gauge contribution to $\beta_{\lambda_{\rm eff}}$ at one-loop  scales as $(9/8) g^4$ (instead of $6g^4$ in Eq.~(2.17) of  Ref.~\cite{Conaci:2024tlc}), which would modify the PT dynamics and significantly shrink the available parameter space,  compared to the Abelian case discussed here.}
	\bea
	\label{eq:beta-eff}
	\beta_{\lambda_\textrm{eff}} = \frac{\partial \lambda_\textrm{eff}}{\partial \log \mu} =  \frac{1}{16 \pi^2} \left[8 \lambda_1 \lambda_2 + 6 g^4 \cos^4 \theta - 6 y^4 \sin^4 \theta  +  \lambda_{h2}^2 \sin^4\theta \right] \,.
	\eea

At finite temperature, thermal corrections lift the flatness of the potential near the origin and play a crucial role in the PT dynamics. Following the general discussion in Sec.~\ref{sec:finite-T-effects}, the coefficient `$\tilde{a}$' [cf.~Eqs.~(\ref{eq:free-energy-flat-direction}) and (\ref{eq:approxpot})], corresponding to the quadratic terms in the high-temperature regime is given as
\bea
\label{eq:coeff-a}
\tilde{a} = 3 g^2 \cos^2 \theta + 3 y^2 \sin^2 \theta + 2 \sqrt{\lambda_1 \lambda_2} + \frac{\lambda_{h2}}{2} \sin^2\theta \,.
\eea
Note that, for our choice of $\lambda_{h2} > 0$, the Higgs contribution to $\tilde{a}$ is positive, implying that the Higgs field increases the thermal mass of $\sigma$ near the origin. As a result, the symmetric phase is further stabilised at finite temperature, reinforcing the characteristic supercooling behaviour of radiatively broken conformal theories~\cite{Levi:2022bzt,Salvio:2023qgb, Salvio:2026bco}. In this work, we primarily focus on the supercooling regime~\cite{DelleRose:2019pgi,Conaci:2024tlc}, 
and in this limit, the relevant PT parameters, i.e, $T_n, \, \beta/H$ and $\alpha$, are accurately captured by the analytic approximations outlined in Sec.~\ref{sec:finite-T-effects}.
Furthermore, we specialise to the parameter region in which the PT is predominantly driven by the hidden gauge sector~\cite{DelleRose:2019pgi,Conaci:2024tlc}, namely, $\lambda_1 \simeq \lambda_2 \ll g^2$ and $y \simeq 0$.\footnote{In our Higgs-portal scenario, this choice would imply that the secluded sector fields $\psi$ and $\tau$ are much lighter [cf.~Eq.~\eqref{eq:masses}] and can, in principle, contribute to invisible Higgs decay. However, we have enough freedom in the model parameters to ensure that the predicted Higgs invisible branching ratio remains consistent with the experimental limit~\cite{ATLAS:2023tkt}, without affecting the PT dynamics.} The Higgs-portal coupling $\lambda_{h2}$ is nevertheless retained in Eq.~(\ref{eq:coeff-a}), as it can contribute non-trivially to the thermal corrections and hence influence the PT dynamics. 

We close the model discussion with a comment on the validity regime. We assume efficient reheating, $T_{\rm reh} \simeq T_{\rm inf}$ (see Sec.~\ref{sec:FOPT-ALP}), for our PT analysis, which may not be a good approximation if the ALP is sufficiently long-lived in parts of the parameter space. A consistent treatment in this regime would require a detailed analysis using Boltzmann equations. Naively, a condition for instantaneous reheating, $\tau_a \lesssim t_{\rm reh}$, where $t_{\rm reh}$ denotes the age of the Universe at reheating, roughly implies $f_a \gtrsim \mathcal{O}(1)~\mathrm{GeV}$ for $m_a\lesssim 0.1f_a$, which is easily satisfied for the parameter region of our interest, thus validating the instantaneous reheating approximation used here.

\section{Primordial Magnetic Field Generation from FOPT}
\label{sec:pmf-from-fopt}
\noindent The generation of primordial magnetic fields during an FOPT has been extensively studied in the literature (see e.g. Refs.~\cite{Vachaspati:1991nm,Baym:1995fk,Sigl:1996dm,Ahonen:1997wh,Diaz-Gil:2007fch,Caprini:2009yp,Zhang:2019vsb,Ellis:2019tjf,Di:2020kbw,Di:2025ncl}). As discussed in the Introduction, an FOPT provides the necessary out-of-equilibrium conditions for magnetogenesis. During the nucleation, expansion and subsequent collision of bubbles, a fraction of the released vacuum energy is converted into bulk plasma motion, driving large Reynolds-number flows that develop into MHD turbulence~\cite{Witten:1984rs, Hogan:1986dsh, Kamionkowski:1993fg, Brandenburg:1996fc, Christensson:2000sp, Kahniashvili:2010gp, Brandenburg:2017neh}. In a conducting primordial plasma, these turbulent motions source electric currents and seed magnetic fields, while additional $\mathcal{CP}$-violating dynamics~\cite{Forbes:2000gr, Brandenburg:2017neh} may generate a net magnetic helicity through changes in the Chern-Simons number~\cite{Vachaspati:2001nb, Copi:2008he, Chu:2011tx}. The subsequent MHD evolution then determines the final magnetic field strength and coherence length, with maximally helical configurations undergoing an inverse cascade that efficiently transfers magnetic energy towards larger length scales. In the present work, we adopt this well-established FOPT magnetogenesis framework and apply it to the radiatively induced ALP phase transition described in the previous section. In addition, since the microscopic origin of the initial helicity is model dependent, our analysis follows the standard treatment in the literature and considers both maximally helical and non-helical magnetic field configurations. The resulting present-day magnetic field spectrum is discussed below.
\subsection{Magnetic field spectrum today}

The magnetic fields generated during an FOPT evolve under the influence of MHD turbulence~\cite{Banerjee:2004df} in a highly conducting primordial plasma. Their subsequent evolution is primarily dictated by the initial helicity and the turbulent properties of the plasma, which determine how magnetic energy is redistributed across different length scales. A central role is played by the magnetic helicity, which describes how twisted and linked the magnetic field lines are. It is defined as $\langle \bm{A} \cdot \bm{B} \rangle$, where $\bm{B}=\nabla \times \bm{A}$, and $\bm{A}$, $\bm{B}$ are the usual vector potential and magnetic field,  respectively. It is approximately conserved in the early Universe due to the large conductivity of the plasma. In the presence of maximum helicity, the system undergoes an inverse cascade, transferring magnetic energy from small to large scales. This leads to a growth of the coherence length accompanied by a relatively slow decay of the magnetic field amplitude. Such dynamics may have been critical for the persistence and evolution of PMFs, enabling their survival and large-scale coherence in the present universe. In contrast, configurations with negligible helicity evolve predominantly through direct turbulent decay, with energy dissipating at small scales through viscous and resistive effects.

During the radiation-dominated era, the decay of MHD turbulence for maximally helical magnetic field follows a power-law dependence on conformal time $\eta$, with the following relations for the magnetic energy and correlation length~\cite{PhysRevLett.83.2195}:
\begin{equation}
B \sim \eta^{-1/3} \, \, \mathrm{and} \, \, \lambda \sim \eta^{2/3}.
\end{equation}
Here, $B\equiv |\bm{B}|$ is the magnetic field strength, while $\lambda$ represents its correlation length. Numerical simulations have indicate that even in the absence of magnetic helicity, an effective transfer of magnetic energy toward larger scales can occur if the plasma possesses a non-vanishing initial kinetic helicity~\cite{Brandenburg:2017rnt}. In such a situation, the magnetic field and its correlation length evolve according to
\begin{equation}
    B \sim \eta^{-1/2} \,\, \, \mathrm{and} \,\, \, \lambda \sim \eta^{1/2} \, .
\end{equation}
These scaling relations apply when the Universe is radiation-dominated, for which the scale factor grows linearly with conformal time, i.e.~$a \sim \eta$. Following recombination, the magnetic field evolves adiabatically with the expansion, red-shifting as $B \sim a^{-2}$. To treat both helical and non-helical cases in a concise way, we introduce the
generalised parameters 
\begin{equation}
q_b = \frac{2}{b+3} \, \, \, \, \mathrm{and} \, \, \, \,  p_b = \frac{2}{b+3}(b+1) \, ,
\end{equation}
yielding the power-law scalings
\begin{equation}
    B \sim \eta^{- p_b / 2} \, \,\, \, \mathrm{and} \,\, \, \, \lambda \sim \eta^{q_b} \, ,
\end{equation}
where the cases $b=0$ and $b=1$ correspond to the maximally helical and non-helical scenarios described above, respectively.

The magnetic field energy density produced from FOPT  can be estimated by~\cite{Ellis:2019tjf, RoperPol:2023bqa}
\begin{equation}
\rho_{B,*} \simeq 0.1\frac{ \kappa_{\rm col} \alpha}{1+ \alpha} \rho_{*} \approx \frac{ \pi^2}{3} T_{\rm reh}^4 \approx 0.1 \, \rho_{\rm vac}\, ,
\label{eq:mag_energy_dens}
\end{equation}
where  $\rho_{*}=\frac{3M_{\rm Pl}^2}{8\pi}H_{*}^2 = \frac{g_{*}(T_\textrm{reh})\pi^2}{30} T_{\rm reh}^4 \approx \rho_{\rm vac}$ denotes the total energy density at the time of percolation~\cite{Ellis:2019tjf}, and $g_{*}(T_{\rm reh})$ corresponds to the total number of relativistic {\it d.o.f.} contributing to the entropy density at the reheating temperature $T_{\rm reh}$. The parameter $\kappa_{\rm col}$ quantifies the fraction of the released vacuum energy that is transferred into the acceleration of the bubble walls. We further assume that a fixed fraction of the resulting bulk motion of the plasma is converted into magnetic fields through MHD turbulence, with an efficiency of order $10 \%$~\cite{Kahniashvili:2009qi,Durrer:2013pga, Brandenburg:2017neh}. In Eq.~\eqref{eq:mag_energy_dens}, we have assumed that $\alpha\gg 1$, as expected in a supercooled PT. 

The present-day magnetic field spectrum can be computed as~\cite{Ellis:2019tjf,Vachaspati:2024vbw,ArteagaTupia:2025awh}
\begin{align}
\label{eq:B0-spectrum}
B_0(\lambda) \equiv B(\lambda,t_0) = \left(\frac{a_{\rm reh}}{a_{\rm rec}}\right)^{p_b/2} \left(\frac{a_{\rm reh}}{a_0}\right)^2 \sqrt{{\frac{10}{17}}\,\rho_{B,*}} \,
\begin{cases}
\left(\frac{\lambda}{\lambda_0}\right)^{-5/2}  & \mbox{ for }~~ \lambda\geq\lambda_0 \, \\    \left(\frac{\lambda}{\lambda_0}\right)^{1/3} & \mbox{ for }~~ \lambda < \lambda_0 \,, 
\end{cases}
\end{align}
which assumes a power-law spectrum for the magnetic field strength. 
The coherence length scale of the magnetic field $\lambda_0$ redshifted to today is~\cite{Ellis:2019tjf}
\begin{equation}
\label{eq:l0}
\lambda_0 \equiv \lambda_B(t_0) = \left(\frac{a_{\rm rec}}{a_{\rm reh}}\right)^{q_b} \left(\frac{a_0}{a_{\rm reh}}\right) \lambda_*, \,
\end{equation}
where the initial correlation length $\lambda_{*}$ is determined by the bubble size $R_{*}$ at percolation~\cite{Caprini:2019egz}, 
\begin{equation}
\label{eq:coherence-l0-betaH}
\lambda_{*} = R_{*} = \frac{(8 \pi)^{1/3}}{H_{*}} \left( \frac{\beta}{H} \right)^{-1},
\end{equation}
where a bubble wall velocity of $v_{w}=1$ is assumed. The peak value of the magnetic field spectrum is denoted as $B_{0} = B_{0}(\lambda_{0})$.
The redshift factors are given by
\begin{align}
\label{eq:red-shift-factors}
&\frac{a_{\rm reh}}{a_0} = 8 \times 10^{-14} \left( \frac{100}{g_{*}(T_{\rm reh})} \right)^{1/3} \left( \frac{1 \GeV}{T_{\rm reh}} \right) \, , \nonumber\\
&\frac{a_{\rm reh}}{a_{\rm rec}} = 8 \times 10^{-11} \left( \frac{100}{g_{*}(T_{\rm reh})} \right)^{1/3} \left( \frac{1 \GeV}{T_{\rm reh}} \right) \, ,
\end{align}
In our analysis, we will assume $g_{*}(T_{\rm reh})=106.75$ as in the case of the SM. It should be noted, however, that this quantity is sensitive to the particle content of the underlying UV-complete theory.

As a final remark, in scenarios where the PT originates in a hidden sector that communicates with the SM via a portal interaction, only a fraction of the released energy is transferred to the visible sector. In particular, in the presence of a Higgs portal interaction [cf.~Eq.~(\ref{eq:elemLag}] in this model, the energy released at the completion of the PT is partially transferred from the ALP sector to the Higgs field, and hence to the SM plasma.
We take into account this effect through an efficiency factor $\kappa_h$ which quantifies the fraction of energy transferred to the Higgs sector. Since the observable magnetic field originates only from the visible plasma, whereas most of the vacuum energy is initially released in the ALP sector, the magnetic energy density is reduced approximately in proportion to the fraction of energy transferred to the SM plasma. Consequently, the magnetic field spectrum $B_0$ in Eq.~(\ref{eq:B0-spectrum}) scales as $\sqrt{\kappa_h}$~\cite{Ellis:2019tjf, Balaji:2024rvo, Balaji:2025tun,ArteagaTupia:2025awh}.
In the present framework, it can be determined by how strongly the Higgs sector participates in the dynamics of the radiatively induced ALP PT. As we have assumed that the Higgs field remains at the origin during the transition, its involvement is entirely mediated by the portal coupling $\lambda_{h2}$. As a result, the fraction of vacuum energy transferred to the SM plasma is expected to scale with the relative strength of Higgs-induced radiative effects compared to those arising from the ALP sector. Therefore, following Eqs.~(\ref{eq:effpot0}) and (\ref{eq:beta-eff}), the efficiency factor $\kappa_h$ in the present model can be approximated as
\bea
\label{eq:kappa-h}
\kappa_h \simeq \frac{\lambda^2_{h2} \sin^4\theta}{6 g^4 \cos^4\theta + \lambda^2_{h2} \sin^4\theta}~.
\eea
Although Eq.~(\ref{eq:kappa-h}) should be regarded as a model-specific estimate of the efficiency factor, it is sufficient for our purpose of assessing the simultaneous production of GWs and PMFs in the present model. 
\subsection{Experimental constraints from blazar emissions}

Blazars are active galactic nuclei with relativistic jets that are roughly
pointed towards us, and they are copious emitters of TeV $\gamma$-rays. These high-energy photons interact with the extragalactic background light, leading to the production of ultra-relativistic $e^{+}e^{-}$ pairs ~\cite{Nikishov:1962rmq, Gould:1966pza, Franceschini:2021wkr}. 
The resulting pairs subsequently up-scatter cosmic microwave background (CMB) photons via the inverse Compton process, generating a secondary population of $\gamma$-rays in the GeV range. This electromagnetic cascade redistributes the initial energy, depleting the TeV flux while enhancing the GeV component of the spectrum. In the presence of IGMF, the charged pairs are deflected by the Lorentz force. For sufficiently large field strengths, this deflection alters the propagation direction of the cascade photons, effectively suppressing the GeV flux along the line of sight~\cite{Vachaspati:2016xji,Vachaspati:2020blt}. The non-observation of the expected GeV signal has therefore been widely used to infer lower limits on the IGMF strength. These limits arise from two complementary effects: the angular broadening of the cascade emission and the time delay between the TeV and secondary GeV signals.

In this work, we adopt representative bounds from primarily two recent analyses of blazar data. Observations by the MAGIC collaboration yield a lower limit of $ B > 1.8 \times 10^{-17} \, \mathrm{G}$ for correlation lengths $\lambda \geq 0.2 \, \mathrm{Mpc}$~\cite{MAGIC:2022piy}. Complementarily, analysis based on \textit{Fermi}-LAT and H.E.S.S. data reports a stronger limit of $ B > 7.1 \times 10^{-16} \, \mathrm{G}$ for $\lambda \geq 1 \, \mathrm{Mpc}$ assuming a characteristic blazar activity timescale of $t_{\rm max} = 10 \, \mathrm{yrs}$~\cite{HESS:2023zwb}. We also include the corresponding blazar limits for duty cycles of $t_{\rm max} = 10^4$ and $10^7$ yrs~\cite{HESS:2023zwb} in our analysis, and these bounds will be employed in the following to delineate the viable parameter space of the model. We also note that a recent revision of the IGMF lower bound from {\it Fermi} and Imaging Atmospheric Cherenkov telescopes (IACT)~\cite{Blunier:2025ddu} suggests a more conservative lower limit of $B_0 \gtrsim 2.1\times10^{-17}~\textrm{G}$ at $\lambda \geq 0.1~\textrm{Mpc}$. In addition to these exclusion limits, we note that a recent likelihood analysis of anisotropic pair halos in 14 years of \textit{Fermi}-LAT data from 21 HBL sources reports a $3.8\sigma$ exclusion of the null IGMF hypothesis, with a best-fit value of $B_0 = 2.8\times10^{-16}$~G and a $99\%$ confidence interval of $[0.9,\,8.9]\times10^{-16}$~G at $\lambda = 1$ Mpc~\cite{Zhang:2026lte}, considering non-helical configuration. Unlike the bounds above, this constitutes a direct measurement and further motivates a primordial, cosmological origin for the IGMF.

\section{Gravitational Waves from ALP Phase Transition}
\label{sec:GWs}

In this section, we briefly recap the FOPT SGWB computations in view of the updates on the SGWB frequency shape recently updated in Ref.~\cite{Caprini:2024hue}. Here, we focus on the regime of strongly supercooled PT, which is the relevant region of interest in our framework involving radiative symmetry breaking; see the discussion below Eq.~(\ref{eq:coeff-a}). During such supercooled phase transitions, one expects the bubble to expand at the speed of light, i.e., $v_w \simeq 1$,\footnote{Particle production on bubble walls may lead to friction and slow the wall, potentially invalidating this assumption~\cite{Shakya:2023kjf}. Here, we assume these effects are negligible, consistent with the strongly supercooled, vacuum-dominated regime where plasma friction is subdominant and the transition proceeds essentially like a simple vacuum decay. GW sourced due to particle production from bubble walls is also neglected as they become relevant only for heavier masses and very very large couplings, see Refs.\cite{Inomata:2024rkt,Ghoshal:2026hev}} and the plasma to play a negligible role in the GW production. In this regime, where the bubble collision mechanism mainly sources the GW signal\footnote{The PQ symmetry breaking with associated axion cosmology can also lead to the formation of topological defects like global cosmic string network and domain wall network, which also contribute to the SGWB~\cite{Gorghetto:2021fsn,Fu:2023nrn,Datta:2025vyu,Ghoshal:2025iil,Zhao:2025hzr}. 
However, it only becomes relevant when the PQ symmetry breaking scale $f_a\gtrsim \mathcal{O}(10^{14})$ GeV.}, the SGWB critical energy density per logarithmic frequency interval is given by~\cite{Caprini:2024hue}
\begin{equation}
h^2\Omega_{\rm GW}(f) \simeq \frac{16 (f/f_p)^{2.4}}{\left[ 1 + (f/f_p)^{1.2} \right]^4} h^2\bar\Omega_{\rm GW} \,,
\label{eq:spectrum}
\end{equation}
where $h \equiv 10^{-2} H_0 \,\text{Mpc} /(\text{km/s})  \simeq 0.674$~\cite{Planck:2018vyg}
denotes the reduced Hubble parameter at present, expressed in terms of the Hubble constant, $H_0$. The quantities denoted by $h^2 \bar\Omega_{\rm GW}$ and $f_p$ correspond to the peak amplitude and peak frequency of the gravitational wave spectrum, respectively, and are given by
\begin{eqnarray}
  h^2 \bar\Omega_{\rm GW} &\simeq& 3.8 \times 10^{-6} \left( \frac{H_\ast}{\beta} \frac{\alpha}{1+\alpha} \right)^2 \frac{1}{\left[g_*(T_{\rm reh})\right]^{1/3}} \,, 
  \label{eq:omegap}\\
  f_p &\simeq& (8.4 \times 10^{-7}~{\rm Hz})  \frac{\beta}{H_\ast} \left(\frac{T_{\rm reh}}{100~{\rm GeV}}\right) \,  \left[g_*(T_{\rm reh})\right]^{1/6} \,.
    \label{eq:fp}
\end{eqnarray}
The resulting SGWB spectrum follows a broken power-law profile in frequency. In the high-frequency regime, $f \gg f_p$, the spectrum scales as $\Omega_{\rm GW}(f) \propto f^{-2.4} $ , showing a significantly steeper fall-off compared to earlier treatments based on Ref.~\cite{Caprini:2019egz}, where the peak spectrum was described by a milder behaviour~$\Omega_{\rm GW}(f) \propto f^{-1} $.
Here, in Eq. (\ref{eq:spectrum}), we adopt the broken power-law template of Ref. \cite{Caprini:2024hue} for strong first-order phase transitions, for which the high-frequency branch is parametrised by $f^{-2.4}$ (cf. Table 1 of Ref.~\cite{Caprini:2024hue}). This exponent is therefore an effective fit to the simulation-based template for bubble collisions and highly relativistic fluid shells, rather than a universal asymptotic prediction.

A wide range of current and future experiments can probe the parameter space spanned by $\{h^2 \bar\Omega_{\rm GW}, f_p\}$. Among them, one important class of constraints arises indirectly from observations of the CMB and BBN, which are sensitive to the total radiation energy density in the Universe. These bounds are usually expressed in terms of the effective number of relativistic species, or relativistic degrees of freedom, denoted by $\Delta N_{\rm eff}=N_{\rm eff}-N_{\rm eff}^{\rm SM}$, parametrising any additional BSM relativistic {\it d.o.f.} at the time of recombination. Given this, one may translate into a limit on the radiation energy density of the form $\Omega_{N_{\rm eff}} \lesssim 5.6\times10^{-6}\;\Delta N_\text{eff}$, with $N_{\rm eff}^{\rm SM}=3.0440(2)$~\cite{Akita:2020szl,Froustey:2020mcq,Drewes:2024wbw}. Since the SGWB contributes as an additional (dark) radiation component, it is subject to the same constraint. 
A recent analysis combining data from BBN, CMB and BAO gives $N_{\rm eff}=2.990\pm 0.070$~\cite{Goldstein:2026iuu}, which can be directly interpreted as an upper bound on the GW energy density today. For the spectrum given in Eq.~\eqref{eq:spectrum}, this translates into the following constraint on $h^2 \bar\Omega_{\rm GW}$:
\begin{align}
    &\int_{f_\text{min}}^{\infty} \frac{\text{d}f}{f}  h^2 \Omega_\text{GW}(f)\simeq 2.22\, h^2\bar\Omega_{\rm GW}  \leq 5.6\times10^{-6}\;\Delta N_\text{eff} \,, \label{eq:darkrad}
\end{align}
where the lower integration limit is set by $f_\text{min}\sim 10^{-10}\, \text{Hz}$ for the BBN constraints and $f_\text{min}\sim 10^{-18}\, \text{Hz}$ for the CMB observations. In practice, for spectra peaking at much higher frequencies, the result is largely insensitive to the precise choice of $f_\text{min}$. Eq.~\eqref{eq:darkrad} yields an upper bound of  
\begin{align}
h^2\bar\Omega_{\rm GW}\leq 1.8\times 10^{-7} \, .
\label{eq:darkrad2}
\end{align}
Next-generation CMB experiments such as CMB-S4~\cite{CMB-S4:2020lpa}, CMB-HD~\cite{Sehgal:2019ewc}, COrE~\cite{COrE:2011bfs} and EUCLID~\cite{EUCLID:2011zbd} are expected to significantly improve the sensitivity to additional radiation components, reaching $\Delta N_{\rm eff}\sim 0.06$ (CMB-S4), 0.027 (CMB-HD), 0.013 (COrE/EUCLID) at 95\% C.L. These projected sensitivities correspond to upper limits of $h^2\bar\Omega_{\rm GW}<1.5\times 10^{-7}$ (CMB-S4), $6.8\times 10^{-8}$ (CMB-HD), $3.3\times 10^{-8}$ (COrE/EUCLID).

Finally, the detectability of an SGWB is typically quantified through the signal-to-noise ratio (SNR), defined as~\cite{Allen:1997ad,Kudoh:2005as,Thrane:2013oya,Caprini:2019pxz,Brzeminski:2022haa}
    \begin{equation}
	(\mathrm{SNR})^2 = t_{\rm obs} \int df\left(\frac{\Omega_{\rm GW}^2}{\Omega_{\rm sens}^2 + 2\Omega_{\rm GW}\Omega_{\rm sens}+	2\Omega_{\rm GW}^2}\right) \, ,
	\label{eq:snr}
	\end{equation}
where $t_{\rm obs}$ denotes the observation time and $\Omega_{\rm sens}$ characterises the detector sensitivity. In the parameter region of interest of our scenario, GWs sourced by a supercooled FOPT lead to signals that can be probed across a broad range of frequencies. Space-based interferometers, such as LISA~\cite{LISA:2017pwj},  DECIGO~\cite{Kawamura:2020pcg}  and BBO~\cite{Corbin:2005ny}
are sensitive to the mHz-dHz regime, while ground-based detectors like the Einstein Telescope (ET)~\cite{Punturo:2010zz} and Cosmic Explorer (CE)~\cite{Reitze:2019iox}
will probe higher frequencies, typically in the $10$-$10^3$ Hz range, and  PTA experiments, including the Square Kilometre Array (SKA)~\cite{Weltman:2018zrl}, THEIA~\cite{Garcia-Bellido:2021zgu} and $\mu$-ARES~\cite{Sesana:2019vho} provide complementary coverage at lower frequencies in the nHz-$\mu$Hz range. Throughout this analysis, we assume a benchmark observation time of $t_{\rm obs}=4$ years for interferometer-based detectors and $t_{\rm obs}=20$ years for PTA-based detectors and SNR $ >10$ to be a GW detection threshold.

\section{Results}
\label{sec:results}
In this section, we present the phenomenological implications of the supercooled PT in the ALP model described in Sec.~\ref{sec:the-model}. We first present our numerical results for the primordial magnetogenesis and GW signal in the ALP model under consideration and then point out the complementarity with existing laboratory constraints on ALPs and future detection prospects.
\subsection{Primordial Magnetic Fields and Gravitational Waves}
\label{subsec:results:PMF-GW}
We compute the properties of the PMFs created during the FOPT, such as their strength and coherence length as outlined in Sec.~\ref{sec:pmf-from-fopt}, and also evaluate the GW signal generated in this transition following Sec.~\ref{sec:GWs}. 
\begin{figure}[!htpb]
	\centering
    {\includegraphics[height=8.0cm,width=11.0cm]{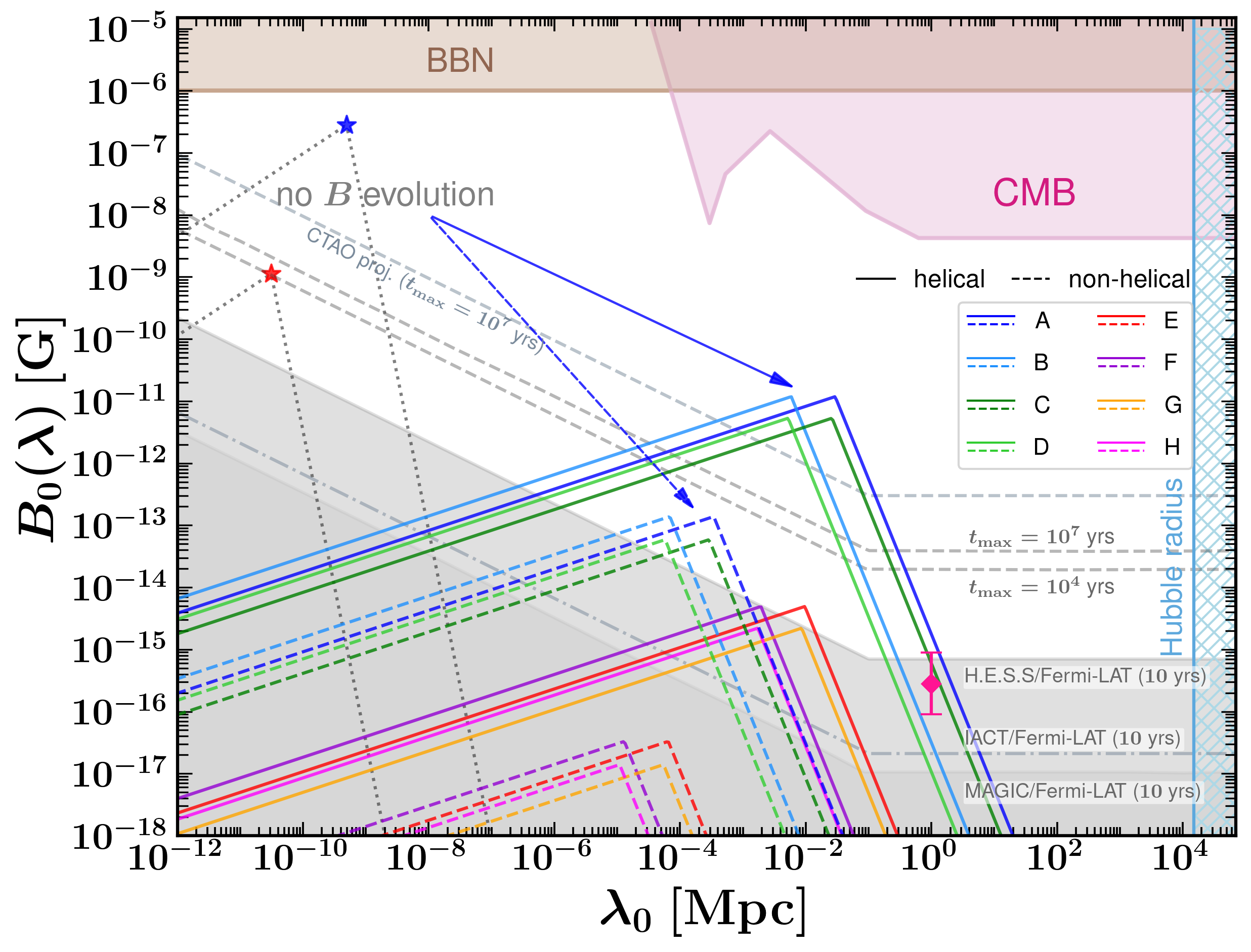}}
	\caption{The predicted PMF strength $B_0$ as a function of its coherence length $\lambda_0$ (today) produced by our ALP-assisted strong FOPT scenario for different choices of parameters given in Table \ref{tab:table-BPs}. The solid (dashed) coloured curves show the cosmological magnetic evolution in the helical (non-helical) limit, while the dotted grey curves (representing BP-A and E) neglect such evolution. The shaded regions on top are excluded by upper limits set from  BBN and CMB observations. The grey-shaded regions on the bottom half correspond to the lower bound on the magnetic field set by MAGIC, H.E.S.S. and {\it Fermi}-LAT~\cite{MAGIC:2022piy, HESS:2023zwb} for a maximum duty cycle of $t_{\rm max} = 10$ yrs (with the dot-dashed grey line showing a reassessment of the limit from {\it Fermi}-LAT and IACT observations~\cite{Blunier:2025ddu}), whereas the grey dashed lines show the uncertainty on the blazar emission timescale up to $t_{\rm max} = 10^4$ and $10^7$ yrs. The CTAO projection~\cite{CTA:2020hii} for $t_{\rm max}=10^7$ yrs is also shown. The deep-pink diamond marks the best-fit non-helical IGMF strength $B_0 = 2.8\times10^{-16}$~G at $\lambda_0 = 1$~Mpc from a recent pair-halo analysis~\cite{Zhang:2026lte}, with the error bar indicating the $99\%$ confidence interval $[0.9,\,8.9]\times10^{-16}$~G. }  
	\label{fig:fig-BPs-no-B-evolution}
\end{figure}

In Fig.~\ref{fig:fig-BPs-no-B-evolution} we show the present-day PMF amplitude, $B_0$, obtained from Eq.~(\ref{eq:B0-spectrum}), as a function of the comoving coherence length $\lambda_0$ for both maximally helical ($b=0$) and non-helical ($b=1$) configurations. We see that the calculated field strength exhibits generically a peak at a coherence scale given by Eq.~(\ref{eq:l0}). The representative benchmark points (BPs) [cf.~Table~\ref{tab:table-BPs}] span different regions of the parameter space by varying the axion decay constant $f_a$, the gauge coupling $g$, and the inverse PT duration parameter $\beta/H$. The dotted grey curves (representing BP-A `$\textcolor{blue}{\bigstar}$' and BP-E `$\textcolor{red}{\bigstar}$') show the magnetic field spectra in the idealised limit where magnetic interactions with the plasma at $a < a_{\rm rec}$ is neglected. In contrast, the coloured solid (dashed) curves include plasma effects and correspond to maximally helical (non-helical) evolution. Helical fields undergo an inverse cascade that transfers power to larger scales, leading to a simultaneous growth of $\lambda_0$ and a slower decay of $B_0$. Consequently, these solutions populate the upper-right region of the plane and yield significantly stronger large-scale fields. Non-helical fields, lacking magnetic helicity conservation, experience direct turbulent decay and therefore remain comparatively weaker.

The grey-shaded areas of Fig.~\ref{fig:fig-BPs-no-B-evolution} correspond to the current lower bounds on the IGMF set by MAGIC, H.E.S.S. and {\it Fermi}-LAT~\cite{MAGIC:2022piy,HESS:2023zwb} assuming a blazar activity timescale of $t_{\rm max}=10$ yrs. Since there is a large uncertainty on this timescale, we also show the corresponding {\it Fermi}-LAT/H.E.S.S. bounds for larger blazar emission timescales up to ${t_{\rm max}} = 10^4$ and $10^7$ yrs~\cite{HESS:2023zwb}. We further include the recent reassessment of the IGMF lower bound based on \textit{Fermi}-LAT and IACT data~\cite{Blunier:2025ddu}. The projected sensitivity from CTAO~\cite{CTA:2020hii} for $t_{\rm max} = 10^7$ yrs is also shown (similar IGMF sensitivity is expected from GRAINE~\cite{Takahashi:2016xsf, Zhang:2026lte}).  
A subset of the helical benchmarks intersect or approach these regions, demonstrating that an ALP-assisted supercooled PT can naturally generate IGMFs compatible with current observational limits, whereas the non-helical scenarios typically fall short. 

We also show the recent $3.8\sigma$ evidence for a non-zero  IGMF~\cite{Zhang:2026lte}, based on anisotropic pair-halo searches in 14 years of \textit{Fermi}-LAT data from 21 HBL sources, which provides a direct measurement rather than a lower bound, favouring a best-fit value of $B_{0} = 2.8 \times 10^{-16} \text{ G}$ at $\lambda_0 = 1$ Mpc. We display this value by `$\textcolor{deeppink}{\blacklozenge}$' in Fig.~\ref{fig:fig-BPs-no-B-evolution} and the corresponding $97\%$ confidence interval, i.e., $[0.5, 10] \times 10^{-16}~\text{G}$, is indicated by the error bar. This measurement lends further observational support to a cosmological origin of the IGMF, consistent with its generation due to an FOPT. It is important to emphasise that Ref.~\cite{Zhang:2026lte} does not currently provide a scaling relation of $B_0$ with coherence length. Extending this result to include a scaling behaviour, as traditionally seen in blazar exclusion limits (e.g., Eq.~(4) of Ref.~\cite{Blunier:2025ddu}), could potentially bring a subset of our non-helical benchmark configurations within the corresponding confidence interval. Exploring such scaling behaviour would be an interesting direction for future work.

We also display in the same plot the observational constraints on the magnetic field from CMB (light purple-shaded region)~\cite{Yamazaki:2012pg, Planck:2015zrl}\footnote{Stronger CMB limits have been derived in Ref.~\cite{Jedamzik:2018itu} using baryonic density fluctuations induced by PMFs. In fact, a recent analysis combining different cosmological datasets finds a mild to moderate preference for present-day total magnetic field strengths of approximately 5-10 pG~\cite{Jedamzik:2025cax}, which could potentially relieve the Hubble tension~\cite{Jedamzik:2020krr}. However, Ref.~\cite{Thiele:2021okz} does not find any evidence of baryon clumping.} and BBN (light brown-shaded band)~\cite{Grasso:2000wj,Kawasaki:2012va}  measurements. We note that there is no strict upper bound on the IGMF coherence length, in particular if it is of inflationary origin~\cite{Durrer:2013pga}. Nevertheless, correlations cannot be probed on scales exceeding the present Hubble radius, shown by the ``Hubble radius'' line in Fig.~\ref{fig:fig-BPs-no-B-evolution}. 
\begin{table}[!htpb]
\hspace*{-0.75cm}
\small
\centering
\renewcommand{\arraystretch}{1.5}
\setlength{\tabcolsep}{6.5pt}
\begin{tabular}{|c|c|c|c|c|c|c|c|c|}
\hline 
\multirow{2}{*}{BP} & \multirow{2}{*}{$f_a$ [GeV] } & \multirow{2}{*}{$g$} & \multirow{2}{*}{$\beta/H$} & \multirow{2}{*}{$f_p$ [Hz]} & \multicolumn{2}{c|}{peak $B_0$ [G]} & \multicolumn{2}{c|}{peak $\lambda_0$ [Mpc]} \\ \cline{6-9}
& & & & & ~~~helical~~~ & non-helical & ~~~helical~~~ & non-helical\\
\hline
A & $10^{3.0}$ & 0.85 & 10 & $1.72 \times 10^{-5}$ & $8.91 \times 10^{-12}$ & $1.04 \times 10^{-13}$ & $2.71 \times 10^{-2}$ & $3.23 \times 10^{-4}$ \\
B & $10^{3.0}$ & 0.85 & 50 & $8.57 \times 10^{-5}$ & $8.97 \times 10^{-12}$ & $1.02 \times 10^{-13}$ & $5.51 \times 10^{-3}$ & $6.14 \times 10^{-5}$ \\
C & $10^{3.0}$ & 1.20 & 10 & $2.42 \times 10^{-5}$ & $4.06 \times 10^{-12}$ & $4.34 \times 10^{-14}$ & $2.53 \times 10^{-2}$ & $2.62 \times 10^{-4}$ \\
D & $10^{3.0}$ & 1.20 & 50 & $1.21 \times 10^{-4}$ & $3.99 \times 10^{-12}$ & $4.37 \times 10^{-14}$ & $4.80 \times 10^{-3}$ & $5.34 \times 10^{-5}$ \\
E & $10^{4.5}$ & 0.85 & 10 & $5.42 \times 10^{-4}$ & $4.90 \times 10^{-15}$ & $3.24 \times 10^{-17}$ & $9.59 \times 10^{-3}$ & $6.14 \times 10^{-5}$ \\
F & $10^{4.5}$ & 0.85 & 50 & $2.71 \times 10^{-3}$ & $4.83 \times 10^{-15}$ & $3.26 \times 10^{-17}$ & $1.95 \times 10^{-3}$ & $1.25 \times 10^{-5}$ \\
G & $10^{4.5}$ & 1.20 & 10 & $7.60 \times 10^{-4}$ & $2.18 \times 10^{-15}$ & $1.38 \times 10^{-17}$ & $8.35 \times 10^{-3}$ & $5.34 \times 10^{-5}$ \\
H & $10^{4.5}$ & 1.20 & 50 & $3.83 \times 10^{-3}$ & $2.19 \times 10^{-15}$ & $1.39 \times 10^{-17}$ & $1.70 \times 10^{-3}$ & $1.09 \times 10^{-5}$ \\
\hline
\end{tabular}
\caption{Representative benchmark points and their corresponding peak values of $B_0$ and $\lambda_0$ for both helical and non-helical configurations. The peak GW frequency $f_p$ is also provided. In addition to $(f_a, g)$, the parameter $\beta/H$ is treated as an input to assess its impact on the magnetic field spectrum (see Fig.~\ref{fig:fig-BPs-no-B-evolution}). 
}
\label{tab:table-BPs}
\end{table}

From Fig.~\ref{fig:fig-BPs-no-B-evolution}, it can be seen that for a fixed value of the parameters $(f_a,g)$, varying the inverse duration parameter $\beta/H$ (see Table \ref{tab:table-BPs})  primarily shifts the location of the spectral peak along the coherence-length axis, while leaving the peak amplitude $B_0$ nearly unchanged for both helical and non-helical configurations. This behaviour is expected because $\beta/H$ mainly controls the characteristic length scale of the source [cf.~Eq.~(\ref{eq:coherence-l0-betaH})], whereas the total fraction of released energy converted into magnetic fields is only weakly affected. Consequently, faster transitions (larger $\beta/H$) yield spectra peaked at smaller $\lambda_0$, while slower transitions move the peak toward larger coherence scales.
\begin{figure}[!htpb]
	\centering
    \subfigure[\label{fig:fig-fa-g-blazar-a}]{\includegraphics[height=8.0cm,width=11.5cm]{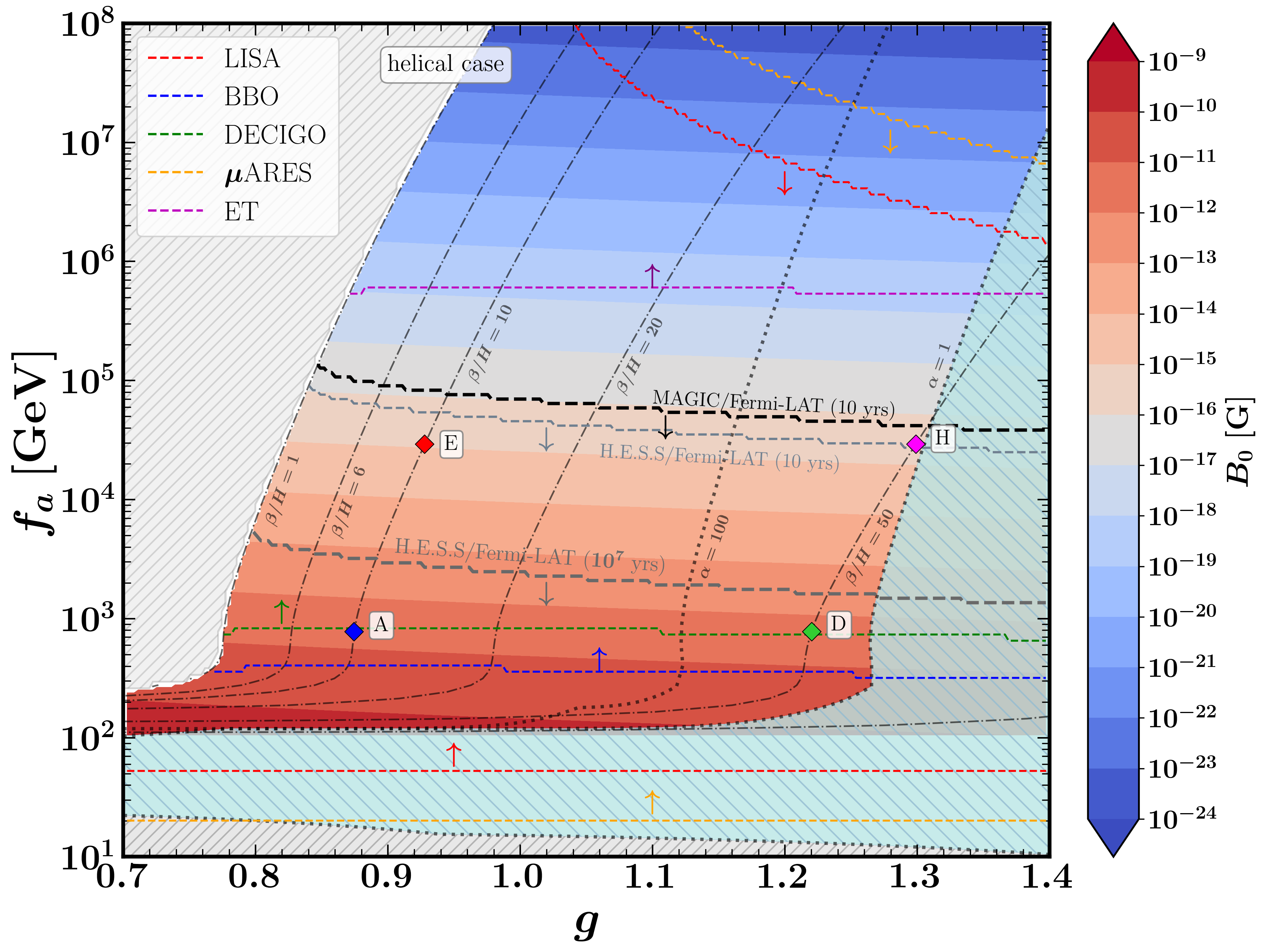}}
    \hspace*{0.10cm}
    \subfigure[\label{fig:fig-fa-g-blazar-b}]{\includegraphics[height=8.0cm,width=11.5cm]{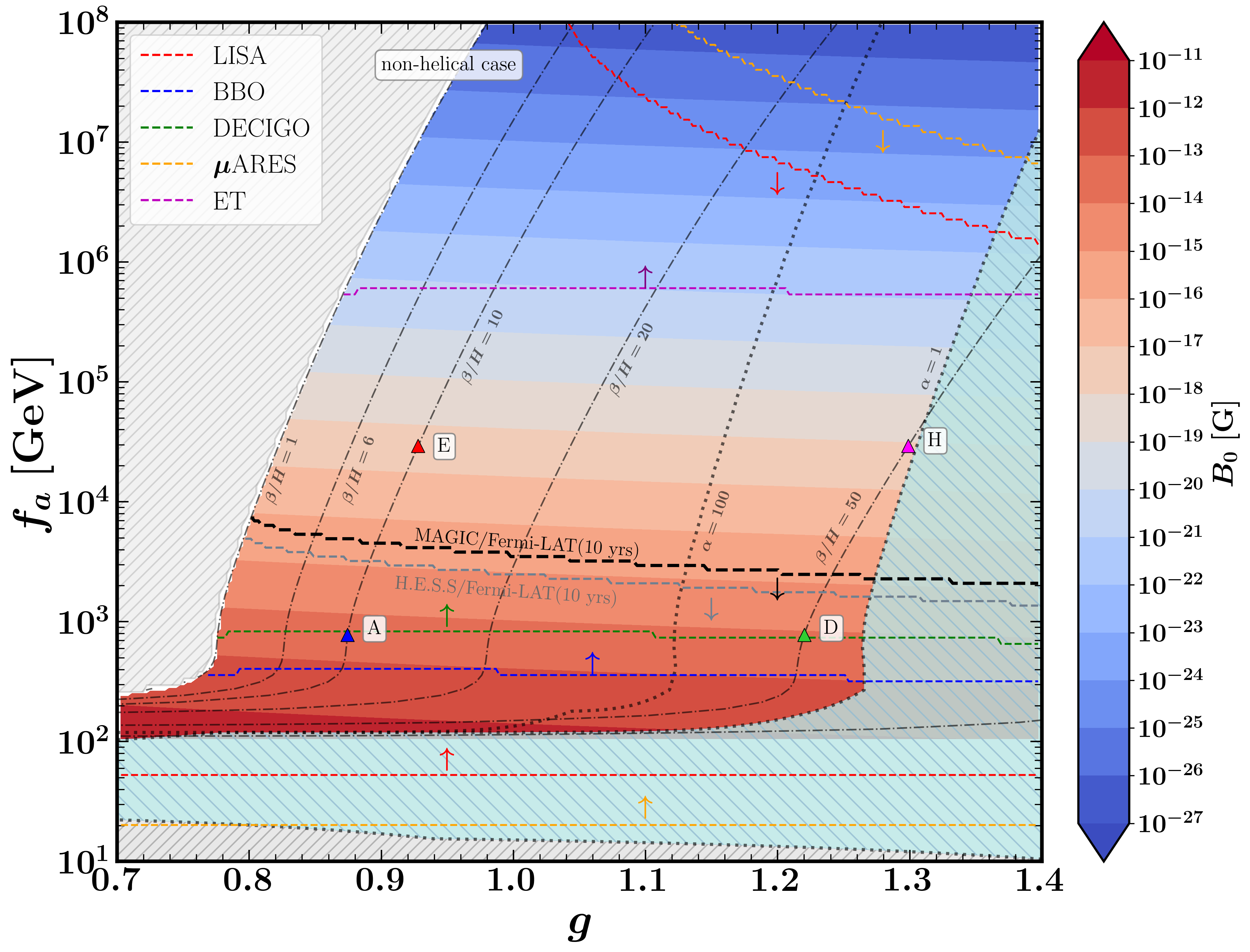}}
	\caption{PMF and GW predictions in our scale-invariant ALP model, shown in the $(g, f_a)$ plane. The coloured contours indicate the present-day magnetic field strength $B_0(\lambda_0)$ generated by the ALP-assisted FOPT. The field strength, computed assuming helical (\textbf{upper panel}) and non-helical (\textbf{lower panel}) evolution, is sufficient to account for the blazar observations below the black (MAGIC/{\it Fermi}-LAT~\cite{MAGIC:2022piy}) and grey (H.E.S.S./{\it Fermi}-LAT~\cite{HESS:2023zwb}) thick curves. The associated GW signals, yielding SNR $> 10$, can be observed at various GW detectors in the regions denoted by coloured dashed lines and arrows, with the corresponding experiments indicated in the legend. Representative contours of $\alpha$ and $\beta/H$, as well as selected BPs from Table \ref{tab:table-BPs}, are also shown. The light-cyan and grey-shaded regions correspond to $\alpha<1$ and absence of an  FOPT, respectively. See text for further details.
    }
	\label{fig:fig-fa-g-blazar}
\end{figure}

On the other hand, increasing the axion decay constant $f_a$ leads to a pronounced suppression of the peak field strength. This follows directly from the scaling $B_0 \propto \sqrt{\kappa_h}$ together with the dependence of the efficiency factor $\kappa_h$ on $f_a$ via $\lambda_{h2}$ [cf.~Eq.~(\ref{eq:kappa-h})]. A larger $f_a$ corresponds to a smaller $\lambda_{h2}$, which reduces $\kappa_h$, and hence lowers the efficiency of gauge-field amplification. The resulting decrease in magnetic energy density manifests as a sharp downward shift of the entire spectrum, most visibly at the peak. The gauge coupling $g$ produces a qualitatively similar but milder effect. Increasing $g$ also suppresses the final magnetic amplitude, however the dependence is less steep than for $f_a$, leading to a moderate reduction of $B_0$. Therefore, among the parameters, the spectrum is most sensitive to variations in $f_a$, while $g$ acts as a subdominant correction and $\beta/H$ predominantly sets the characteristic length scale.

Fig.~\ref{fig:fig-fa-g-blazar} represents the parameter space of the ALP model projected onto the plane defined by the gauge coupling constant $g$ and the ALP decay constant $f_a$. The coloured contours indicate the values of the present-day PMF strength, $B_0 (\lambda_0)$, generated by the ALP-assisted FOPT. Fig.~\ref{fig:fig-fa-g-blazar-a} corresponds to the evolution of the magnetic field with maximal magnetic helicity ($b=0$), while Fig.~\ref{fig:fig-fa-g-blazar-b} is for a negligible or non-helical magnetic helicity ($b = 1$) configuration. For both panels, we also display the contour lines of several $\beta/H$ values as dash-dotted grey curves. The grey-shaded regions in Fig.~\ref{fig:fig-fa-g-blazar} correspond to parameter choices for which the transition to the global minimum never completes, such that the system remains trapped in the metastable, inflating phase. These regions are therefore excluded from further consideration. We also overlay contours for $\alpha = 10^2$ and $\alpha = 1$ (black dotted lines). The $\alpha = 1$ contour approximately delineates the boundary below which the supercooling assumption breaks down. Since our GW forecasts rely on a supercooled FOPT, the region with $\alpha < 1$, indicated by the light-cyan shaded region, lies outside the regime of validity of our analysis. Consistency with the supercooling assumption imposes a lower bound on $f_a \gtrsim 10^2$ GeV, as evident from the figure. Thus, it is difficult in our model to explain the recent observation of stochastic GW by PTAs~\cite{NANOGrav:2023gor, EPTA:2023fyk}, for which the best-fit value of $f_a$ lies below 1 GeV~\cite{Conaci:2024tlc}. 

We further account for the constraints from the effective dark radiation bounds during BBN and CMB decoupling [cf.~Eq.~\eqref{eq:darkrad2}].
Projecting this constraint onto the $(g, f_a)$ parameter space of our framework, we found that the only potentially relevant restriction arises from the future CMB-HD, and only in the regime $\beta/H \lesssim 2$. However, this parameter region nearly coincides with the domain in which the FOPT fails to complete. As discussed earlier, such configurations are already excluded from our analysis.

We also project the SGWB detection reach of future GW experiments onto the $(g,f_a)$ plane. In particular, we show the sensitivity curves for LISA (red dashed line), BBO (blue dashed line), DECIGO (green dashed line), $\mu$ARES (orange dashed line) and ET (magenta dashed line), and compute the regions of the plane ($g, f_a$) that satisfy ${\rm SNR} > 10$, following Eq.~(\ref{eq:snr}). The parameter region that is favoured by the SNR criterion is represented by upward and downward arrows, as shown in the figure. 

Finally, we display the lower limits on the present-day magnetic field strength $B_0(\lambda_0)$, i.e., the peak value of $B_0(\lambda)$ [cf.~Eq.~\eqref{eq:B0-spectrum} and Fig.~\ref{fig:fig-BPs-no-B-evolution}] for any given choice of $(g,f_a)$. The bounds set by the combined MAGIC and {\it Fermi}-LAT~\cite{MAGIC:2022piy} are shown as the thick black dashed lines, while those obtained from H.E.S.S. and {\it Fermi}-LAT~\cite{HESS:2023zwb} observations are indicated by the thin light-grey dashed lines. In both cases, the maximal duty cycle of observation is taken as $t_{\rm max} = 10$ yrs. The corresponding constraint assuming $t_{\rm max} = 10^7$ yrs~\cite{HESS:2023zwb} is indicated by the thick light-grey dashed line in the upper panel plot only.

From Fig.~\ref{fig:fig-fa-g-blazar}, we can see that the coloured contours representing the present-day magnetic field strength $B_0$ are nearly horizontal. This is because $B_0$ is determined by the reheating temperature [cf.~Eqs.~(\ref{eq:B0-spectrum}) and (\ref{eq:red-shift-factors})], which in the current model approximately scales as $T_{\rm reh} \propto \left( 6 g^4 + \lambda^2_{h2} \right)^{1/4} f_a$. Assuming a helical magnetic field configuration (Fig. \ref{fig:fig-fa-g-blazar-a}), the maximum strength of the generated magnetic field can reach up-to $B_0^{\rm max} \approx 10^{-9}$ G within the allowed parameter space of the ($g, f_a$) plane. In contrast, for a non-helical configuration (Fig.~\ref{fig:fig-fa-g-blazar-b}), the maximal achievable field strength is significantly smaller, $B_0^{\rm max} \approx 10^{-11}$ G.

The plots clearly demonstrate that blazar observations impose complementary constraints on the ALP parameter space that are accessible to future GW experiments. In particular, the lower bounds on the IGMF inferred from blazar data severely restrict the viable region in the ($g, f_a$) plane. Quantitatively, in the maximally helical case (upper panel) the blazar bounds require $f_a \lesssim 10^5$ GeV, while in the non-helical configuration (lower panel), it corresponds to $f_a \lesssim 10^4$ GeV. In addition, we find the corresponding present-day peak coherence length spans approximately from $\sim 10^{-3}$ to $\sim 10^{-1}$ Mpc for helical magnetic field configuration, while in the non-helical case it lies between $\sim 10^{-5}$ and $\sim 10^{-2}$ Mpc. 

Forthcoming space-based interferometer detectors, including LISA, BBO, DECIGO, and $\mu$ARES, are capable of probing substantial portions of the parameter space. In particular, LISA and $\mu$ARES can test essentially the entire region compatible with the blazar bounds in both the helical and non-helical scenarios. BBO and DECIGO are sensitive to regions with $f_a \gtrsim 10^{2.4}$ GeV and $f_a \gtrsim 10^{2.8}$ GeV, respectively, although their accessible parameter space is significantly reduced in the non-helical configuration (Fig.~\ref{fig:fig-fa-g-blazar-b}) compared to the maximally helical case (Fig.~\ref{fig:fig-fa-g-blazar-a}). By contrast, future ground-based interferometers, such as ET, are sensitive primarily to regions with $f_a \gtrsim 10^6$ GeV. However, these values lie well above the range allowed by the blazar constraints, and therefore, these detectors cannot probe the parameter region that simultaneously accounts for the blazar observations, in either the helical or non-helical scenarios.

We have also displayed a few BPs of Table \ref{tab:table-BPs} in Fig.~\ref{fig:fig-fa-g-blazar}, chosen to represent characteristic regions of the parameter space. In particular, the sets (A, D) and (E, H) probe two distinct scales of $f_a$, while the paired choices within each set isolate the effect of varying ($g, \beta/H$). In the maximally helical case (Fig.~\ref{fig:fig-fa-g-blazar-a}), these benchmarks lie within or close to the region compatible with current blazar bounds and future GW sensitivities. In contrast, in the non-helical scenario (Fig.~\ref{fig:fig-fa-g-blazar-b}), benchmarks with larger $f_a$ (E and H) yield suppressed magnetic fields irrespective of $(g, \beta/H)$, placing them outside the blazar lower limit. We also note that the points D and H lie close to the $\alpha = 1$ contour, marking the boundary of the supercooling regime, and hence should be interpreted as configurations approaching the limit of validity of our supercooled PT analysis. The inclusion of these BPs in Fig.~\ref{fig:fig-fa-g-blazar} therefore provides a direct correspondence between the spectral features shown in Fig.~\ref{fig:fig-BPs-no-B-evolution} and their embedding in the underlying ALP parameter space. In doing so, they also explicitly illustrate the parametric trends governing the simultaneous realisation of observable magnetic fields and GW signals as discussed earlier.

These observations allow us to conclude that GW production and PMF generation are strongly correlated in this framework. In particular, for a maximally helical configuration and within the parameter range of $10^3~{\rm GeV} \lesssim f_a \lesssim 10^5~{\rm GeV}$ and $0.80 \lesssim g \lesssim 1.25$, future space-based GW experiments, such as  LISA, BBO, DECIGO, and $\mu$ARES, will be able to probe essentially the entire region of parameter space in which the generated magnetic field is sufficiently strong to satisfy the blazar constraints. For the non-helical case, it limits $f_a$ within a narrower range, $10^3~{\rm GeV} \lesssim f_a \lesssim 10^4~{\rm GeV}$.

\subsection{Prospects from precision Higgs data at future colliders}
In addition to the constraints from GW searches and PMF observations, precision measurements of the SM Higgs trilinear self-coupling at high-energy colliders provide an additional and complementary probe of our Higgs-portal ALP parameter space. The relative deviation in the Higgs cubic coupling, $\delta \kappa_\lambda \equiv (\lambda_3-\lambda_3^{\rm SM})/\lambda_3^{\rm SM}$, in our framework can be obtained from the temperature-independent effective potential, after integrating out the $\sigma$-field, and is given as\footnote{Here, we have taken the same assumptions and approximations on the model parameters as discussed in Sec.~\ref{sec:the-model}.}~\cite{Dev:2019njv}
\bea
\label{eq:cubic-coupling-deviation}
\delta \kappa_\lambda \simeq \frac{16 \pi^2}{3} \frac{m^2_h v^2}{g^4 f^4_a},
\eea
where the SM contribution is $\lambda^{\rm SM}_3 = 3 m^2_h/v$~\cite{Yao:2013ika,Huang:2015tdv}. Eq.~(\ref{eq:cubic-coupling-deviation}) exhibits a decoupling behaviour scaling as $f_a^{-4}$, and therefore, the deviation $\delta\kappa_\lambda$ decreases rapidly with increasing $f_a$.
\begin{figure}[!t]
	\centering
    {\includegraphics[height=6.8cm,width=9.3cm]{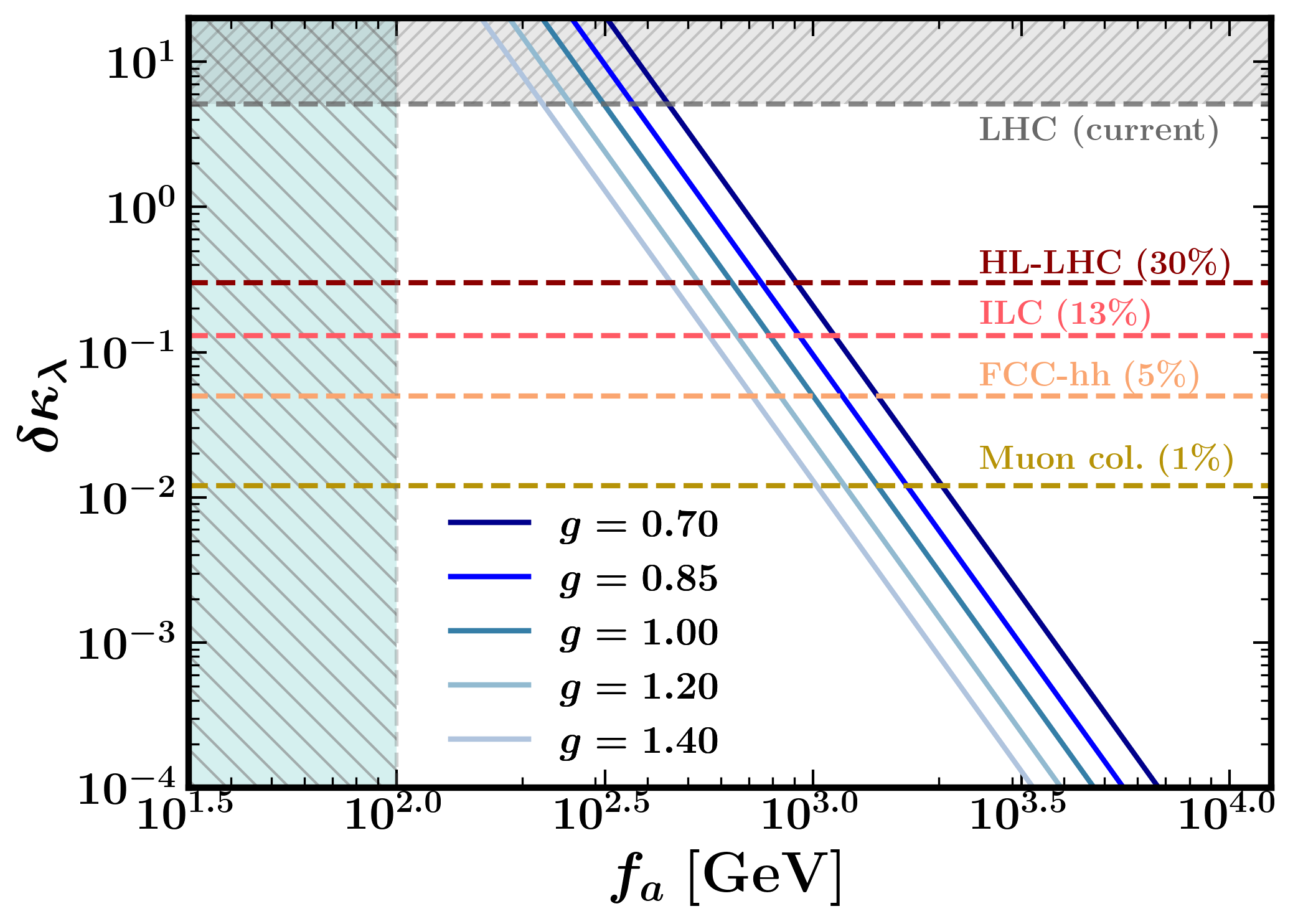}}
	\caption{Deviations in the trilinear self-coupling of the SM Higgs boson as predicted in our model [cf.~Eq.~(\ref{eq:cubic-coupling-deviation})] as a function of $f_a$ for different values of the hidden gauge coupling (coloured solid lines).  Projected sensitivities are shown for HL-LHC ($\sqrt{s}=14~\text{TeV},~3~\text{ab}^{-1}$)~\cite{ATLAS:2013rir, CidVidal:2018eel}, the ILC ($\sqrt{s}=1~\text{TeV},~2.5~\text{ab}^{-1}$)~~\cite{Asner:2013psa, Tian:2013yda}, the FCC-hh ($\sqrt{s}=100~\text{TeV},~30~\text{ab}^{-1}$)~\cite{Barr:2014sga}, and the Muon collider ($\sqrt{s}=30~\text{TeV},~90~\text{ab}^{-1}$)~\cite{Aime:2022flm,Castelli:2025mqk} with coloured dashed lines. For comparison, we also indicate the current constraint from the combined ATLAS and CMS analysis with $\sqrt{s}=13~\text{TeV},~140~\text{fb}^{-1}$ data~\cite{CMS:2026nuu} indicated by the grey-shaded horizontal region.  The light-cyan shaded vertical region with $f_a \leq 10^{2}$ GeV corresponds to $\alpha < 1$ (see Fig.~\ref{fig:fig-fa-g-blazar}), where supercooling may not happen efficiently.}
	\label{fig:trilinear-Higgs-probe}
\end{figure}

As shown in Fig.~\ref{fig:trilinear-Higgs-probe}, for relatively small values of the decay constant, $f_a \sim 10^{2.5} - 10^{3.0}$ GeV, the deviation can be sizeable, reaching $\mathcal{O}(1)$ values depending on the gauge coupling $g$. Such large deviations are already excluded by current LHC Higgs data (i.e., light-grey shaded region with $\delta \kappa_\lambda > 5.1$)~\cite{CMS:2026nuu} and will be further probed by future colliders, including the HL-LHC ($\sim 30-50\%$)~\cite{ATLAS:2013rir, CidVidal:2018eel}, 
ILC ($\sim 13-27\%$)~\cite{Asner:2013psa, Tian:2013yda}, CLIC ($\sim 10–15\%$)~\cite{Abramowicz:2016zbo}, FCC-hh ($\sim 5\%$)~\cite{Barr:2014sga} and Muon collider ($\sim 1\%$)~\cite{Aime:2022flm,Castelli:2025mqk}. However, as $f_a$ increases, the deviation rapidly decreases, falling below $10^{-2}$ for $f_a \gtrsim \mathcal{O}(10^3)$ GeV, and reaching values well below $10^{-4}$ in the parameter region relevant for simultaneously observable magnetic field and GW signals (see Fig.~\ref{fig:fig-fa-g-blazar}).

Therefore, while precision Higgs measurements can probe and already constrain the low-$f_a$ ($\lesssim \mathcal{O}(100~\text{GeV})$)  region of the parameter space, they do not impose competitive bounds on the region relevant for magnetogenesis and GW production in this framework. Consequently, cosmological observables such as PMF and SGWB  remain the dominant probes of the phenomenologically viable parameter space.
\subsection{Complementary with laboratory and astrophysics ALP searches}

Although in our model framework, the ALP {\it a} has no direct coupling to the SM Higgs $H$ and the complex scalar singlet $\phi_2$ at the renormalisable level [cf.~Eq.~(\ref{eq:elemLag})], at low energies, effective interactions with SM fields can generically be induced through higher-dimensional operators. For instance, below the weak scale, the leading $\mathcal{CP}$-conserving effective couplings of {\it a} to the SM photon, gluon and fermion can be written as\footnote{In general, the ALP may also couple to other gauge bosons ($W, Z$), and can admit $\mathcal{CP}$-violating interactions, e.g.,~\cite{Bauer:2017ris,Bisal:2025jwv}. For definiteness, we restrict ourselves to the $\mathcal{CP}$-conserving ALP couplings to photons, gluons and fermions in the present study.}~\cite{Dev:2019njv,Bauer:2020jbp}
\bea
\label{eq:ALP-SM-couplings}
\mathcal{L}_{a} \supset - C_{{\it a}\gamma \gamma} \frac{\alpha_{\rm em}}{8 \pi} \frac{a}{f_a} F_{\mu \nu} \Tilde{F}^{\mu \nu} - C_{{\it a}gg} \frac{\alpha_s}{8 \pi} \frac{a}{f_a} G^a_{\mu \nu} \Tilde{G}^{\mu \nu, a} + \frac{\partial^{\mu} a}{2 f_a} \sum_{f} C_{{\it a}f f} \left( \bar{f} \gamma_\mu \gamma_5 f \right).
\eea
Here, $F_{\mu \nu}$ and $G^a_{\mu \nu}$ denote the electromagnetic and $SU(3)_c$ field-strength tensors, respectively. Their dual tensors are defined as $\Tilde{X}^{\mu \nu} = \tfrac{1}{2} \epsilon^{\mu \nu \rho \sigma} X_{\rho \sigma}$,  with $X = F, G$ and $\epsilon^{0123} = 1$ being the Levi-Civita tensor. The parameters $\alpha_{\rm em}$ and $\alpha_s$ correspond to the electromagnetic fine-structure constant and strong coupling constant respectively, while $C_{{\it a} \gamma \gamma}, C_{{\it a} g g}$ and $C_{{\it a} f f }$ are model-dependent dimensionless Wilson coefficients. For phenomenological convenience, we can define the corresponding effective ALP couplings appearing in Eq.~(\ref{eq:ALP-SM-couplings}), which follow an explicit dependence on the scale $f_a$. We define the effective ALP-photon coupling as $g_{{\it a} \gamma \gamma} = C_{{\it a} \gamma \gamma} {\alpha_{\rm em}}/{2 \pi f_a}$~\cite{Dev:2019njv}, while the ALP-gluon coupling can be written as $g_{{\it a} g g} = C_{{\it a} g g} / f_G$, with $f_G = 4 \pi^2 f_a$~\cite{Kelly:2020dda}, and the ALP-fermion coupling is given by $g_{{\it a} f f} = C_{{\it a} f f}/ f_a$~\cite{Ferber:2022rsf}.
In general, the Wilson coefficients $C_{{\it a} \gamma \gamma}, C_{{\it a} g g}, C_{{\it a} f f}$ are expected to be $\mathcal{O}(1)$ in typical QCD axion UV constructions, such as the KSVZ~\cite{Kim:1979if,Shifman:1978by}, DFSZ~\cite{Dine:1981rt,Zhitnitsky:1980tq} and composite axion~\cite{Choi:1985cb} models (see e.g., Ref.~\cite{DiLuzio:2020wdo} for a review). Therefore, for simplicity, we also set them to one\footnote{Ref.~\cite{Ferber:2022rsf} includes the RG evolution of $C_{aff}/f_a$ when deriving experimental limits in the $(m_a, g_{aff})$ plane. We adopt their bounds on $g_{aff}$, but consistently fix $C_{aff}=1$ when projecting our GW and PMF limits onto the $(m_a, g_{aff})$ plane (see Fig.~\ref{fig:lab-search:full-ALPfermion-close-up}), consistent with the treatment in Ref.~\cite{Alimena:2025kjv}.} throughout our analysis~\cite{Dev:2019njv, Kelly:2020dda, Alimena:2025kjv}.

\begin{figure}[!t]
	\centering
	\includegraphics[width=\textwidth]{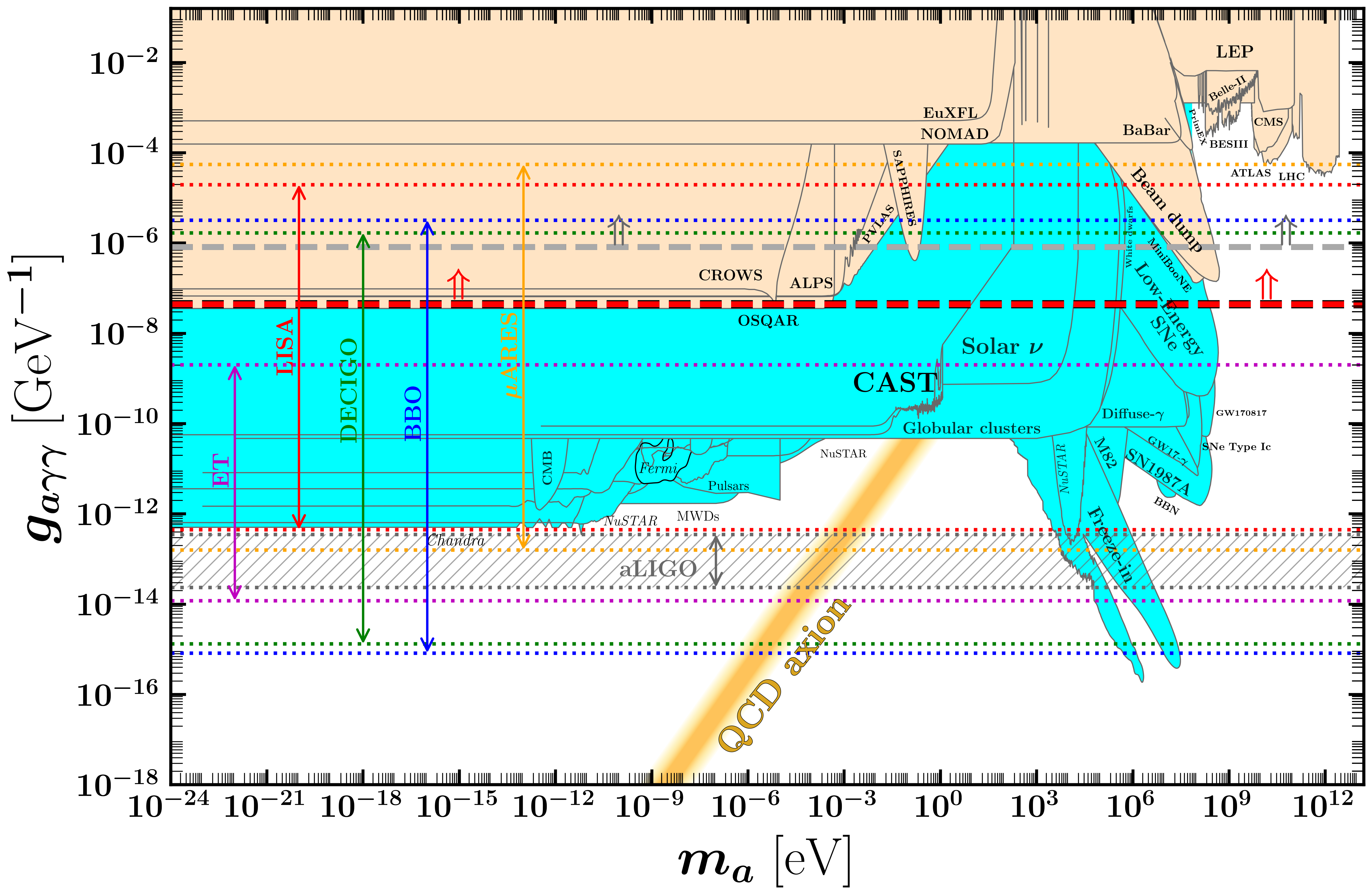} 
	\caption{Complementarity between IGMF observations, projected GW sensitivities, and constraints on ALP-photon coupling $g_{a\gamma\gamma}$ from laboratory, astrophysical and cosmological probes. The region above the red (light-grey) dashed line is favoured by the blazar observations on the magnetic field strength set by Ref.~\cite{HESS:2023zwb} for $t_{\rm max} = 10~(10^7)$ yrs, assuming a maximally helical magnetic field configuration. The GW prospects are indicated by the differently coloured dotted lines, while the hatched grey region is already excluded by aLIGO. Existing ALP constraints, taken from Ref.~\cite{AxionLimits}, are displayed as filled coloured regions. See text for more details.
    }
	\label{fig:lab-search:full-ALP-noDM}
\end{figure}

\begin{figure}[t!]
	\centering
	\includegraphics[width=\textwidth]{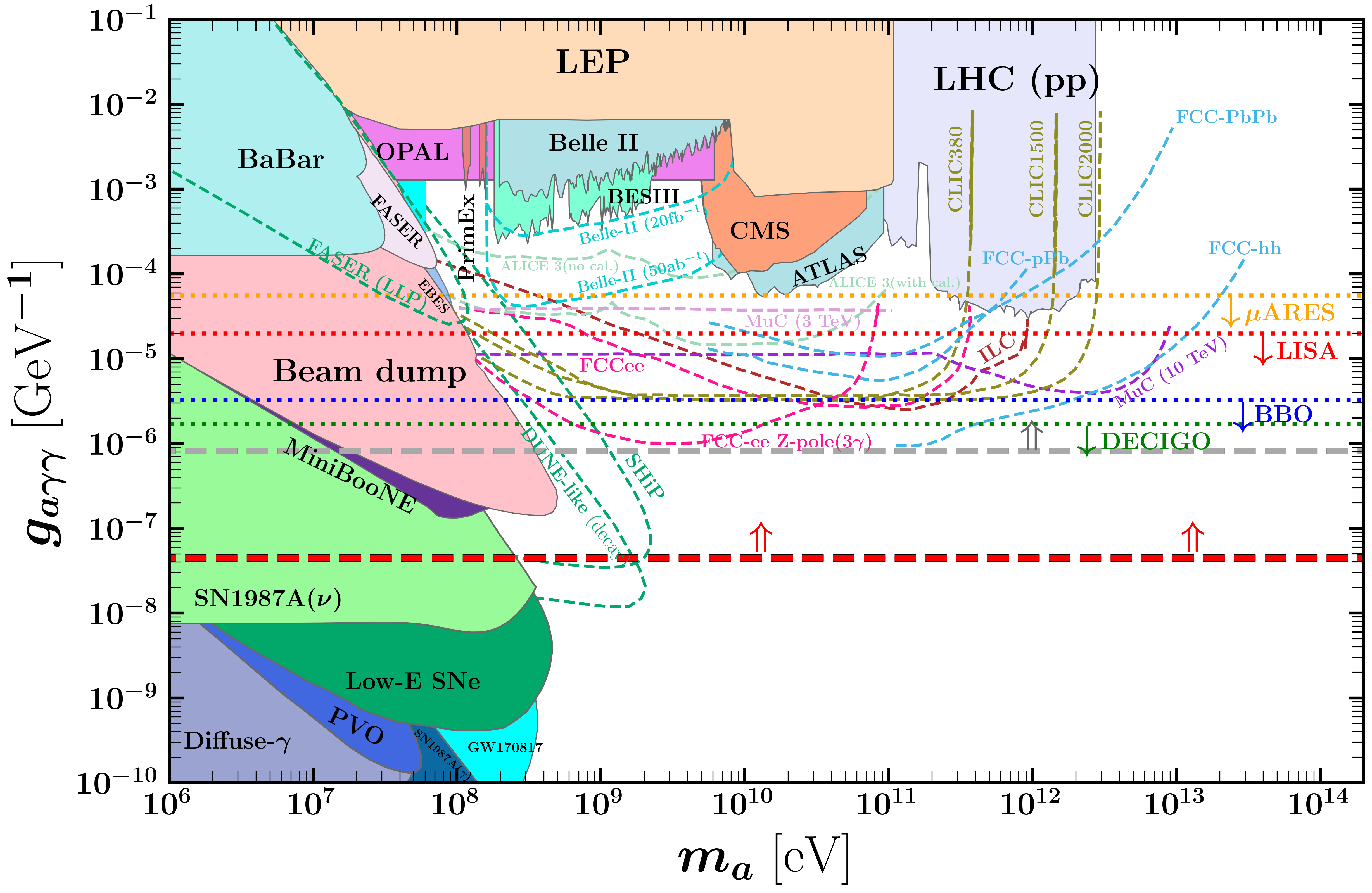} 
	\caption{A zoomed-in view of Fig.~\ref{fig:lab-search:full-ALP-noDM} in the heavier ALP mass region. Also shown are the current bounds and future projections for ALP-photon coupling at $90\%$ C.L. (except for FCC-ee/hh, which are shown at $95\%$ C.L.). Current bounds are represented as filled coloured areas (as in Fig.~\ref{fig:lab-search:full-ALP-noDM}), while future projections are indicated by coloured dashed lines. The limits are primarily compiled from Refs.~\cite{AxionLimits,Alimena:2025kjv,deBlas:2025gyz}. 
    }
	\label{fig:lab-search:full-ALPphoton-close-up}
\end{figure}

\begin{figure}[!t]
	\centering
	\includegraphics[width=\textwidth]{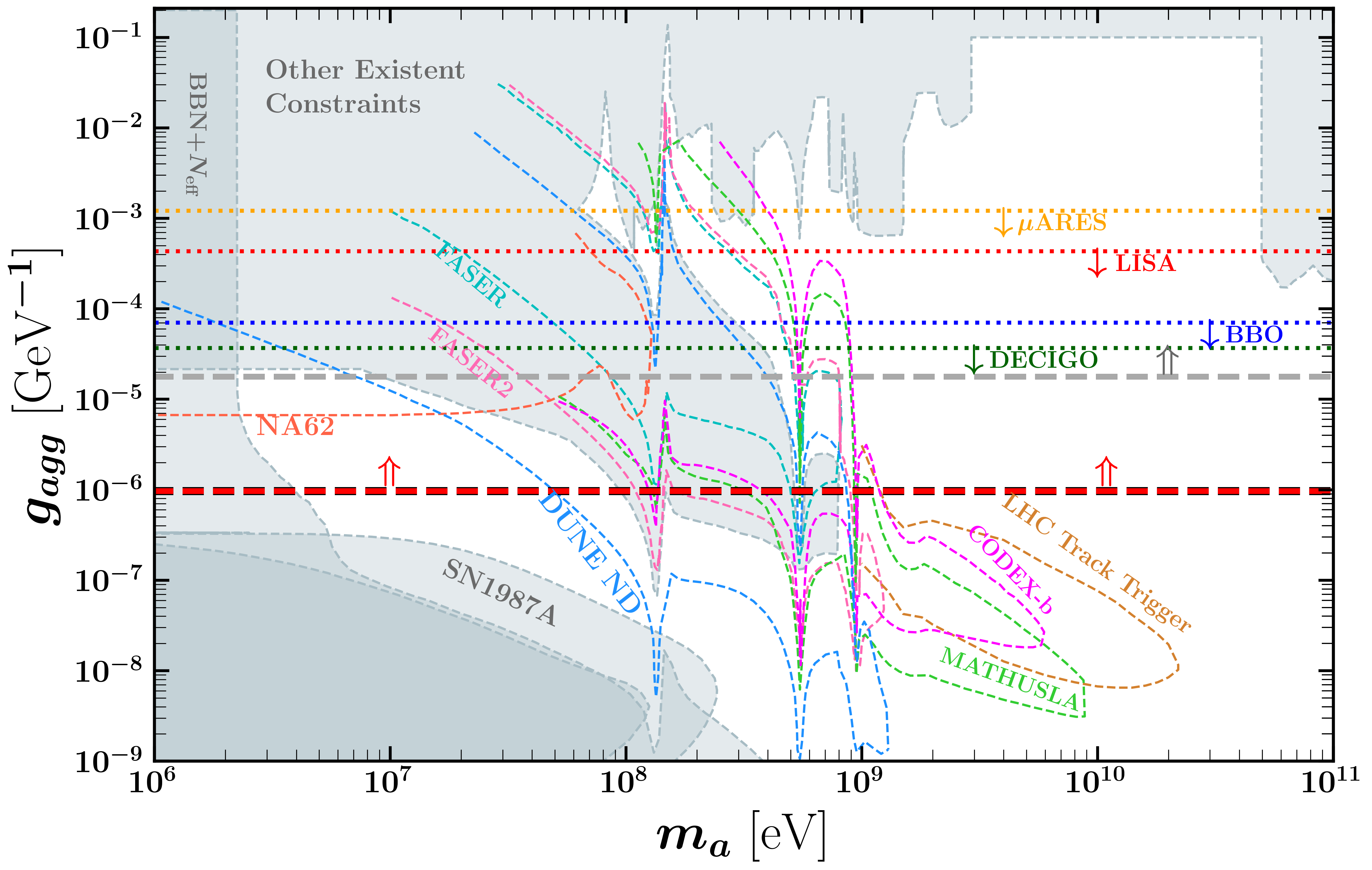} 
	\caption{Similar to Fig.~\ref{fig:lab-search:full-ALPphoton-close-up} but for the ALP coupling to gluons ($g_{agg}$). Filled grey coloured areas correspond to bounds coming from interpretation of existing datasets, astrophysical observations, or current experiments at $90\%$ C.L., while coloured dashed curves indicate projected sensitivities at $90\%$ C.L. The limits are taken from Refs.~\cite{Kelly:2020dda, Alimena:2025kjv}. 
    }
	\label{fig:lab-search:full-ALPgluon-close-up}
\end{figure}

\begin{figure}[!t]
	\centering
	\includegraphics[width=\textwidth]{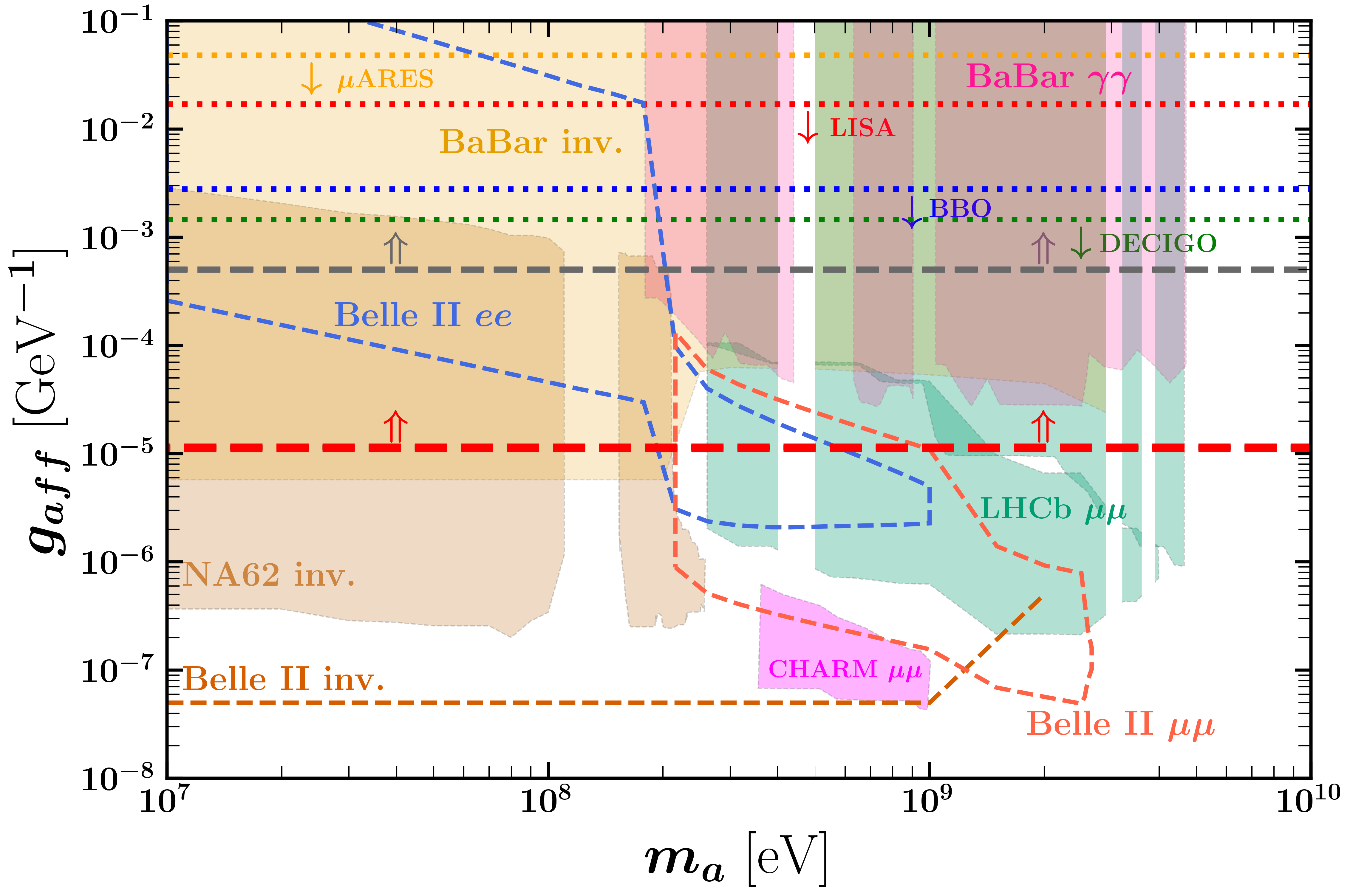} 
	\caption{Similar to Fig.~\ref{fig:lab-search:full-ALPphoton-close-up} but for the ALP coupling to SM fermions ($g_{aff}$). The limits are primarily compiled from Refs.~\cite{Ferber:2022rsf, Alimena:2025kjv}.
    }
	\label{fig:lab-search:full-ALPfermion-close-up}
\end{figure}

With this definition, the effective ALP couplings scale explicitly with the inverse of the axion decay constant $f_a$. In addition, as shown in Fig.~\ref{fig:fig-fa-g-blazar}, the SGWB and IGMF observations constrain $f_a$ independently of the ALP mass $m_a$. Therefore, these bounds can be directly translated into corresponding limits on $g_{a\gamma\gamma}$, $g_{agg}$, and $g_{aff}$. Thus, GW and magnetic field observations provide a complementary avenue to laboratory, astrophysical and cosmological searches targeting the effective ALP couplings to SM particles~\cite{AxionLimits}. We summarise our results in Figs.~\ref{fig:lab-search:full-ALP-noDM} (with zoomed-in view in Fig.~\ref{fig:lab-search:full-ALPphoton-close-up} for clarity), \ref{fig:lab-search:full-ALPgluon-close-up}, and \ref{fig:lab-search:full-ALPfermion-close-up}, which display the constraints in the $(m_a, g_{{\it a} \gamma \gamma}), (m_a, g_{{\it a} g g})$, and $(m_a, g_{{\it a} f f})$ planes, respectively.

Fig.~\ref{fig:lab-search:full-ALP-noDM} demonstrates the existing constraints on the ALP-photon interaction strength, based on the collection of experimental and observational results summarised in Ref.~\cite{AxionLimits}. These limits arise predominantly from astrophysical observations as well as laboratory and collider searches. Since we do not assume that the ALP constitutes the dark matter content in the present framework, bounds derived from dedicated ALP dark matter searches are not included in the figure. The excluded parameter space is separated into two distinct categories: the light–orange shaded region corresponds to terrestrial laboratory bounds, while the light–cyan shaded region mainly arises from astrophysical and cosmological constraints. The laboratory constraints include helioscopes, light-shining-through-wall (LSW) experiments, beam-dump searches, and collider probes. At low masses $m_a \lesssim \text{eV}$, heliscope searches (e.g., CAST~\cite{CAST:2017uph}) probe solar ALPs produced via the Primakoff process in the solar core, excluding couplings above $g_{{\it a} \gamma \gamma} \sim \mathcal{O}(10^{-10})~\text{GeV}^{-1}$, while LSW experiments (e.g. ALPS~\cite{Ehret:2010mh}, CROWS~\cite{Betz:2013dza}, OSQAR~\cite{OSQAR:2015qdv}) constrain similar masses but with weaker sensitivity. At higher masses $m_a \gtrsim 10^{4}~\text{eV}$, beam-dump experiments and collider searches (e.g., LEP~\cite{OPAL:2000puu, L3:2003yon, DELPHI:2003dlq}, BaBar~\cite{BaBar:2010eww}, LHC~\cite{CMS:2012lmn, ATLAS:2012ezx, ATLAS:2014kci}) probe ALPs through rare meson decays or direct production followed by diphoton decays, excluding relatively large couplings over the MeV–GeV mass range.

In addition, we have shown cosmological constraints from CMB~\cite{Mondino:2024rif,Goldstein:2024mfp} and  BBN~\cite{Depta:2020wmr}, as well as astrophysical bounds on the ALP–photon coupling from stellar evolution~\cite{Friedland:2012hj,Aoyama:2015asa,Dominguez:2017yhy, Dolan:2022kul, Candon:2024eah,Buckley:2024ldr}, supernova 1987A~\cite{Caputo:2022mah,Diamond:2023scc,Muller:2023vjm}, low-energy supernovae (e.g., \cite{Fiorillo:2025yzf}), type Ic supernova~\cite{Candon:2025ypl} and neutron-star merger GW170817~\cite{Dev:2023hax, Diamond:2023cto}, where excessive ALP emission would enhance energy loss and/or alter the multi-messenger photon signal. They provide some of the most stringent limits in the low-mass regime and constrain $g_{{\it a} \gamma \gamma}$ up to $\sim \mathcal{O}(10^{-11})~\text{GeV}^{-1}$. The plot also includes limits from $X$-ray observations of compact objects and galaxy clusters (e.g., Chandra~\cite{Reynes:2021bpe}, NuSTAR~\cite{Ning:2024eky,Candon:2024eah}), pulsar polar cap~\cite{Noordhuis:2022ljw}, as well as irreducible freeze-in considerations~\cite{Langhoff:2022bij}. The golden-yellow band is where the QCD axion lives (since $m_af_a\approx m_\pi f_\pi$)~\cite{Gorghetto:2018ocs}. For more details, see Refs.~\cite{AxionLimits, Caputo:2024oqc}.

The projected GW sensitivities to $g_{{\it a} \gamma \gamma}$ are shown as horizontal bands, reflecting the fact that the GW signal in our setup constrains the coupling essentially independently of $m_a$. The dotted bordered regions correspond to the sensitivity ranges of future interferometers: LISA (red), BBO (blue), DECIGO (green), $\mu$ARES (orange), and ET (purple), while the grey-hatched region indicates the current sensitivity of advanced-LIGO, which already places upper limits on the SGWB. Taken together, these GW detectors are sensitive to the range of
\bea
10^{-15}~\text{GeV}^{-1} \lesssim g_{{\it a} \gamma \gamma} \lesssim 10^{-4} ~\text{GeV}^{-1}.
\eea
As evident from Fig.~\ref{fig:lab-search:full-ALP-noDM}, parts of this interval are already excluded by laboratory, astrophysical and cosmological bounds. However, a substantial window remains open. In particular, the region $10^{-15}~\text{GeV}^{-1} \lesssim g_{{\it a} \gamma \gamma} \lesssim 10^{-12} ~\text{GeV}^{-1}$ is largely unconstrained over a broad mass range, including for ultra-light ALPs. On the other hand, heavier states with $m_a \gtrsim 10^8$ eV can still survive for larger $g_{{\it a} \gamma \gamma}$ coupling. In the absence of a detected SGWB signal at these future GW observatories, the corresponding null result would translate into upper bounds on 
$g_{{\it a} \gamma \gamma}$, thereby excluding additional regions of the ALP parameter space beyond the current laboratory and astrophysical limits.

We also display the lower limits on $g_{{\it a} \gamma \gamma}$ derived from blazar observations of IGMF in Fig.~\ref{fig:lab-search:full-ALP-noDM}, set by H.E.S.S. and {\it Fermi}-LAT~\cite{HESS:2023zwb}. For illustrative purpose, we only present the maximally helical configuration ($b=0$) case [cf.~Fig.~\ref{fig:fig-fa-g-blazar-a}]\footnote{The non-helical case ($b=1$) yields comparatively weaker magnetic field amplitudes (see Fig.~\ref{fig:fig-fa-g-blazar-b}) in our ALP framework and results in more stringent constraints on the effective ALP couplings. As this does not qualitatively change our conclusions but merely tightens the bounds, we show the conservative results for the maximally helical configuration.}. The lower limits are indicated by thick red-dashed and thick light–grey dashed lines corresponding to observation times $t_{\rm max} = 10$ yrs and $t_{\rm max} = 10^7$ yrs, respectively. These lines are obtained by translating the bounds on the decay constant $f_a$ (see Fig.~\ref{fig:fig-fa-g-blazar}) required for generating sufficiently strong PMFs during the supercooled FOPT. The up-arrow denotes the region that is favoured by the blazar observations~\cite{HESS:2023zwb}.
Since the magnetic field strength in this framework depends primarily on $f_a$, and not directly on $m_a$, the resulting limits appear as horizontal lines, analogous to the GW sensitivities. Fig.~\ref{fig:lab-search:full-ALP-noDM} clearly indicates that only relatively heavy ALPs, $m_a \gtrsim 10^8$ eV, can consistently account for the IGMFs within this scenario, as light or ultra-light ALPs are tightly constrained by laboratory and astrophysical constraints. As mentioned earlier (see subsection \ref{subsec:results:PMF-GW}), the parameter region favoured by blazar observations lies largely beyond the sensitivity of ground-based GW detectors such as ET, CE, and aLIGO. In contrast, space-based interferometers (i.e., LISA, BBO, DECIGO, etc.) remain well-positioned to probe the heavy-ALP mass regime, consistent with the observed magnetic fields within our framework. 

For further clarity on this region and to illustrate the complementarity with the GW experiments, we present a zoomed-in view of the projected sensitivities from current and future laboratory searches in Fig.~\ref{fig:lab-search:full-ALPphoton-close-up}. In the heavy-ALP mass regime displayed here, several current limits and and future projections from fixed-target, beam-dump (including recent EBES~\cite{Fusayasu:2026nto}), and collider facilities are shown that probe sizeable portions of the parameter space relevant to our magnetogenesis and GW analysis. For $m_a \lesssim \mathcal{O}(1-100)~\text{GeV}$, beam-dump experiments such as SHiP~\cite{Aberle:2022SHiP,Ovchynnikov:2023cry}, along with FASER (LLP)~\cite{FASER:2018eoc,Feng:2022inv, FASER:2024bbl}, DUNE~\cite{Brdar:2020dpr}, and proposed lepton-collider dump modes~\cite{Asai:2021ehn,Cesarotti:2023sje} can test small photon couplings through displaced decays and missing-energy signatures. Prompt diphoton searches are already pursued at Belle II~\cite{Acanfora:2024spi} and will be significantly enhanced at future facilities~\cite{deBlas:2025gyz}. In parallel, future high-luminosity lepton colliders (FCC-ee, CLIC, MuC)~\cite{Bauer:2018uxu,Casarsa:2021rud,Bao:2022onq,Bhattacharya:2025hme,Bao:2025tqs}) and hadron colliders (HL-LHC upgrades, FCC-hh)~\cite{deBlas:2025gyz} will probe prompt diphoton resonances and associated production channels, extending sensitivity both to smaller couplings and to masses well above the EW scale, potentially up to $m_a \sim 10^4$ GeV. For further details, we refer the reader to Refs.~\cite{AxionLimits,Alimena:2025kjv,deBlas:2025gyz} and the references therein.

As clearly observed in Fig.~\ref{fig:lab-search:full-ALPphoton-close-up}, the projected sensitivities of forthcoming laboratory and collider experiments extend into the parameter region preferred by blazar-motivated magnetic field considerations, corresponding to the heavy-ALP window $m_a \gtrsim 10^8$ eV. Therefore, while blazar observations set a lower bound on $g_{{\it a} \gamma \gamma}$ and space-based GW detectors probe the same region (indicated by differently coloured horizontal dotted lines with downward arrows), future laboratory and collider searches can independently test this parameter space via direct production. This establishes a threefold complementarity: magnetogenesis requirement favours the heavy ALP mass regime, space-based GW experiments constrain the ALP-photon coupling from the cosmological side, and next-generation intensity- and energy-frontier experiments provide direct terrestrial probes of the same region.

Fig.~\ref{fig:lab-search:full-ALPgluon-close-up} summarises the parameter space in the plane of the ALP–gluon coupling and mass $m_a$. We present our results considering the gluon-dominance scenario following Ref.~\cite{Kelly:2020dda}. The filled grey regions denote existing constraints from laboratory, collider, and astrophysical probes, while the coloured contours indicate projected sensitivities of existing and upcoming experiments. Current bounds are coming primarily from interpretation of older beam-dump data and astrophysical observations, as well as from dedicated laboratory searches. Representative constraints include those from SN1987A~\cite{Chang:2018rso, Ertas:2020xcc} and cosmology~\cite{Depta:2020wmr}, as well as a variety of beam-dump, rare meson, and prompt decay searches (collectively denoted as ``other existent constraints'' in the plot), including electron beam dump~\cite{Dobrich:2015jyk,Dolan:2017osp, NA64:2020qwq}, CHARM~\cite{CHARM:1985anb, FASER:2018eoc}, NuCal~\cite{Blumlein:2013cua}, and LHC dijet searches~\cite{Mariotti:2017vtv}, among others. It is worth noting that, compared to existing searches which primarily target ALPs coupled dominantly to photons or EW gauge bosons, the gluon-dominated scenario considered here remains comparatively less constrained. Future facilities are expected to substantially extend the experimental reach in this parameter space. Beam-dump and fixed-target experiments such as DUNE near detector (ND)~\cite{DUNE:2020ypp} and NA62 upgrades~\cite{Ertas:2020xcc} will extend the sensitivity in the MeV-GeV regime, while collider-based searches (e.g. FASER/FASER2~\cite{FASER:2018eoc}, CODEX-b~\cite{Gligorov:2017nwh,CODEX-b:2019jve}, MATHUSLA~\cite{Chou:2016lxi, CODEX-b:2019jve} and LHC track-trigger strategies~\cite{Hook:2019qoh}) can probe heavier masses with couplings down to $g_{{\it a} g g} \sim \mathcal{O}(10^{-8})~\text{GeV}^{-1}$.

Superimposed on these experimental limits, we also display the parameter region relevant for our magnetogenesis scenario and the associated SGWB forecasts, analogous to Fig.~\ref{fig:lab-search:full-ALPphoton-close-up}. The allowed region lies close to (and above) $g_{{\it a} g g} \sim 10^{-6}~\text{GeV}^{-1}$ for $t_{\rm max} = 10$ yrs over a MeV-GeV ALP mass range. As seen in Fig.~\ref{fig:lab-search:full-ALPgluon-close-up}, this region is largely consistent with existing laboratory and astrophysical bounds, while overlapping with the projected sensitivities of several future searches. In particular, beam-dump and fixed-target experiments such as the DUNE ND and NA62 upgrades will probe part of the parameter space at sub-GeV masses, whereas forward and long-lived particle detectors at the LHC can explore complementary regions of larger ALP-gluon couplings. Consequently, similar to the case shown in Fig.~\ref{fig:lab-search:full-ALPphoton-close-up}, we find that a sizeable fraction of the parameter space favoured by magnetogenesis and the associated GW signal in our framework can be tested through direct laboratory searches sensitive to the ALP–gluon interaction.

Finally, in Fig.~\ref{fig:lab-search:full-ALPfermion-close-up} we show the constraints on the ALP–fermion coupling $g_{{\it a} f f}$ together with the region favoured by blazar observations and upcoming space-based GW experiments. Existing bounds arise primarily from flavour and fixed-target experiments probing rare meson decays and long-lived particle signatures. In particular, invisible decay searches at BaBar~\cite{BaBar:2013npw} and NA62~\cite{NA62:2020xlg,NA62:2021zjw} constrain the parameter space at lower ALP masses$\lesssim \mathcal{O}(100~{\rm MeV})$, while di-photon searches at BaBar~\cite{BaBar:2021ich}, and displaced di-muon searches at LHCb~\cite{LHCb:2015nkv,LHCb:2016awg} and CHARM~\cite{Dobrich:2018jyi} probe larger masses at $\sim \mathcal{O}({\rm GeV})$ scale. Future measurements will significantly extend this reach: invisible searches at Belle II as well as displaced di-lepton signatures at LHCb and Belle II are expected to probe couplings well below current bounds over a broad sub-GeV mass range (see Ref.~\cite{Ferber:2022rsf} and references therein). As illustrated in Fig.~\ref{fig:lab-search:full-ALPfermion-close-up}, the parameter region relevant for our magnetogenesis scenario, typically around (and above) $g_{{\it a} f f} \sim 10^{-5}~\text{GeV}^{-1}$ for $t_{\rm max} = 10$ yrs, is subject to stringent constraints from existing laboratory and flavour experiments. Nevertheless, it partially overlaps with the projected sensitivities of several upcoming searches, leaving a narrow mass window where the scenario can be tested through complementary terrestrial probes of the ALP–fermion interaction.

\section{Summary and Conclusions}
\label{sec:summary}
In this work, we investigated the cosmological consequences of an FOPT occurring in an ALP framework with radiative symmetry breaking. The model is based on a conformal scalar sector in which a global $U(1)$ symmetry is dynamically broken through the Coleman–Weinberg mechanism, generating the ALP, which is associated with the phase of a complex scalar field. The ALP sector is coupled to the SM through a Higgs-portal interaction, which allows the vacuum energy released during the PT to partially transfer to the SM plasma. This setup naturally realises a strongly supercooled PT, providing a suitable environment for both stochastic GW production and PMF generation.

We first analysed the dynamics of the PT within the radiatively broken conformal potential. Due to the approximate scale invariance of the model, the tunnelling rate depends only logarithmically on the temperature, leading generically to significant supercooling and a vacuum-dominated transition. We then studied the generation and subsequent evolution of PMFs sourced by bubble collisions and MHD turbulence during the PT. Two representative scenarios were considered: maximally helical ($b=0$) and non-helical ($b=1$) magnetic fields evolving in the plasma with kinetic helicity. After accounting for the inverse cascade of magnetic energy and cosmological redshifting, we computed the present-day magnetic field amplitude ($B_0$) as a function of the coherence length ($\lambda_0$). We find that our model can generate magnetic fields with strengths up to $B_0 \sim 10^{-9}$ G for maximally helical configurations, with peak coherence lengths in the range $\lambda_0 \sim 10^{-3} - 10^{-1}$ Mpc, which are compatible with the lower limits on IGMFs inferred from blazar observations, such as {\it Fermi}-LAT/H.E.S.S and {\it Fermi}-LAT/MAGIC (see Figs.~\ref{fig:fig-BPs-no-B-evolution} and \ref{fig:fig-fa-g-blazar}). In contrast, non-helical configurations typically yield weaker fields, with peak amplitudes around $B_0 \sim 10^{-11}$ G with $\lambda_0 \sim 10^{-5} - 10^{-2}$ Mpc. These results demonstrate that an ALP-assisted supercooled PT can naturally account for the origin of large-scale IGMFs. Furthermore, the recent $3.8\sigma$ evidence for an IGMF from anisotropic pair-halo searches in \textit{Fermi}-LAT data~\cite{Zhang:2026lte} strengthens the case for a cosmological origin of the IGMF, such as from FOPT of the kind studied here. Extending the IGMF measurement to a coherence-length-dependent scaling would be a valuable step toward more directly confronting such primordial magnetogenesis scenarios.

In addition, we evaluated the SGWB spectrum generated by the same FOPT. We find that the predicted GW signals, with SNR $> 10$, fall within the sensitivity range of several future space-based detectors, including LISA, BBO, DECIGO and $\mu$ARES. In particular, LISA and $\mu$ARES can probe essentially the entire region of parameter space that simultaneously explains the blazar-inferred magnetic fields (see Fig.~\ref{fig:fig-fa-g-blazar}). Ground-based detectors such as ET and CE are instead sensitive mainly to higher symmetry-breaking scales ($f_a \gtrsim 10^6$ GeV) and therefore do not significantly overlap with the region favoured by magnetic field observations.

Combining the GW forecasts with the magnetic field predictions allows us to delineate the viable parameter space of the model in terms of the ALP decay constant $f_a$ and the hidden gauge coupling $g$. In the maximally helical case, the parameter region with $10^3~\text{GeV} \lesssim f_a \lesssim 10^5~\text{GeV}$ and $0.8 \lesssim g \lesssim 1.25$ simultaneously yields sufficiently strong PMFs and detectable GW signals at future experiments (see Fig.~\ref{fig:fig-fa-g-blazar}). For non-helical configurations, the viable range of $f_a$ is somewhat narrower, $10^3~\text{GeV} \lesssim f_a \lesssim 10^4~\text{GeV}$. These results highlight a strong correlation between primordial magnetogenesis and SGWB production in this ALP-assisted FOPT framework.

We also examined the complementarity between cosmological probes and laboratory searches for ALPs. Since the effective ALP couplings to SM particles scale inversely with the decay constant $f_a$, the bounds obtained from magnetic field observations and GW experiments can be translated directly into limits on the effective ALP interactions, such as with photons ($g_{{\it a} \gamma \gamma}$), gluons ($g_{{\it a} g g}$), and fermions ($g_{{\it a} f f}$). Considering the maximally helical magnetic field configuration, we find that the cosmological observables explored in this work can provide a complementary probe of the relatively heavier ALP parameter space ($m_a \gtrsim 10^8$ eV) currently unconstrained by laboratory and astrophysical searches (see Figs.~\ref{fig:lab-search:full-ALPphoton-close-up}, \ref{fig:lab-search:full-ALPgluon-close-up} and \ref{fig:lab-search:full-ALPfermion-close-up}).

Overall, our results demonstrate that supercooled PT in the ALP framework provides a compelling and testable mechanism linking early Universe dynamics with present-day cosmological observables. The simultaneous generation of IGMF and SGWB offers a distinctive multi-messenger signature of such models. Future GW observations, together with improved measurements of cosmic magnetic fields and laboratory ALP searches, will therefore play a crucial role in testing this scenario and probing the thermal history of the early Universe.


\section*{Acknowledgements}

P.B. thanks Angela Conaci for useful initial discussions.  B.D. thanks Manel Errando and Olive Zhang for useful discussions on IGMF and for sharing an unpublished version of their paper~\cite{Zhang:2026lte} with us.  A.G. thanks Subir Sarkar for very helpful discussion regarding the plasma instability of magnetic fields. A.G. also acknowledges discussion with Pratyay Pal. P.B. acknowledges the financial support from the Indian Institute of Technology, Delhi (IITD) as an Early-Doc fellow (IITD/IRD/MI02585G/EDF723/430439) and the computing facility at the MS-516 HEP-PH Laboratory, Department of Physics, IITD. The work of B.D. was partly supported by the US Department of Energy under grant No. DE-SC0017987.
 
\bibliographystyle{JHEP}
\bibliography{ALPmagneto}

@article{LIGOScientific:2016jlg,
    author = "Abbott, Benjamin P. and others",
    collaboration = "LIGO Scientific, Virgo",
    title = "{Upper Limits on the Stochastic Gravitational-Wave Background from Advanced LIGO{\textquoteright}s First Observing Run}",
    eprint = "1612.02029",
    archivePrefix = "arXiv",
    primaryClass = "gr-qc",
    doi = "10.1103/PhysRevLett.118.121101",
    journal = "Phys. Rev. Lett.",
    volume = "118",
    number = "12",
    pages = "121101",
    year = "2017",
    note = "[Erratum: Phys.Rev.Lett. 119, 029901 (2017)]"
}

@article{Dev:2019njv,
    author = "Dev, P. S. Bhupal and Ferrer, Francesc and Zhang, Yiyang and Zhang, Yongchao",
    title = "{Gravitational Waves from First-Order Phase Transition in a Simple Axion-Like Particle Model}",
    eprint = "1905.00891",
    archivePrefix = "arXiv",
    primaryClass = "hep-ph",
    doi = "10.1088/1475-7516/2019/11/006",
    journal = "JCAP",
    volume = "11",
    pages = "006",
    year = "2019"
}

@article{Ehret:2010mh,
    author = "Ehret, Klaus and others",
    title = "{New ALPS Results on Hidden-Sector Lightweights}",
    eprint = "1004.1313",
    archivePrefix = "arXiv",
    primaryClass = "hep-ex",
    reportNumber = "DESY-10-030, MPP-2010-27",
    doi = "10.1016/j.physletb.2010.04.066",
    journal = "Phys. Lett. B",
    volume = "689",
    pages = "149--155",
    year = "2010"
}

@article{Betz:2013dza,
    author = "Betz, M. and Caspers, F. and Gasior, M. and Thumm, M. and Rieger, S. W.",
    title = "{First results of the CERN Resonant Weakly Interacting sub-eV Particle Search (CROWS)}",
    eprint = "1310.8098",
    archivePrefix = "arXiv",
    primaryClass = "physics.ins-det",
    doi = "10.1103/PhysRevD.88.075014",
    journal = "Phys. Rev. D",
    volume = "88",
    number = "7",
    pages = "075014",
    year = "2013"
}

@article{OSQAR:2015qdv,
    author = "Ballou, R. and others",
    collaboration = "OSQAR",
    title = "{New exclusion limits on scalar and pseudoscalar axionlike particles from light shining through a wall}",
    eprint = "1506.08082",
    archivePrefix = "arXiv",
    primaryClass = "hep-ex",
    doi = "10.1103/PhysRevD.92.092002",
    journal = "Phys. Rev. D",
    volume = "92",
    number = "9",
    pages = "092002",
    year = "2015"
}

@article{CAST:2017uph,
    author = "Anastassopoulos, V. and others",
    collaboration = "CAST",
    title = "{New CAST Limit on the Axion-Photon Interaction}",
    eprint = "1705.02290",
    archivePrefix = "arXiv",
    primaryClass = "hep-ex",
    doi = "10.1038/nphys4109",
    journal = "Nature Phys.",
    volume = "13",
    pages = "584--590",
    year = "2017"
}

@article{Caputo:2022mah,
    author = "Caputo, Andrea and Janka, Hans-Thomas and Raffelt, Georg and Vitagliano, Edoardo",
    title = "{Low-Energy Supernovae Severely Constrain Radiative Particle Decays}",
    eprint = "2201.09890",
    archivePrefix = "arXiv",
    primaryClass = "astro-ph.HE",
    doi = "10.1103/PhysRevLett.128.221103",
    journal = "Phys. Rev. Lett.",
    volume = "128",
    number = "22",
    pages = "221103",
    year = "2022"
}

@article{Diamond:2023scc,
    author = "Diamond, Melissa and Fiorillo, Damiano F. G. and Marques-Tavares, Gustavo and Vitagliano, Edoardo",
    title = "{Axion-sourced fireballs from supernovae}",
    eprint = "2303.11395",
    archivePrefix = "arXiv",
    primaryClass = "hep-ph",
    doi = "10.1103/PhysRevD.107.103029",
    journal = "Phys. Rev. D",
    volume = "107",
    number = "10",
    pages = "103029",
    year = "2023",
    note = "[Erratum: Phys.Rev.D 108, 049902 (2023)]"
}

@article{Muller:2023vjm,
    author = {M{\"u}ller, Eike and Calore, Francesca and Carenza, Pierluca and Eckner, Christopher and Marsh, M. C. David},
    title = "{Investigating the gamma-ray burst from decaying MeV-scale axion-like particles produced in supernova explosions}",
    eprint = "2304.01060",
    archivePrefix = "arXiv",
    primaryClass = "astro-ph.HE",
    doi = "10.1088/1475-7516/2023/07/056",
    journal = "JCAP",
    volume = "07",
    pages = "056",
    year = "2023"
}

@article{Dolan:2022kul,
    author = "Dolan, Matthew J. and Hiskens, Frederick J. and Volkas, Raymond R.",
    title = "{Advancing globular cluster constraints on the axion-photon coupling}",
    eprint = "2207.03102",
    archivePrefix = "arXiv",
    primaryClass = "hep-ph",
    doi = "10.1088/1475-7516/2022/10/096",
    journal = "JCAP",
    volume = "10",
    pages = "096",
    year = "2022"
}

@article{Noordhuis:2022ljw,
    author = "Noordhuis, Dion and Prabhu, Anirudh and Witte, Samuel J. and Chen, Alexander Y. and Cruz, F{\'a}bio and Weniger, Christoph",
    title = "{Novel Constraints on Axions Produced in Pulsar Polar-Cap Cascades}",
    eprint = "2209.09917",
    archivePrefix = "arXiv",
    primaryClass = "hep-ph",
    doi = "10.1103/PhysRevLett.131.111004",
    journal = "Phys. Rev. Lett.",
    volume = "131",
    number = "11",
    pages = "111004",
    year = "2023"
}

@article{Gorghetto:2018ocs,
    author = "Gorghetto, Marco and Villadoro, Giovanni",
    title = "{Topological Susceptibility and QCD Axion Mass: QED and NNLO corrections}",
    eprint = "1812.01008",
    archivePrefix = "arXiv",
    primaryClass = "hep-ph",
    doi = "10.1007/JHEP03(2019)033",
    journal = "JHEP",
    volume = "03",
    pages = "033",
    year = "2019"
}

@article{LIGOScientific:2019vic,
    author = "Abbott, B. P. and others",
    collaboration = "LIGO Scientific, Virgo",
    title = "{Search for the isotropic stochastic background using data from Advanced LIGO{\textquoteright}s second observing run}",
    eprint = "1903.02886",
    archivePrefix = "arXiv",
    primaryClass = "gr-qc",
    reportNumber = "LIGO-P1800248",
    doi = "10.1103/PhysRevD.100.061101",
    journal = "Phys. Rev. D",
    volume = "100",
    number = "6",
    pages = "061101",
    year = "2019"
}

@article{KAGRA:2021kbb,
    author = "Abbott, R. and others",
    collaboration = "KAGRA, Virgo, LIGO Scientific",
    title = "{Upper limits on the isotropic gravitational-wave background from Advanced LIGO and Advanced Virgo{\textquoteright}s third observing run}",
    eprint = "2101.12130",
    archivePrefix = "arXiv",
    primaryClass = "gr-qc",
    reportNumber = "LIGO-DCC-P2000314",
    doi = "10.1103/PhysRevD.104.022004",
    journal = "Phys. Rev. D",
    volume = "104",
    number = "2",
    pages = "022004",
    year = "2021"
}

@article{NANOGrav:2023hvm,
    author = "Afzal, Adeela and others",
    collaboration = "NANOGrav",
    title = "{The NANOGrav 15 yr Data Set: Search for Signals from New Physics}",
    eprint = "2306.16219",
    archivePrefix = "arXiv",
    primaryClass = "astro-ph.HE",
    reportNumber = "FERMILAB-PUB-23-589-T",
    doi = "10.3847/2041-8213/acdc91",
    journal = "Astrophys. J. Lett.",
    volume = "951",
    number = "1",
    pages = "L11",
    year = "2023",
    note = "[Erratum: Astrophys.J.Lett. 971, L27 (2024), Erratum: Astrophys.J. 971, L27 (2024)]"
}

@article{EPTA:2023fyk,
    author = "Antoniadis, J. and others",
    collaboration = "EPTA, InPTA:",
    title = "{The second data release from the European Pulsar Timing Array - III. Search for gravitational wave signals}",
    eprint = "2306.16214",
    archivePrefix = "arXiv",
    primaryClass = "astro-ph.HE",
    doi = "10.1051/0004-6361/202346844",
    journal = "Astron. Astrophys.",
    volume = "678",
    pages = "A50",
    year = "2023"
}

@article{Reardon:2023gzh,
    author = "Reardon, Daniel J. and others",
    title = "{Search for an Isotropic Gravitational-wave Background with the Parkes Pulsar Timing Array}",
    eprint = "2306.16215",
    archivePrefix = "arXiv",
    primaryClass = "astro-ph.HE",
    doi = "10.3847/2041-8213/acdd02",
    journal = "Astrophys. J. Lett.",
    volume = "951",
    number = "1",
    pages = "L6",
    year = "2023"
}

@article{Xu:2023wog,
    author = "Xu, Heng and others",
    title = "{Searching for the Nano-Hertz Stochastic Gravitational Wave Background with the Chinese Pulsar Timing Array Data Release I}",
    eprint = "2306.16216",
    archivePrefix = "arXiv",
    primaryClass = "astro-ph.HE",
    doi = "10.1088/1674-4527/acdfa5",
    journal = "Res. Astron. Astrophys.",
    volume = "23",
    number = "7",
    pages = "075024",
    year = "2023"
}

@article{NANOGrav:2023hfp,
    author = "Agazie, Gabriella and others",
    collaboration = "NANOGrav",
    title = "{The NANOGrav 15 yr Data Set: Constraints on Supermassive Black Hole Binaries from the Gravitational-wave Background}",
    eprint = "2306.16220",
    archivePrefix = "arXiv",
    primaryClass = "astro-ph.HE",
    doi = "10.3847/2041-8213/ace18b",
    journal = "Astrophys. J. Lett.",
    volume = "952",
    number = "2",
    pages = "L37",
    year = "2023"
}

@article{EPTA:2023xxk,
    author = "Antoniadis, J. and others",
    collaboration = "EPTA, InPTA",
    title = "{The second data release from the European Pulsar Timing Array - IV. Implications for massive black holes, dark matter, and the early Universe}",
    eprint = "2306.16227",
    archivePrefix = "arXiv",
    primaryClass = "astro-ph.CO",
    doi = "10.1051/0004-6361/202347433",
    journal = "Astron. Astrophys.",
    volume = "685",
    pages = "A94",
    year = "2024"
}

@article{Weltman:2018zrl,
    author = "Weltman, A. and others",
    title = "{Fundamental physics with the Square Kilometre Array}",
    eprint = "1810.02680",
    archivePrefix = "arXiv",
    primaryClass = "astro-ph.CO",
    doi = "10.1017/pasa.2019.42",
    journal = "Publ. Astron. Soc. Austral.",
    volume = "37",
    pages = "e002",
    year = "2020"
}

@article{Garcia-Bellido:2021zgu,
    author = "Garcia-Bellido, Juan and Murayama, Hitoshi and White, Graham",
    title = "{Exploring the early Universe with Gaia and Theia}",
    eprint = "2104.04778",
    archivePrefix = "arXiv",
    primaryClass = "hep-ph",
    reportNumber = "IFT-UAM/CSIC-2021-038",
    doi = "10.1088/1475-7516/2021/12/023",
    journal = "JCAP",
    volume = "12",
    number = "12",
    pages = "023",
    year = "2021"
}

@article{MAGIS-100:2021etm,
    author = "Abe, Mahiro and others",
    collaboration = "MAGIS-100",
    title = "{Matter-wave Atomic Gradiometer Interferometric Sensor (MAGIS-100)}",
    eprint = "2104.02835",
    archivePrefix = "arXiv",
    primaryClass = "physics.atom-ph",
    reportNumber = "FERMILAB-PUB-21-031-AD-DI-FESS-QIS-T",
    doi = "10.1088/2058-9565/abf719",
    journal = "Quantum Sci. Technol.",
    volume = "6",
    number = "4",
    pages = "044003",
    year = "2021"
}

@article{Badurina:2019hst,
    author = "Badurina, L. and others",
    title = "{AION: An Atom Interferometer Observatory and Network}",
    eprint = "1911.11755",
    archivePrefix = "arXiv",
    primaryClass = "astro-ph.CO",
    reportNumber = "AION-2019-001, CERN-TH-2019-199",
    doi = "10.1088/1475-7516/2020/05/011",
    journal = "JCAP",
    volume = "05",
    pages = "011",
    year = "2020"
}

@article{AEDGE:2019nxb,
    author = "El-Neaj, Yousef Abou and others",
    collaboration = "AEDGE",
    title = "{AEDGE: Atomic Experiment for Dark Matter and Gravity Exploration in Space}",
    eprint = "1908.00802",
    archivePrefix = "arXiv",
    primaryClass = "gr-qc",
    reportNumber = "KCL-PH-TH/2019-65, CERN-TH-2019-126",
    doi = "10.1140/epjqt/s40507-020-0080-0",
    journal = "EPJ Quant. Technol.",
    volume = "7",
    pages = "6",
    year = "2020"
}

@article{Sesana:2019vho,
    author = "Sesana, Alberto and others",
    title = "{Unveiling the gravitational universe at $\mu$-Hz frequencies}",
    eprint = "1908.11391",
    archivePrefix = "arXiv",
    primaryClass = "astro-ph.IM",
    doi = "10.1007/s10686-021-09709-9",
    journal = "Exper. Astron.",
    volume = "51",
    number = "3",
    pages = "1333--1383",
    year = "2021"
}

@article{LISA:2017pwj,
    author = "Amaro-Seoane, Pau and others",
    collaboration = "LISA",
    title = "{Laser Interferometer Space Antenna}",
    eprint = "1702.00786",
    archivePrefix = "arXiv",
    primaryClass = "astro-ph.IM",
    month = "2",
    year = "2017"
}

@article{TianQin:2015yph,
    author = "Luo, Jun and others",
    collaboration = "TianQin",
    title = "{TianQin: a space-borne gravitational wave detector}",
    eprint = "1512.02076",
    archivePrefix = "arXiv",
    primaryClass = "astro-ph.IM",
    doi = "10.1088/0264-9381/33/3/035010",
    journal = "Class. Quant. Grav.",
    volume = "33",
    number = "3",
    pages = "035010",
    year = "2016"
}

@article{Ruan:2018tsw,
    author = "Ruan, Wen-Hong and Guo, Zong-Kuan and Cai, Rong-Gen and Zhang, Yuan-Zhong",
    title = "{Taiji program: Gravitational-wave sources}",
    eprint = "1807.09495",
    archivePrefix = "arXiv",
    primaryClass = "gr-qc",
    doi = "10.1142/S0217751X2050075X",
    journal = "Int. J. Mod. Phys. A",
    volume = "35",
    number = "17",
    pages = "2050075",
    year = "2020"
}

@article{Kawamura:2020pcg,
    author = "Kawamura, Seiji and others",
    title = "{Current status of space gravitational wave antenna DECIGO and B-DECIGO}",
    eprint = "2006.13545",
    archivePrefix = "arXiv",
    primaryClass = "gr-qc",
    doi = "10.1093/ptep/ptab019",
    journal = "PTEP",
    volume = "2021",
    number = "5",
    pages = "05A105",
    year = "2021"
}

@article{Corbin:2005ny,
    author = "Corbin, Vincent and Cornish, Neil J.",
    title = "{Detecting the cosmic gravitational wave background with the big bang observer}",
    eprint = "gr-qc/0512039",
    archivePrefix = "arXiv",
    doi = "10.1088/0264-9381/23/7/014",
    journal = "Class. Quant. Grav.",
    volume = "23",
    pages = "2435--2446",
    year = "2006"
}

@article{Punturo:2010zz,
    author = "Punturo, M. and others",
    editor = "Ricci, Fulvio",
    title = "{The Einstein Telescope: A third-generation gravitational wave observatory}",
    doi = "10.1088/0264-9381/27/19/194002",
    journal = "Class. Quant. Grav.",
    volume = "27",
    pages = "194002",
    year = "2010"
}

@article{Reitze:2019iox,
    author = "Reitze, David and others",
    title = "{Cosmic Explorer: The U.S. Contribution to Gravitational-Wave Astronomy beyond LIGO}",
    eprint = "1907.04833",
    archivePrefix = "arXiv",
    primaryClass = "astro-ph.IM",
    reportNumber = "LIGO-P1900316",
    journal = "Bull. Am. Astron. Soc.",
    volume = "51",
    number = "7",
    pages = "035",
    year = "2019"
}

@article{Aggarwal:2020olq,
    author = "Aggarwal, Nancy and others",
    title = "{Challenges and opportunities of gravitational-wave searches at MHz to GHz frequencies}",
    eprint = "2011.12414",
    archivePrefix = "arXiv",
    primaryClass = "gr-qc",
    reportNumber = "CERN-TH-2020-185, HIP-2020-28/TH, DESY 20-195, CERN-TH-2020-185, HIP-2020-28/TH, DESY 20-195",
    doi = "10.1007/s41114-021-00032-5",
    journal = "Living Rev. Rel.",
    volume = "24",
    number = "1",
    pages = "4",
    year = "2021"
}

@article{Berlin:2021txa,
    author = {Berlin, Asher and Blas, Diego and Tito D'Agnolo, Raffaele and Ellis, Sebastian A. R. and Harnik, Roni and Kahn, Yonatan and Sch{\"u}tte-Engel, Jan},
    title = "{Detecting high-frequency gravitational waves with microwave cavities}",
    eprint = "2112.11465",
    archivePrefix = "arXiv",
    primaryClass = "hep-ph",
    reportNumber = "FERMILAB-PUB-21-724-SQMS-T",
    doi = "10.1103/PhysRevD.105.116011",
    journal = "Phys. Rev. D",
    volume = "105",
    number = "11",
    pages = "116011",
    year = "2022"
}

@article{Herman:2022fau,
    author = "Herman, Nicolas and Lehoucq, L{\'e}onard and F{\'{u}}zfa, Andr{\'e}",
    title = "{Electromagnetic antennas for the resonant detection of the stochastic gravitational wave background}",
    eprint = "2203.15668",
    archivePrefix = "arXiv",
    primaryClass = "gr-qc",
    doi = "10.1103/PhysRevD.108.124009",
    journal = "Phys. Rev. D",
    volume = "108",
    number = "12",
    pages = "124009",
    year = "2023"
}

@article{Bringmann:2023gba,
    author = "Bringmann, Torsten and Domcke, Valerie and Fuchs, Elina and Kopp, Joachim",
    title = "{High-frequency gravitational wave detection via optical frequency modulation}",
    eprint = "2304.10579",
    archivePrefix = "arXiv",
    primaryClass = "hep-ph",
    reportNumber = "CERN-TH-2023-065, MITP-23-017",
    doi = "10.1103/PhysRevD.108.L061303",
    journal = "Phys. Rev. D",
    volume = "108",
    number = "6",
    pages = "L061303",
    year = "2023"
}

@article{Valero:2024ncz,
    author = "Valero, Jos{\'e} Reina and Madrid, Jose R. Navarro and Blas, Diego and Morcillo, Alejandro D{\'\i}az and Irastorza, Igor Garc{\'\i}a and Gimeno, Benito and Cabrera, Juan Monz{\'o}",
    title = "{High-frequency gravitational waves detection with the BabyIAXO haloscopes}",
    eprint = "2407.20482",
    archivePrefix = "arXiv",
    primaryClass = "hep-ex",
    doi = "10.1103/PhysRevD.111.043024",
    journal = "Phys. Rev. D",
    volume = "111",
    number = "4",
    pages = "043024",
    year = "2025"
}

@article{Witten:1984rs,
    author = "Witten, Edward",
    title = "{Cosmic Separation of Phases}",
    reportNumber = "PRINT-84-0400 (IAS,PRINCETON)",
    doi = "10.1103/PhysRevD.30.272",
    journal = "Phys. Rev. D",
    volume = "30",
    pages = "272--285",
    year = "1984"
}

@article{Hogan:1986dsh,
    author = "Hogan, C. J.",
    title = "{Gravitational radiation from cosmological phase transitions}",
    doi = "10.1093/mnras/218.4.629",
    journal = "Mon. Not. Roy. Astron. Soc.",
    volume = "218",
    number = "4",
    pages = "629--636",
    year = "1986"
}

@article{LISACosmologyWorkingGroup:2022jok,
    author = "Auclair, Pierre and others",
    collaboration = "LISA Cosmology Working Group",
    title = "{Cosmology with the Laser Interferometer Space Antenna}",
    eprint = "2204.05434",
    archivePrefix = "arXiv",
    primaryClass = "astro-ph.CO",
    reportNumber = "LISA CosWG-22-03, FERMILAB-PUB-22-349-SCD",
    doi = "10.1007/s41114-023-00045-2",
    journal = "Living Rev. Rel.",
    volume = "26",
    number = "1",
    pages = "5",
    year = "2023"
}

@article{Kajantie:1996mn,
    author = "Kajantie, K. and Laine, M. and Rummukainen, K. and Shaposhnikov, Mikhail E.",
    title = "{Is there a~ hot electroweak phase transition at $m_H \gtrsim m_W$?}",
    eprint = "hep-ph/9605288",
    archivePrefix = "arXiv",
    reportNumber = "CERN-TH-96-126, HD-THEP-96-15, IUHET-333",
    doi = "10.1103/PhysRevLett.77.2887",
    journal = "Phys. Rev. Lett.",
    volume = "77",
    pages = "2887--2890",
    year = "1996"
}

@article{Laine:2015kra,
    author = "Laine, M. and Meyer, M.",
    title = "{Standard Model thermodynamics across the electroweak crossover}",
    eprint = "1503.04935",
    archivePrefix = "arXiv",
    primaryClass = "hep-ph",
    doi = "10.1088/1475-7516/2015/07/035",
    journal = "JCAP",
    volume = "07",
    pages = "035",
    year = "2015"
}

@article{Bhattacharya:2014ara,
    author = "Bhattacharya, Tanmoy and others",
    title = "{QCD Phase Transition with Chiral Quarks and Physical Quark Masses}",
    eprint = "1402.5175",
    archivePrefix = "arXiv",
    primaryClass = "hep-lat",
    reportNumber = "BNL-103837-2014-JA, CU-TP-1205, INT-PUB-14-003, LLNL-JRNL-650194",
    doi = "10.1103/PhysRevLett.113.082001",
    journal = "Phys. Rev. Lett.",
    volume = "113",
    number = "8",
    pages = "082001",
    year = "2014"
}

@article{Meissner:2006zh,
    author = "Meissner, Krzysztof A. and Nicolai, Hermann",
    title = "{Conformal Symmetry and the Standard Model}",
    eprint = "hep-th/0612165",
    archivePrefix = "arXiv",
    doi = "10.1016/j.physletb.2007.03.023",
    journal = "Phys. Lett. B",
    volume = "648",
    pages = "312--317",
    year = "2007"
}

@article{Coleman:1973jx,
    author = "Coleman, Sidney R. and Weinberg, Erick J.",
    title = "{Radiative Corrections as the Origin of Spontaneous Symmetry Breaking}",
    doi = "10.1103/PhysRevD.7.1888",
    journal = "Phys. Rev. D",
    volume = "7",
    pages = "1888--1910",
    year = "1973"
}

@article{Witten:1980ez,
    author = "Witten, Edward",
    title = "{Cosmological Consequences of a Light Higgs Boson}",
    reportNumber = "HUTP-80/A040",
    doi = "10.1016/0550-3213(81)90182-6",
    journal = "Nucl. Phys. B",
    volume = "177",
    pages = "477--488",
    year = "1981"
}

@article{DelleRose:2019pgi,
    author = "Delle Rose, Luigi and Panico, Giuliano and Redi, Michele and Tesi, Andrea",
    title = "{Gravitational Waves from Supercool Axions}",
    eprint = "1912.06139",
    archivePrefix = "arXiv",
    primaryClass = "hep-ph",
    doi = "10.1007/JHEP04(2020)025",
    journal = "JHEP",
    volume = "04",
    pages = "025",
    year = "2020"
}

@article{VonHarling:2019rgb,
    author = "Von Harling, Benedict and Pomarol, Alex and Pujol{\`a}s, Oriol and Rompineve, Fabrizio",
    title = "{Peccei-Quinn Phase Transition at LIGO}",
    eprint = "1912.07587",
    archivePrefix = "arXiv",
    primaryClass = "hep-ph",
    doi = "10.1007/JHEP04(2020)195",
    journal = "JHEP",
    volume = "04",
    pages = "195",
    year = "2020"
}

@article{Ghoshal:2020vud,
    author = "Ghoshal, Anish and Salvio, Alberto",
    title = "{Gravitational waves from fundamental axion dynamics}",
    eprint = "2007.00005",
    archivePrefix = "arXiv",
    primaryClass = "hep-ph",
    doi = "10.1007/JHEP12(2020)049",
    journal = "JHEP",
    volume = "12",
    pages = "049",
    year = "2020"
}

@article{MAGIC:2022piy,
    author = "Acciari, V. A. and others",
    collaboration = "MAGIC",
    title = "{A lower bound on intergalactic magnetic fields from time variability of 1ES 0229+200 from MAGIC and Fermi/LAT observations}",
    eprint = "2210.03321",
    archivePrefix = "arXiv",
    primaryClass = "astro-ph.HE",
    doi = "10.1051/0004-6361/202244126",
    journal = "Astron. Astrophys.",
    volume = "670",
    pages = "A145",
    year = "2023"
}

@article{HESS:2023zwb,
    author = "Aharonian, F. and others",
    collaboration = "H.E.S.S., Fermi-LAT",
    title = "{Constraints on the Intergalactic Magnetic Field Using Fermi-LAT and H.E.S.S. Blazar Observations}",
    eprint = "2306.05132",
    archivePrefix = "arXiv",
    primaryClass = "astro-ph.HE",
    doi = "10.3847/2041-8213/acd777",
    journal = "Astrophys. J. Lett.",
    volume = "950",
    number = "2",
    pages = "L16",
    year = "2023"
}

@article{Neronov:2010gir,
    author = "Neronov, A. and Vovk, I.",
    title = "{Evidence for strong extragalactic magnetic fields from Fermi observations of TeV blazars}",
    eprint = "1006.3504",
    archivePrefix = "arXiv",
    primaryClass = "astro-ph.HE",
    doi = "10.1126/science.1184192",
    journal = "Science",
    volume = "328",
    pages = "73--75",
    year = "2010"
}

@article{AlvesBatista:2021sln,
    author = "Alves Batista, Rafael and Saveliev, Andrey",
    title = "{The Gamma-ray Window to Intergalactic Magnetism}",
    eprint = "2105.12020",
    archivePrefix = "arXiv",
    primaryClass = "astro-ph.HE",
    doi = "10.3390/universe7070223",
    journal = "Universe",
    volume = "7",
    number = "7",
    pages = "223",
    year = "2021"
}

@article{Dolag:2010ni,
    author = "Dolag, K. and Kachelriess, M. and Ostapchenko, S. and Tomas, R.",
    title = "{Lower limit on the strength and filling factor of extragalactic magnetic fields}",
    eprint = "1009.1782",
    archivePrefix = "arXiv",
    primaryClass = "astro-ph.HE",
    doi = "10.1088/2041-8205/727/1/L4",
    journal = "Astrophys. J. Lett.",
    volume = "727",
    pages = "L4",
    year = "2011"
}

@article{Vachaspati:1991nm,
    author = "Vachaspati, T.",
    title = "{Magnetic fields from cosmological phase transitions}",
    doi = "10.1016/0370-2693(91)90051-Q",
    journal = "Phys. Lett. B",
    volume = "265",
    pages = "258--261",
    year = "1991"
}

@article{Ellis:2019tjf,
    author = "Ellis, John and Fairbairn, Malcolm and Lewicki, Marek and Vaskonen, Ville and Wickens, Alastair",
    title = "{Intergalactic Magnetic Fields from First-Order Phase Transitions}",
    eprint = "1907.04315",
    archivePrefix = "arXiv",
    primaryClass = "astro-ph.CO",
    reportNumber = "KCL-PH-TH/2019-60, CERN-TH-2019-104",
    doi = "10.1088/1475-7516/2019/09/019",
    journal = "JCAP",
    volume = "09",
    pages = "019",
    year = "2019"
}

@article{Sigl:1996dm,
    author = "Sigl, Guenter and Olinto, Angela V. and Jedamzik, Karsten",
    title = "{Primordial magnetic fields from cosmological first order phase transitions}",
    eprint = "astro-ph/9610201",
    archivePrefix = "arXiv",
    doi = "10.1103/PhysRevD.55.4582",
    journal = "Phys. Rev. D",
    volume = "55",
    pages = "4582--4590",
    year = "1997"
}

@article{Tevzadze:2012kk,
    author = "Tevzadze, Alexander G. and Kisslinger, Leonard and Brandenburg, Axel and Kahniashvili, Tina",
    title = "{Magnetic Fields from QCD Phase Transitions}",
    eprint = "1207.0751",
    archivePrefix = "arXiv",
    primaryClass = "astro-ph.CO",
    reportNumber = "NORDITA-2012-52",
    doi = "10.1088/0004-637X/759/1/54",
    journal = "Astrophys. J.",
    volume = "759",
    pages = "54",
    year = "2012"
}

@article{Vachaspati:2001nb,
    author = "Vachaspati, Tanmay",
    title = "{Estimate of the primordial magnetic field helicity}",
    eprint = "astro-ph/0101261",
    archivePrefix = "arXiv",
    doi = "10.1103/PhysRevLett.87.251302",
    journal = "Phys. Rev. Lett.",
    volume = "87",
    pages = "251302",
    year = "2001"
}

@article{Copi:2008he,
    author = "Copi, Craig J. and Ferrer, Francesc and Vachaspati, Tanmay and Achucarro, Ana",
    title = "{Helical Magnetic Fields from Sphaleron Decay and Baryogenesis}",
    eprint = "0801.3653",
    archivePrefix = "arXiv",
    primaryClass = "astro-ph",
    doi = "10.1103/PhysRevLett.101.171302",
    journal = "Phys. Rev. Lett.",
    volume = "101",
    pages = "171302",
    year = "2008"
}

@article{Kamionkowski:1993fg,
    author = "Kamionkowski, Marc and Kosowsky, Arthur and Turner, Michael S.",
    title = "{Gravitational radiation from first order phase transitions}",
    eprint = "astro-ph/9310044",
    archivePrefix = "arXiv",
    reportNumber = "IASSNS-HEP-93-44, FERMILAB-PUB-93-235-A",
    doi = "10.1103/PhysRevD.49.2837",
    journal = "Phys. Rev. D",
    volume = "49",
    pages = "2837--2851",
    year = "1994"
}

@article{Brandenburg:1996fc,
    author = "Brandenburg, Axel and Enqvist, Kari and Olesen, Poul",
    title = "{Large scale magnetic fields from hydromagnetic turbulence in the very early universe}",
    eprint = "astro-ph/9602031",
    archivePrefix = "arXiv",
    reportNumber = "NORDITA-96-6-A",
    doi = "10.1103/PhysRevD.54.1291",
    journal = "Phys. Rev. D",
    volume = "54",
    pages = "1291--1300",
    year = "1996"
}

@article{Christensson:2000sp,
    author = "Christensson, Mattias and Hindmarsh, Mark and Brandenburg, Axel",
    title = "{Inverse cascade in decaying 3-D magnetohydrodynamic turbulence}",
    eprint = "astro-ph/0011321",
    archivePrefix = "arXiv",
    doi = "10.1103/PhysRevE.64.056405",
    journal = "Phys. Rev. E",
    volume = "64",
    pages = "056405",
    year = "2001"
}

@article{Kahniashvili:2010gp,
    author = "Kahniashvili, Tina and Brandenburg, Axel and Tevzadze, Alexander G. and Ratra, Bharat",
    title = "{Numerical simulations of the decay of primordial magnetic turbulence}",
    eprint = "1004.3084",
    archivePrefix = "arXiv",
    primaryClass = "astro-ph.CO",
    reportNumber = "NORDITA-PREPRINT-NORDITA-2010-20",
    doi = "10.1103/PhysRevD.81.123002",
    journal = "Phys. Rev. D",
    volume = "81",
    pages = "123002",
    year = "2010"
}

@article{Brandenburg:2017neh,
    author = "Brandenburg, Axel and Kahniashvili, Tina and Mandal, Sayan and Roper Pol, Alberto and Tevzadze, Alexander G. and Vachaspati, Tanmay",
    title = "{Evolution of hydromagnetic turbulence from the electroweak phase transition}",
    eprint = "1711.03804",
    archivePrefix = "arXiv",
    primaryClass = "astro-ph.CO",
    reportNumber = "NORDITA-2017-116",
    doi = "10.1103/PhysRevD.96.123528",
    journal = "Phys. Rev. D",
    volume = "96",
    number = "12",
    pages = "123528",
    year = "2017"
}

@article{Chu:2011tx,
    author = "Chu, Yi-Zen and Dent, James B. and Vachaspati, Tanmay",
    title = "{Magnetic Helicity in Sphaleron Debris}",
    eprint = "1105.3744",
    archivePrefix = "arXiv",
    primaryClass = "hep-th",
    doi = "10.1103/PhysRevD.83.123530",
    journal = "Phys. Rev. D",
    volume = "83",
    pages = "123530",
    year = "2011"
}

@article{Kahniashvili:2012uj,
    author = "Kahniashvili, Tina and Tevzadze, Alexander G. and Brandenburg, Axel and Neronov, Andrii",
    title = "{Evolution of Primordial Magnetic Fields from Phase Transitions}",
    eprint = "1212.0596",
    archivePrefix = "arXiv",
    primaryClass = "astro-ph.CO",
    reportNumber = "NORDITA-2012-96",
    doi = "10.1103/PhysRevD.87.083007",
    journal = "Phys. Rev. D",
    volume = "87",
    number = "8",
    pages = "083007",
    year = "2013"
}

@article{Forbes:2000gr,
    author = "Forbes, Michael McNeil and Zhitnitsky, Ariel R.",
    title = "{Primordial galactic magnetic fields from domain walls at the QCD phase transition}",
    eprint = "hep-ph/0004051",
    archivePrefix = "arXiv",
    doi = "10.1103/PhysRevLett.85.5268",
    journal = "Phys. Rev. Lett.",
    volume = "85",
    pages = "5268--5271",
    year = "2000"
}

@article{Sakharov:1967dj,
    author = "Sakharov, A. D.",
    title = "{Violation of CP Invariance, C asymmetry, and baryon asymmetry of the universe}",
    doi = "10.1070/PU1991v034n05ABEH002497",
    journal = "Pisma Zh. Eksp. Teor. Fiz.",
    volume = "5",
    pages = "32--35",
    year = "1967"
}

@article{Brandenburg:2017rnt,
    author = "Brandenburg, Axel and Kahniashvili, Tina and Mandal, Sayan and Roper Pol, Alberto and Tevzadze, Alexander G. and Vachaspati, Tanmay",
    title = "{The dynamo effect in decaying helical turbulence}",
    eprint = "1710.01628",
    archivePrefix = "arXiv",
    primaryClass = "physics.flu-dyn",
    reportNumber = "NORDITA-2017-099",
    doi = "10.1103/PhysRevFluids.4.024608",
    journal = "Phys. Rev. Fluids.",
    volume = "4",
    pages = "024608",
    year = "2019"
}

@article{Banerjee:2004df,
    author = "Banerjee, Robi and Jedamzik, Karsten",
    title = "{The Evolution of cosmic magnetic fields: From the very early universe, to recombination, to the present}",
    eprint = "astro-ph/0410032",
    archivePrefix = "arXiv",
    doi = "10.1103/PhysRevD.70.123003",
    journal = "Phys. Rev. D",
    volume = "70",
    pages = "123003",
    year = "2004"
}

@article{RoperPol:2023bqa,
    author = "Roper Pol, A. and Neronov, A. and Caprini, C. and Boyer, T. and Semikoz, D.",
    title = "{LISA and $\gamma$-ray telescopes as multi-messenger probes of a first-order cosmological phase transition}",
    eprint = "2307.10744",
    archivePrefix = "arXiv",
    primaryClass = "astro-ph.CO",
    month = "7",
    year = "2023"
}

@article{Kahniashvili:2009qi,
    author = "Kahniashvili, Tina and Tevzadze, Alexander G. and Ratra, Bharat",
    title = "{Phase Transition Generated Cosmological Magnetic Field at Large Scales}",
    eprint = "0907.0197",
    archivePrefix = "arXiv",
    primaryClass = "astro-ph.CO",
    doi = "10.1088/0004-637X/726/2/78",
    journal = "Astrophys. J.",
    volume = "726",
    pages = "78",
    year = "2011"
}

@article{Durrer:2013pga,
    author = "Durrer, Ruth and Neronov, Andrii",
    title = "{Cosmological Magnetic Fields: Their Generation, Evolution and Observation}",
    eprint = "1303.7121",
    archivePrefix = "arXiv",
    primaryClass = "astro-ph.CO",
    doi = "10.1007/s00159-013-0062-7",
    journal = "Astron. Astrophys. Rev.",
    volume = "21",
    pages = "62",
    year = "2013"
}

@article{Caprini:2019egz,
    author = "Caprini, Chiara and others",
    title = "{Detecting gravitational waves from cosmological phase transitions with LISA: an update}",
    eprint = "1910.13125",
    archivePrefix = "arXiv",
    primaryClass = "astro-ph.CO",
    reportNumber = "DESY-19-159, IPPP/19/27, HIP-2019-14/TH, MITP/19-066, IFT-UAM/CSIC-19-139",
    doi = "10.1088/1475-7516/2020/03/024",
    journal = "JCAP",
    volume = "03",
    pages = "024",
    year = "2020"
}

@article{Salvio:2023qgb,
    author = "Salvio, Alberto",
    title = "{Model-independent radiative symmetry breaking and gravitational waves}",
    eprint = "2302.10212",
    archivePrefix = "arXiv",
    primaryClass = "hep-ph",
    doi = "10.1088/1475-7516/2023/04/051",
    journal = "JCAP",
    volume = "04",
    pages = "051",
    year = "2023"
}

@article{Vachaspati:2016xji,
    author = "Vachaspati, Tanmay",
    title = "{Fundamental Implications of Intergalactic Magnetic Field Observations}",
    eprint = "1606.06186",
    archivePrefix = "arXiv",
    primaryClass = "astro-ph.CO",
    doi = "10.1103/PhysRevD.95.063505",
    journal = "Phys. Rev. D",
    volume = "95",
    number = "6",
    pages = "063505",
    year = "2017"
}

@article{Vachaspati:2020blt,
    author = "Vachaspati, Tanmay",
    title = "{Progress on cosmological magnetic fields}",
    eprint = "2010.10525",
    archivePrefix = "arXiv",
    primaryClass = "astro-ph.CO",
    doi = "10.1088/1361-6633/ac03a9",
    journal = "Rept. Prog. Phys.",
    volume = "84",
    number = "7",
    pages = "074901",
    year = "2021"
}

@article{Gildener:1976ih,
    author = "Gildener, Eldad and Weinberg, Steven",
    title = "{Symmetry Breaking and Scalar Bosons}",
    reportNumber = "PRINT-76-0068 (HARVARD)",
    doi = "10.1103/PhysRevD.13.3333",
    journal = "Phys. Rev. D",
    volume = "13",
    pages = "3333",
    year = "1976"
}

@article{Conaci:2024tlc,
    author = "Conaci, Angela and Delle Rose, Luigi and Dev, P. S. Bhupal and Ghoshal, Anish",
    title = "{Slaying axion-like particles via gravitational waves and primordial black holes from supercooled phase transition}",
    eprint = "2401.09411",
    archivePrefix = "arXiv",
    primaryClass = "astro-ph.CO",
    doi = "10.1007/JHEP12(2024)196",
    journal = "JHEP",
    volume = "12",
    pages = "196",
    year = "2024"
}

@article{Hambye:2013dgv,
    author = "Hambye, Thomas and Strumia, Alessandro",
    title = "{Dynamical generation of the weak and Dark Matter scale}",
    eprint = "1306.2329",
    archivePrefix = "arXiv",
    primaryClass = "hep-ph",
    doi = "10.1103/PhysRevD.88.055022",
    journal = "Phys. Rev. D",
    volume = "88",
    pages = "055022",
    year = "2013"
}

@article{Iso:2017uuu,
    author = "Iso, Satoshi and Serpico, Pasquale D. and Shimada, Kengo",
    title = "{QCD-Electroweak First-Order Phase Transition in a Supercooled Universe}",
    eprint = "1704.04955",
    archivePrefix = "arXiv",
    primaryClass = "hep-ph",
    reportNumber = "KEK-TH-1969, LAPTH-008-17",
    doi = "10.1103/PhysRevLett.119.141301",
    journal = "Phys. Rev. Lett.",
    volume = "119",
    number = "14",
    pages = "141301",
    year = "2017"
}

@article{Hambye:2018qjv,
    author = "Hambye, Thomas and Strumia, Alessandro and Teresi, Daniele",
    title = "{Super-cool Dark Matter}",
    eprint = "1805.01473",
    archivePrefix = "arXiv",
    primaryClass = "hep-ph",
    reportNumber = "ULB-TH/18-06, CERN-TH-2018-110, IFUP-TH/2018, ULB-TH-18-06",
    doi = "10.1007/JHEP08(2018)188",
    journal = "JHEP",
    volume = "08",
    pages = "188",
    year = "2018"
}

@article{Allen:1997ad,
    author = "Allen, Bruce and Romano, Joseph D.",
    title = "{Detecting a stochastic background of gravitational radiation: Signal processing strategies and sensitivities}",
    eprint = "gr-qc/9710117",
    archivePrefix = "arXiv",
    reportNumber = "WISC-MILW-97-TH-14",
    doi = "10.1103/PhysRevD.59.102001",
    journal = "Phys. Rev. D",
    volume = "59",
    pages = "102001",
    year = "1999"
}

@article{Kudoh:2005as,
    author = "Kudoh, Hideaki and Taruya, Atsushi and Hiramatsu, Takashi and Himemoto, Yoshiaki",
    title = "{Detecting a gravitational-wave background with next-generation space interferometers}",
    eprint = "gr-qc/0511145",
    archivePrefix = "arXiv",
    reportNumber = "UTAP-544, RESCEU-37-05",
    doi = "10.1103/PhysRevD.73.064006",
    journal = "Phys. Rev. D",
    volume = "73",
    pages = "064006",
    year = "2006"
}

@article{Thrane:2013oya,
    author = "Thrane, Eric and Romano, Joseph D.",
    title = "{Sensitivity curves for searches for gravitational-wave backgrounds}",
    eprint = "1310.5300",
    archivePrefix = "arXiv",
    primaryClass = "astro-ph.IM",
    doi = "10.1103/PhysRevD.88.124032",
    journal = "Phys. Rev. D",
    volume = "88",
    number = "12",
    pages = "124032",
    year = "2013"
}

@article{Caprini:2019pxz,
    author = "Caprini, Chiara and Figueroa, Daniel G. and Flauger, Raphael and Nardini, Germano and Peloso, Marco and Pieroni, Mauro and Ricciardone, Angelo and Tasinato, Gianmassimo",
    title = "{Reconstructing the spectral shape of a stochastic gravitational wave background with LISA}",
    eprint = "1906.09244",
    archivePrefix = "arXiv",
    primaryClass = "astro-ph.CO",
    reportNumber = "LISA-CosWG-19-02",
    doi = "10.1088/1475-7516/2019/11/017",
    journal = "JCAP",
    volume = "11",
    pages = "017",
    year = "2019"
}

@article{Brzeminski:2022haa,
    author = "Brzeminski, Dawid and Hook, Anson and Marques-Tavares, Gustavo",
    title = "{Precision early universe cosmology from stochastic gravitational waves}",
    eprint = "2203.13842",
    archivePrefix = "arXiv",
    primaryClass = "hep-ph",
    doi = "10.1007/JHEP11(2022)061",
    journal = "JHEP",
    volume = "11",
    pages = "061",
    year = "2022"
}

@article{Caprini:2024hue,
    author = "Caprini, Chiara and Jinno, Ryusuke and Lewicki, Marek and Madge, Eric and Merchand, Marco and Nardini, Germano and Pieroni, Mauro and Roper Pol, Alberto and Vaskonen, Ville",
    collaboration = "LISA Cosmology Working Group",
    title = "{Gravitational waves from first-order phase transitions in LISA: reconstruction pipeline and physics interpretation}",
    eprint = "2403.03723",
    archivePrefix = "arXiv",
    primaryClass = "astro-ph.CO",
    reportNumber = "LISA-COSWG-24-01, CERN-TH-2024-029",
    doi = "10.1088/1475-7516/2024/10/020",
    journal = "JCAP",
    volume = "10",
    pages = "020",
    year = "2024"
}

@article{Akita:2020szl,
    author = "Akita, Kensuke and Yamaguchi, Masahide",
    title = "{A precision calculation of relic neutrino decoupling}",
    eprint = "2005.07047",
    archivePrefix = "arXiv",
    primaryClass = "hep-ph",
    doi = "10.1088/1475-7516/2020/08/012",
    journal = "JCAP",
    volume = "08",
    pages = "012",
    year = "2020"
}

@article{Froustey:2020mcq,
    author = "Froustey, Julien and Pitrou, Cyril and Volpe, Maria Cristina",
    title = "{Neutrino decoupling including flavour oscillations and primordial nucleosynthesis}",
    eprint = "2008.01074",
    archivePrefix = "arXiv",
    primaryClass = "hep-ph",
    doi = "10.1088/1475-7516/2020/12/015",
    journal = "JCAP",
    volume = "12",
    pages = "015",
    year = "2020"
}

@article{CMB-S4:2020lpa,
    author = "Abazajian, Kevork and others",
    collaboration = "CMB-S4",
    title = "{CMB-S4: Forecasting Constraints on Primordial Gravitational Waves}",
    eprint = "2008.12619",
    archivePrefix = "arXiv",
    primaryClass = "astro-ph.CO",
    reportNumber = "FERMILAB-PUB-20-468-AE-SCD",
    doi = "10.3847/1538-4357/ac1596",
    journal = "Astrophys. J.",
    volume = "926",
    number = "1",
    pages = "54",
    year = "2022"
}

@article{Sehgal:2019ewc,
    author = "Sehgal, Neelima and others",
    title = "{CMB-HD: An Ultra-Deep, High-Resolution Millimeter-Wave Survey Over Half the Sky}",
    eprint = "1906.10134",
    archivePrefix = "arXiv",
    primaryClass = "astro-ph.CO",
    journal = "Bull. Am. Astron. Soc.",
    volume = "51",
    number = "7",
    pages = "1--23",
    year = "2019"
}

@article{Bauer:2020jbp,
    author = "Bauer, Martin and Neubert, Matthias and Renner, Sophie and Schnubel, Marvin and Thamm, Andrea",
    title = "{The Low-Energy Effective Theory of Axions and ALPs}",
    eprint = "2012.12272",
    archivePrefix = "arXiv",
    primaryClass = "hep-ph",
    reportNumber = "IPPP/20/69, MITP/20-070 SISSA 30/2020/FISI, ZH-TH-47/20",
    doi = "10.1007/JHEP04(2021)063",
    journal = "JHEP",
    volume = "04",
    pages = "063",
    year = "2021"
}

@article{Kelly:2020dda,
    author = "Kelly, Kevin J. and Kumar, Soubhik and Liu, Zhen",
    title = "{Heavy axion opportunities at the DUNE near detector}",
    eprint = "2011.05995",
    archivePrefix = "arXiv",
    primaryClass = "hep-ph",
    reportNumber = "FERMILAB-PUB-20-581-T",
    doi = "10.1103/PhysRevD.103.095002",
    journal = "Phys. Rev. D",
    volume = "103",
    number = "9",
    pages = "095002",
    year = "2021"
}

@article{Ferber:2022rsf,
    author = {Ferber, Torben and Filimonova, Anastasiia and Sch{\"a}fer, Ruth and Westhoff, Susanne},
    title = "{Displaced or invisible? ALPs from B decays at Belle II}",
    eprint = "2201.06580",
    archivePrefix = "arXiv",
    primaryClass = "hep-ph",
    reportNumber = "P3H-22-005, Nikhef 2022-001",
    doi = "10.1007/JHEP04(2023)131",
    journal = "JHEP",
    volume = "04",
    pages = "131",
    year = "2023"
}

@article{Nikishov:1962rmq,
    author = "Nikishov, A. I.",
    title = "{Absorption of High-Energy Photons in the Universe}",
    journal = "Sov. Phys. JETP",
    volume = "14",
    pages = "393--394",
    year = "1962"
}

@article{Gould:1966pza,
    author = "Gould, Robert and Schr{\'e}der, Gerald",
    title = "{Opacity of the Universe to High-Energy Photons}",
    doi = "10.1103/PhysRevLett.16.252",
    journal = "Phys. Rev. Lett.",
    volume = "16",
    number = "6",
    pages = "252--254",
    year = "1966"
}

@article{Franceschini:2021wkr,
    author = "Franceschini, Alberto",
    title = "{Photon{\textendash}Photon Interactions and the Opacity of the Universe in Gamma Rays}",
    doi = "10.3390/universe7050146",
    journal = "Universe",
    volume = "7",
    number = "5",
    pages = "146",
    year = "2021"
}

@article{Planck:2018vyg,
    author = "Aghanim, N. and others",
    collaboration = "Planck",
    title = "{Planck 2018 results. VI. Cosmological parameters}",
    eprint = "1807.06209",
    archivePrefix = "arXiv",
    primaryClass = "astro-ph.CO",
    doi = "10.1051/0004-6361/201833910",
    journal = "Astron. Astrophys.",
    volume = "641",
    pages = "A6",
    year = "2020",
    note = "[Erratum: Astron.Astrophys. 652, C4 (2021)]"
}

@article{Drewes:2024wbw,
    author = "Drewes, Marco and Georis, Yannis and Klasen, Michael and Wiggering, Luca Paolo and Wong, Yvonne Y. Y.",
    title = "{Towards a precision calculation of N $_{eff}$ in the Standard Model. Part III. Improved estimate of NLO contributions to the collision integral}",
    eprint = "2402.18481",
    archivePrefix = "arXiv",
    primaryClass = "hep-ph",
    reportNumber = "CPPC-2024-01, MS-TP-24-06",
    doi = "10.1088/1475-7516/2024/06/032",
    journal = "JCAP",
    volume = "06",
    pages = "032",
    year = "2024"
}

@article{Goldstein:2026iuu,
    author = "Goldstein, Samuel and Hill, J. Colin",
    title = "{A 2{\%} determination of $N_{\rm eff}$ from primordial element abundance, cosmic microwave background, and baryon acoustic oscillation measurements}",
    eprint = "2603.13226",
    archivePrefix = "arXiv",
    primaryClass = "astro-ph.CO",
    month = "3",
    year = "2026"
}

@inproceedings{Alimena:2025kjv,
    author = "Alimena, J. and others",
    title = "{Feebly-Interacting Particles: FIPs at LHCb {\textemdash} Workshop Report 2025 Edition}",
    booktitle = "{LHCb FIP Physics Workshop 2025}",
    eprint = "2510.05257",
    archivePrefix = "arXiv",
    primaryClass = "hep-ph",
    month = "10",
    year = "2025"
}

@article{Bauer:2017ris,
    author = "Bauer, Martin and Neubert, Matthias and Thamm, Andrea",
    title = "{Collider Probes of Axion-Like Particles}",
    eprint = "1708.00443",
    archivePrefix = "arXiv",
    primaryClass = "hep-ph",
    reportNumber = "MITP-17-047",
    doi = "10.1007/JHEP12(2017)044",
    journal = "JHEP",
    volume = "12",
    pages = "044",
    year = "2017"
}

@article{PhysRev.82.863,
  title = {Cosmic Radiation and Cosmic Magnetic Fields. II. Origin of Cosmic Magnetic Fields},
  author = {Biermann, Ludwig and Schl\"uter, Arnulf},
  journal = {Phys. Rev.},
  volume = {82},
  issue = {6},
  pages = {863--868},
  numpages = {0},
  year = {1951},
  month = {Jun},
  publisher = {American Physical Society},
  doi = {10.1103/PhysRev.82.863},
  url = {https://link.aps.org/doi/10.1103/PhysRev.82.863}
}

@article{Athron:2023xlk,
    author = "Athron, Peter and Bal{\'a}zs, Csaba and Fowlie, Andrew and Morris, Lachlan and Wu, Lei",
    title = "{Cosmological phase transitions: From perturbative particle physics to gravitational waves}",
    eprint = "2305.02357",
    archivePrefix = "arXiv",
    primaryClass = "hep-ph",
    doi = "10.1016/j.ppnp.2023.104094",
    journal = "Prog. Part. Nucl. Phys.",
    volume = "135",
    pages = "104094",
    year = "2024"
}

@misc{AxionLimits,
  author =        {Ciaran O'Hare},
  howpublished =  {\url{https://cajohare.github.io/AxionLimits/}},
  publisher =     {Zenodo},
  title =         {AxionLimits},
  year =          {2020},
  doi =           {10.5281/zenodo.3932430},
}

@article{OPAL:2000puu,
    author = "Abbiendi, G. and others",
    collaboration = "OPAL",
    title = "{Photonic events with missing energy in e+ e- collisions at $\sqrt{s} = 189$ GeV}",
    eprint = "hep-ex/0005002",
    archivePrefix = "arXiv",
    reportNumber = "CERN-EP-2000-050",
    doi = "10.1007/s100520000522",
    journal = "Eur. Phys. J. C",
    volume = "18",
    pages = "253--272",
    year = "2000"
}

@article{L3:2003yon,
    author = "Achard, P. and others",
    collaboration = "L3",
    title = "{Single photon and multiphoton events with missing energy in $e^{+} e^{-}$ collisions at LEP}",
    eprint = "hep-ex/0402002",
    archivePrefix = "arXiv",
    reportNumber = "CERN-EP-2003-068",
    doi = "10.1016/j.physletb.2004.01.010",
    journal = "Phys. Lett. B",
    volume = "587",
    pages = "16--32",
    year = "2004"
}

@article{DELPHI:2003dlq,
    author = "Abdallah, J. and others",
    collaboration = "DELPHI",
    title = "{Photon events with missing energy in e+ e- collisions at $\sqrt{s} = 130$ GeV to $209$ GeV}",
    eprint = "hep-ex/0406019",
    archivePrefix = "arXiv",
    reportNumber = "CERN-EP-2003-093",
    doi = "10.1140/epjc/s2004-02051-8",
    journal = "Eur. Phys. J. C",
    volume = "38",
    pages = "395--411",
    year = "2005"
}

@article{BaBar:2010eww,
    author = "del Amo Sanchez, P. and others",
    collaboration = "BaBar",
    title = "{Search for Production of Invisible Final States in Single-Photon Decays of $\Upsilon(1S)$}",
    eprint = "1007.4646",
    archivePrefix = "arXiv",
    primaryClass = "hep-ex",
    reportNumber = "SLAC-PUB-14204, BABAR-PUB-10-23, BABAR-PUB-10-023",
    doi = "10.1103/PhysRevLett.107.021804",
    journal = "Phys. Rev. Lett.",
    volume = "107",
    pages = "021804",
    year = "2011"
}

@article{CMS:2012lmn,
    author = "Chatrchyan, Serguei and others",
    collaboration = "CMS",
    title = "{Search for Dark Matter and Large Extra Dimensions in pp Collisions Yielding a Photon and Missing Transverse Energy}",
    eprint = "1204.0821",
    archivePrefix = "arXiv",
    primaryClass = "hep-ex",
    reportNumber = "CMS-EXO-11-096, CERN-PH-EP-2012-084",
    doi = "10.1103/PhysRevLett.108.261803",
    journal = "Phys. Rev. Lett.",
    volume = "108",
    pages = "261803",
    year = "2012"
}

@article{ATLAS:2012ezx,
    author = "Aad, Georges and others",
    collaboration = "ATLAS",
    title = "{Search for dark matter candidates and large extra dimensions in events with a photon and missing transverse momentum in $pp$ collision data at $\sqrt{s}=7$ TeV with the ATLAS detector}",
    eprint = "1209.4625",
    archivePrefix = "arXiv",
    primaryClass = "hep-ex",
    reportNumber = "CERN-PH-EP-2012-209",
    doi = "10.1103/PhysRevLett.110.011802",
    journal = "Phys. Rev. Lett.",
    volume = "110",
    number = "1",
    pages = "011802",
    year = "2013"
}

@article{ATLAS:2014kci,
    author = "Aad, Georges and others",
    collaboration = "ATLAS",
    title = "{Search for new phenomena in events with a photon and missing transverse momentum in $pp$ collisions at $\sqrt{s}=8$ TeV with the ATLAS detector}",
    eprint = "1411.1559",
    archivePrefix = "arXiv",
    primaryClass = "hep-ex",
    reportNumber = "CERN-PH-EP-2014-245",
    doi = "10.1103/PhysRevD.91.012008",
    journal = "Phys. Rev. D",
    volume = "91",
    number = "1",
    pages = "012008",
    year = "2015",
    note = "[Erratum: Phys.Rev.D 92, 059903 (2015)]"
}

@article{Friedland:2012hj,
    author = "Friedland, Alexander and Giannotti, Maurizio and Wise, Michael",
    title = "{Constraining the Axion-Photon Coupling with Massive Stars}",
    eprint = "1210.1271",
    archivePrefix = "arXiv",
    primaryClass = "hep-ph",
    reportNumber = "LA-UR-12-25074",
    doi = "10.1103/PhysRevLett.110.061101",
    journal = "Phys. Rev. Lett.",
    volume = "110",
    number = "6",
    pages = "061101",
    year = "2013"
}

@article{Aoyama:2015asa,
    author = "Aoyama, Shohei and Suzuki, Takeru K.",
    title = "{Effects of axions on Nucleosynthesis in massive stars}",
    eprint = "1502.02357",
    archivePrefix = "arXiv",
    primaryClass = "astro-ph.SR",
    doi = "10.1103/PhysRevD.92.063016",
    journal = "Phys. Rev. D",
    volume = "92",
    number = "6",
    pages = "063016",
    year = "2015"
}

@article{Dominguez:2017yhy,
    author = "Dom{\i}nguez, I. and Giannotti, M. and Mirizzi, A. and Straniero, O.",
    editor = "Karakas, A. and Ventura, P. and Dell'Agli, F. and Di Criscienzo, M.",
    title = "{On the influence of axions on Mup}",
    journal = "Mem. Soc. Ast. It.",
    volume = "88",
    number = "3",
    pages = "270--273",
    year = "2017"
}

@article{Reynes:2021bpe,
    author = "Reyn{\'e}s, J{\'u}lia Sisk and Matthews, James H. and Reynolds, Christopher S. and Russell, Helen R. and Smith, Robyn N. and Marsh, M. C. David",
    title = "{New constraints on light axion-like particles using Chandra transmission grating spectroscopy of the powerful cluster-hosted quasar H1821+643}",
    eprint = "2109.03261",
    archivePrefix = "arXiv",
    primaryClass = "astro-ph.HE",
    doi = "10.1093/mnras/stab3464",
    journal = "Mon. Not. Roy. Astron. Soc.",
    volume = "510",
    number = "1",
    pages = "1264--1277",
    year = "2021"
}

@article{Ning:2024eky,
    author = "Ning, Orion and Safdi, Benjamin R.",
    title = "{Leading Axion-Photon Sensitivity with NuSTAR Observations of M82 and M87}",
    eprint = "2404.14476",
    archivePrefix = "arXiv",
    primaryClass = "hep-ph",
    doi = "10.1103/PhysRevLett.134.171003",
    journal = "Phys. Rev. Lett.",
    volume = "134",
    number = "17",
    pages = "171003",
    year = "2025"
}

@article{Candon:2024eah,
    author = "Cand{\'o}n, Francisco R. and Fiorillo, Damiano F. G. and Lucente, Giuseppe and Vitagliano, Edoardo and Vogel, Julia K.",
    title = "{NuSTAR Bounds on Radiatively Decaying Particles from M82}",
    eprint = "2412.03660",
    archivePrefix = "arXiv",
    primaryClass = "hep-ph",
    doi = "10.1103/PhysRevLett.134.171004",
    journal = "Phys. Rev. Lett.",
    volume = "134",
    number = "17",
    pages = "171004",
    year = "2025"
}

@article{Mondino:2024rif,
    author = "Mondino, Cristina and P{\^\i}rvu, Dalila and Huang, Junwu and Johnson, Matthew C.",
    title = "{Axion-induced patchy screening of the Cosmic Microwave Background}",
    eprint = "2405.08059",
    archivePrefix = "arXiv",
    primaryClass = "hep-ph",
    doi = "10.1088/1475-7516/2024/10/107",
    journal = "JCAP",
    volume = "10",
    pages = "107",
    year = "2024"
}

@article{Goldstein:2024mfp,
    author = "Goldstein, Samuel and McCarthy, Fiona and Mondino, Cristina and Hill, J. Colin and Huang, Junwu and Johnson, Matthew C.",
    title = "{Constraints on Axions from Patchy Screening of the Cosmic Microwave Background}",
    eprint = "2409.10514",
    archivePrefix = "arXiv",
    primaryClass = "astro-ph.CO",
    doi = "10.1103/PhysRevLett.134.081001",
    journal = "Phys. Rev. Lett.",
    volume = "134",
    number = "8",
    pages = "081001",
    year = "2025"
}

@article{Depta:2020wmr,
    author = "Depta, Paul Frederik and Hufnagel, Marco and Schmidt-Hoberg, Kai",
    title = "{Robust cosmological constraints on axion-like particles}",
    eprint = "2002.08370",
    archivePrefix = "arXiv",
    primaryClass = "hep-ph",
    reportNumber = "DESY-20-003, DESY 20-003",
    doi = "10.1088/1475-7516/2020/05/009",
    journal = "JCAP",
    volume = "05",
    pages = "009",
    year = "2020"
}

@article{Langhoff:2022bij,
    author = "Langhoff, Kevin and Outmezguine, Nadav Joseph and Rodd, Nicholas L.",
    title = "{Irreducible Axion Background}",
    eprint = "2209.06216",
    archivePrefix = "arXiv",
    primaryClass = "hep-ph",
    reportNumber = "CERN-TH-2022-148",
    doi = "10.1103/PhysRevLett.129.241101",
    journal = "Phys. Rev. Lett.",
    volume = "129",
    number = "24",
    pages = "241101",
    year = "2022"
}

@article{Ovchynnikov:2023cry,
    author = "Ovchynnikov, Maksym and Tastet, Jean-Loup and Mikulenko, Oleksii and Bondarenko, Kyrylo",
    title = "{Sensitivities to feebly interacting particles: Public and unified calculations}",
    eprint = "2305.13383",
    archivePrefix = "arXiv",
    primaryClass = "hep-ph",
    doi = "10.1103/PhysRevD.108.075028",
    journal = "Phys. Rev. D",
    volume = "108",
    number = "7",
    pages = "075028",
    year = "2023"
}

@misc{Aberle:2022SHiP,
  author        = "Aberle, O. and others",
  collaboration = "SHiP",
  title         = "{BDF/SHiP at the ECN3 high-intensity beam facility}",
  institution   = "CERN",
  howpublished  = "CERN-SPSC-2022-032, SPSC-I-258",
  address       = "Geneva",
  year          = "2022",
  url           = "https://cds.cern.ch/record/2839677"
}

@article{FASER:2018eoc,
    author = "Ariga, Akitaka and others",
    collaboration = "FASER",
    title = "{FASER{\textquoteright}s physics reach for long-lived particles}",
    eprint = "1811.12522",
    archivePrefix = "arXiv",
    primaryClass = "hep-ph",
    reportNumber = "UCI-TR-2018-19, KYUSHU-RCAPP-2018-06",
    doi = "10.1103/PhysRevD.99.095011",
    journal = "Phys. Rev. D",
    volume = "99",
    number = "9",
    pages = "095011",
    year = "2019"
}

@article{FASER:2024bbl,
    author = "Mammen Abraham, Roshan and others",
    collaboration = "FASER",
    title = "{Shining light on the dark sector: search for axion-like particles and other new physics in photonic final states with FASER}",
    eprint = "2410.10363",
    archivePrefix = "arXiv",
    primaryClass = "hep-ex",
    reportNumber = "CERN-EP-2024-262",
    doi = "10.1007/JHEP01(2025)199",
    journal = "JHEP",
    volume = "01",
    pages = "199",
    year = "2025"
}

@article{Brdar:2020dpr,
    author = "Brdar, Vedran and Dutta, Bhaskar and Jang, Wooyoung and Kim, Doojin and Shoemaker, Ian M. and Tabrizi, Zahra and Thompson, Adrian and Yu, Jaehoon",
    title = "{Axionlike Particles at Future Neutrino Experiments: Closing the Cosmological Triangle}",
    eprint = "2011.07054",
    archivePrefix = "arXiv",
    primaryClass = "hep-ph",
    reportNumber = "FERMILAB-PUB-20-645-V, MI-TH-2029",
    doi = "10.1103/PhysRevLett.126.201801",
    journal = "Phys. Rev. Lett.",
    volume = "126",
    number = "20",
    pages = "201801",
    year = "2021"
}

@article{Acanfora:2024spi,
    author = "Acanfora, Francesca and Franceschini, Roberto and Mastroddi, Alessio and Redigolo, Diego",
    title = "{Fusing photons into diphoton resonances at Belle II and beyond}",
    eprint = "2406.14614",
    archivePrefix = "arXiv",
    primaryClass = "hep-ph",
    doi = "10.1007/JHEP12(2024)099",
    journal = "JHEP",
    volume = "12",
    pages = "099",
    year = "2024"
}

@article{Feng:2022inv,
    author = "Feng, Jonathan L. and others",
    title = "{The Forward Physics Facility at the High-Luminosity LHC}",
    eprint = "2203.05090",
    archivePrefix = "arXiv",
    primaryClass = "hep-ex",
    reportNumber = "UCI-TR-2022-01, CERN-PBC-Notes-2022-001, INT-PUB-22-006, BONN-TH-2022-04, FERMILAB-PUB-22-094-ND-SCD-T",
    doi = "10.1088/1361-6471/ac865e",
    journal = "J. Phys. G",
    volume = "50",
    number = "3",
    pages = "030501",
    year = "2023"
}

@article{Casarsa:2021rud,
    author = "Casarsa, Massimo and Fabbrichesi, Marco and Gabrielli, Emidio",
    title = "{Monochromatic single photon events at the muon collider}",
    eprint = "2111.13220",
    archivePrefix = "arXiv",
    primaryClass = "hep-ph",
    doi = "10.1103/PhysRevD.105.075008",
    journal = "Phys. Rev. D",
    volume = "105",
    number = "7",
    pages = "075008",
    year = "2022"
}

@article{Bao:2022onq,
    author = "Bao, Yunjia and Fan, JiJi and Li, Lingfeng",
    title = "{Electroweak ALP searches at a muon collider}",
    eprint = "2203.04328",
    archivePrefix = "arXiv",
    primaryClass = "hep-ph",
    doi = "10.1007/JHEP08(2022)276",
    journal = "JHEP",
    volume = "08",
    pages = "276",
    year = "2022"
}

@article{Bao:2025tqs,
    author = "Bao, Shou-shan and Ma, Yang and Wu, Yongcheng and Xie, Keping and Zhang, Hong",
    title = "{Light axion-like particles at future lepton colliders}",
    eprint = "2505.10023",
    archivePrefix = "arXiv",
    primaryClass = "hep-ph",
    reportNumber = "COMETA-2025-02, IRMP-CP3-25-09, MSUHEP-25-002, CPTNP-2025-011",
    doi = "10.1007/JHEP10(2025)122",
    journal = "JHEP",
    volume = "10",
    pages = "122",
    year = "2025"
}

@article{Bauer:2018uxu,
    author = "Bauer, Martin and Heiles, Mathias and Neubert, Matthias and Thamm, Andrea",
    title = "{Axion-Like Particles at Future Colliders}",
    eprint = "1808.10323",
    archivePrefix = "arXiv",
    primaryClass = "hep-ph",
    reportNumber = "CERN-TH-2018-199, MITP/18-075",
    doi = "10.1140/epjc/s10052-019-6587-9",
    journal = "Eur. Phys. J. C",
    volume = "79",
    number = "1",
    pages = "74",
    year = "2019"
}

@article{deBlas:2025gyz,
    author = "de Blas, Jorge and others",
    title = "{Physics Briefing Book: Input for the 2026 update of the European Strategy for Particle Physics}",
    eprint = "2511.03883",
    archivePrefix = "arXiv",
    primaryClass = "hep-ex",
    reportNumber = "CERN--2025-008, CERN-ESU-2025-001",
    doi = "10.17181/CERN.35CH.2O2P",
    month = "11",
    year = "2025"
}

@article{CHARM:1985anb,
    author = "Bergsma, F. and others",
    collaboration = "CHARM",
    title = "{Search for Axion Like Particle Production in 400-{GeV} Proton - Copper Interactions}",
    reportNumber = "CERN-EP-85-38",
    doi = "10.1016/0370-2693(85)90400-9",
    journal = "Phys. Lett. B",
    volume = "157",
    pages = "458--462",
    year = "1985"
}

@article{Blumlein:2013cua,
    author = {Bl{\"u}mlein, Johannes and Brunner, J{\"u}rgen},
    title = "{New Exclusion Limits on Dark Gauge Forces from Proton Bremsstrahlung in Beam-Dump Data}",
    eprint = "1311.3870",
    archivePrefix = "arXiv",
    primaryClass = "hep-ph",
    reportNumber = "DESY-13-202, DO-TH-13-29, SFB-CPP-13-87, LPN-13-087",
    doi = "10.1016/j.physletb.2014.02.029",
    journal = "Phys. Lett. B",
    volume = "731",
    pages = "320--326",
    year = "2014"
}

@article{DUNE:2020ypp,
    author = "Abi, Babak and others",
    collaboration = "DUNE",
    title = "{Deep Underground Neutrino Experiment (DUNE), Far Detector Technical Design Report, Volume II: DUNE Physics}",
    eprint = "2002.03005",
    archivePrefix = "arXiv",
    primaryClass = "hep-ex",
    reportNumber = "FERMILAB-PUB-20-025-ND, FERMILAB-DESIGN-2020-02",
    month = "2",
    year = "2020"
}

@article{Hook:2019qoh,
    author = "Hook, Anson and Kumar, Soubhik and Liu, Zhen and Sundrum, Raman",
    title = "{High Quality QCD Axion and the LHC}",
    eprint = "1911.12364",
    archivePrefix = "arXiv",
    primaryClass = "hep-ph",
    reportNumber = "UMD-PP-019-07",
    doi = "10.1103/PhysRevLett.124.221801",
    journal = "Phys. Rev. Lett.",
    volume = "124",
    number = "22",
    pages = "221801",
    year = "2020"
}

@article{Ertas:2020xcc,
    author = "Ertas, Fatih and Kahlhoefer, Felix",
    title = "{On the interplay between astrophysical and laboratory probes of MeV-scale axion-like particles}",
    eprint = "2004.01193",
    archivePrefix = "arXiv",
    primaryClass = "hep-ph",
    reportNumber = "TTK-20-08",
    doi = "10.1007/JHEP07(2020)050",
    journal = "JHEP",
    volume = "07",
    pages = "050",
    year = "2020"
}

@article{Gligorov:2017nwh,
    author = "Gligorov, Vladimir V. and Knapen, Simon and Papucci, Michele and Robinson, Dean J.",
    title = "{Searching for Long-lived Particles: A Compact Detector for Exotics at LHCb}",
    eprint = "1708.09395",
    archivePrefix = "arXiv",
    primaryClass = "hep-ph",
    doi = "10.1103/PhysRevD.97.015023",
    journal = "Phys. Rev. D",
    volume = "97",
    number = "1",
    pages = "015023",
    year = "2018"
}

@article{CODEX-b:2019jve,
    author = "Aielli, Giulio and others",
    collaboration = "CODEX-b",
    title = "{Expression of interest for the CODEX-b detector}",
    eprint = "1911.00481",
    archivePrefix = "arXiv",
    primaryClass = "hep-ex",
    doi = "10.1140/epjc/s10052-020-08711-3",
    journal = "Eur. Phys. J. C",
    volume = "80",
    number = "12",
    pages = "1177",
    year = "2020"
}

@article{Chou:2016lxi,
    author = "Chou, John Paul and Curtin, David and Lubatti, H. J.",
    title = "{New Detectors to Explore the Lifetime Frontier}",
    eprint = "1606.06298",
    archivePrefix = "arXiv",
    primaryClass = "hep-ph",
    doi = "10.1016/j.physletb.2017.01.043",
    journal = "Phys. Lett. B",
    volume = "767",
    pages = "29--36",
    year = "2017"
}

@article{BaBar:2013npw,
    author = "Lees, J. P. and others",
    collaboration = "BaBar",
    title = "{Search for $B \to K^{(*)} \nu \overline \nu$ and invisible quarkonium decays}",
    eprint = "1303.7465",
    archivePrefix = "arXiv",
    primaryClass = "hep-ex",
    reportNumber = "BABAR-CONF-13-002, SLAC-PUB-15411, BABAR-PUB-13-002",
    doi = "10.1103/PhysRevD.87.112005",
    journal = "Phys. Rev. D",
    volume = "87",
    number = "11",
    pages = "112005",
    year = "2013"
}

@article{NA62:2020xlg,
    author = "Cortina Gil, Eduardo and others",
    collaboration = "NA62",
    title = "{Search for a feebly interacting particle  $X$ in the decay $K^{+}\rightarrow\pi^{+}X$}",
    eprint = "2011.11329",
    archivePrefix = "arXiv",
    primaryClass = "hep-ex",
    reportNumber = "CERN-EP-2020-227",
    doi = "10.1007/JHEP03(2021)058",
    journal = "JHEP",
    volume = "03",
    pages = "058",
    year = "2021"
}

@article{NA62:2021zjw,
    author = "Cortina Gil, Eduardo and others",
    collaboration = "NA62",
    title = "{Measurement of the very rare K$^{+}${\textrightarrow}$ {\pi}^{+}\nu \overline{\nu} $ decay}",
    eprint = "2103.15389",
    archivePrefix = "arXiv",
    primaryClass = "hep-ex",
    doi = "10.1007/JHEP06(2021)093",
    journal = "JHEP",
    volume = "06",
    pages = "093",
    year = "2021"
}

@article{BaBar:2021ich,
    author = "Lees, J. P. and others",
    collaboration = "BaBar",
    title = "{Search for an Axionlike Particle in $B$ Meson Decays}",
    eprint = "2111.01800",
    archivePrefix = "arXiv",
    primaryClass = "hep-ex",
    reportNumber = "BABAR-PUB-21/006, SLAC-PUB-17631",
    doi = "10.1103/PhysRevLett.128.131802",
    journal = "Phys. Rev. Lett.",
    volume = "128",
    number = "13",
    pages = "131802",
    year = "2022"
}

@article{LHCb:2015nkv,
    author = "Aaij, Roel and others",
    collaboration = "LHCb",
    title = "{Search for hidden-sector bosons in $B^0 \!\to K^{*0}\mu^+\mu^-$ decays}",
    eprint = "1508.04094",
    archivePrefix = "arXiv",
    primaryClass = "hep-ex",
    reportNumber = "CERN-PH-EP-2015-202, LHCB-PAPER-2015-036",
    doi = "10.1103/PhysRevLett.115.161802",
    journal = "Phys. Rev. Lett.",
    volume = "115",
    number = "16",
    pages = "161802",
    year = "2015"
}

@article{LHCb:2016awg,
    author = "Aaij, R. and others",
    collaboration = "LHCb",
    title = "{Search for long-lived scalar particles in $B^+ \to K^+ \chi (\mu^+\mu^-)$ decays}",
    eprint = "1612.07818",
    archivePrefix = "arXiv",
    primaryClass = "hep-ex",
    reportNumber = "CERN-EP-2016-302, LHCB-PAPER-2016-052",
    doi = "10.1103/PhysRevD.95.071101",
    journal = "Phys. Rev. D",
    volume = "95",
    number = "7",
    pages = "071101",
    year = "2017"
}

@article{Dobrich:2018jyi,
    author = {D{\"o}brich, Babette and Ertas, Fatih and Kahlhoefer, Felix and Spadaro, Tommaso},
    title = "{Model-independent bounds on light pseudoscalars from rare B-meson decays}",
    eprint = "1810.11336",
    archivePrefix = "arXiv",
    primaryClass = "hep-ph",
    reportNumber = "TTK-18-45",
    doi = "10.1016/j.physletb.2019.01.064",
    journal = "Phys. Lett. B",
    volume = "790",
    pages = "537--544",
    year = "2019"
}

@article{Huang:2015tdv,
    author = "Huang, Peisi and Joglekar, Aniket and Li, Bing and Wagner, Carlos E. M.",
    title = "{Probing the Electroweak Phase Transition at the LHC}",
    eprint = "1512.00068",
    archivePrefix = "arXiv",
    primaryClass = "hep-ph",
    reportNumber = "EFI-15-37",
    doi = "10.1103/PhysRevD.93.055049",
    journal = "Phys. Rev. D",
    volume = "93",
    number = "5",
    pages = "055049",
    year = "2016"
}

@article{Ghoshal:2026hev,
    author = "Ghoshal, Anish and Pal, Pratyay",
    title = "{Cosmic Collider Gravitational Waves sourced by Right-handed Neutrino production from Bubbles: Testing Seesaw, Leptogenesis and Dark Matter}",
    eprint = "2601.02458",
    archivePrefix = "arXiv",
    primaryClass = "astro-ph.CO",
    month = "1",
    year = "2026"
}

@article{PhysRevLett.83.2195,
  title = {Decay Laws for Three-Dimensional Magnetohydrodynamic Turbulence},
  author = {Biskamp, Dieter and M\"uller, Wolf-Christian},
  journal = {Phys. Rev. Lett.},
  volume = {83},
  issue = {11},
  pages = {2195--2198},
  numpages = {0},
  year = {1999},
  month = {Sep},
  publisher = {American Physical Society},
  doi = {10.1103/PhysRevLett.83.2195},
  url = {https://link.aps.org/doi/10.1103/PhysRevLett.83.2195}
}

@article{Salvio:2026bco,
    author = "Salvio, Alberto",
    title = "{Supercooled Phase Transitions with Radiative Symmetry Breaking}",
    eprint = "2602.20246",
    archivePrefix = "arXiv",
    primaryClass = "hep-ph",
    month = "2",
    year = "2026"
}

@article{Yamazaki:2012pg,
    author = "Yamazaki, Dai G. and Kajino, Toshitaka and Mathew, Grant J. and Ichiki, Kiyotomo",
    title = "{The Search for a Primordial Magnetic Field}",
    eprint = "1204.3669",
    archivePrefix = "arXiv",
    primaryClass = "astro-ph.CO",
    doi = "10.1016/j.physrep.2012.02.005",
    journal = "Phys. Rept.",
    volume = "517",
    pages = "141--167",
    year = "2012"
}

@article{Planck:2015zrl,
    author = "Ade, P. A. R. and others",
    collaboration = "Planck",
    title = "{Planck 2015 results. XIX. Constraints on primordial magnetic fields}",
    eprint = "1502.01594",
    archivePrefix = "arXiv",
    primaryClass = "astro-ph.CO",
    doi = "10.1051/0004-6361/201525821",
    journal = "Astron. Astrophys.",
    volume = "594",
    pages = "A19",
    year = "2016"
}

@article{Grasso:2000wj,
    author = "Grasso, Dario and Rubinstein, Hector R.",
    title = "{Magnetic fields in the early universe}",
    eprint = "astro-ph/0009061",
    archivePrefix = "arXiv",
    reportNumber = "DFPD-00-TH-35",
    doi = "10.1016/S0370-1573(00)00110-1",
    journal = "Phys. Rept.",
    volume = "348",
    pages = "163--266",
    year = "2001"
}

@article{Kawasaki:2012va,
    author = "Kawasaki, Masahiro and Kusakabe, Motohiko",
    title = "{Updated constraint on a primordial magnetic field during big bang nucleosynthesis and a formulation of field effects}",
    eprint = "1204.6164",
    archivePrefix = "arXiv",
    primaryClass = "astro-ph.CO",
    reportNumber = "ICRR-REPORT-621-2012-10",
    doi = "10.1103/PhysRevD.86.063003",
    journal = "Phys. Rev. D",
    volume = "86",
    pages = "063003",
    year = "2012"
}

@article{Peccei:1977hh,
    author = "Peccei, R. D. and Quinn, Helen R.",
    title = "{CP Conservation in the Presence of Instantons}",
    reportNumber = "ITP-568-STANFORD",
    doi = "10.1103/PhysRevLett.38.1440",
    journal = "Phys. Rev. Lett.",
    volume = "38",
    pages = "1440--1443",
    year = "1977"
}

@article{Peccei:1977ur,
    author = "Peccei, R. D. and Quinn, Helen R.",
    title = "{Constraints Imposed by CP Conservation in the Presence of Instantons}",
    reportNumber = "ITP-572-STANFORD",
    doi = "10.1103/PhysRevD.16.1791",
    journal = "Phys. Rev. D",
    volume = "16",
    pages = "1791--1797",
    year = "1977"
}

@article{Weinberg:1977ma,
    author = "Weinberg, Steven",
    title = "{A New Light Boson?}",
    reportNumber = "HUTP-77/A074",
    doi = "10.1103/PhysRevLett.40.223",
    journal = "Phys. Rev. Lett.",
    volume = "40",
    pages = "223--226",
    year = "1978"
}

@article{Wilczek:1977pj,
    author = "Wilczek, Frank",
    title = "{Problem of Strong  $P$  and  $T$  Invariance in the Presence of Instantons}",
    reportNumber = "Print-77-0939 (COLUMBIA)",
    doi = "10.1103/PhysRevLett.40.279",
    journal = "Phys. Rev. Lett.",
    volume = "40",
    pages = "279--282",
    year = "1978"
}

@article{Graham:2015cka,
    author = "Graham, Peter W. and Kaplan, David E. and Rajendran, Surjeet",
    title = "{Cosmological Relaxation of the Electroweak Scale}",
    eprint = "1504.07551",
    archivePrefix = "arXiv",
    primaryClass = "hep-ph",
    doi = "10.1103/PhysRevLett.115.221801",
    journal = "Phys. Rev. Lett.",
    volume = "115",
    number = "22",
    pages = "221801",
    year = "2015"
}

@article{Freese:1990rb,
    author = "Freese, Katherine and Frieman, Joshua A. and Olinto, Angela V.",
    title = "{Natural inflation with pseudo - Nambu-Goldstone bosons}",
    reportNumber = "FERMILAB-PUB-90-177-A",
    doi = "10.1103/PhysRevLett.65.3233",
    journal = "Phys. Rev. Lett.",
    volume = "65",
    pages = "3233--3236",
    year = "1990"
}

@article{Abbott:1982af,
    author = "Abbott, L. F. and Sikivie, P.",
    editor = "Srednicki, M. A.",
    title = "{A Cosmological Bound on the Invisible Axion}",
    reportNumber = "PRINT-82-0695 (BRANDEIS)",
    doi = "10.1016/0370-2693(83)90638-X",
    journal = "Phys. Lett. B",
    volume = "120",
    pages = "133--136",
    year = "1983"
}

@article{Dine:1982ah,
    author = "Dine, Michael and Fischler, Willy",
    editor = "Srednicki, M. A.",
    title = "{The Not So Harmless Axion}",
    reportNumber = "UPR-0201T",
    doi = "10.1016/0370-2693(83)90639-1",
    journal = "Phys. Lett. B",
    volume = "120",
    pages = "137--141",
    year = "1983"
}

@article{Adams:1992bn,
    author = "Adams, Fred C. and Bond, J. Richard and Freese, Katherine and Frieman, Joshua A. and Olinto, Angela V.",
    title = "{Natural inflation: Particle physics models, power law spectra for large scale structure, and constraints from COBE}",
    eprint = "hep-ph/9207245",
    archivePrefix = "arXiv",
    reportNumber = "FERMILAB-PUB-92-202-A",
    doi = "10.1103/PhysRevD.47.426",
    journal = "Phys. Rev. D",
    volume = "47",
    pages = "426--455",
    year = "1993"
}

@article{Daido:2017wwb,
    author = "Daido, Ryuji and Takahashi, Fuminobu and Yin, Wen",
    title = "{The ALP miracle: unified inflaton and dark matter}",
    eprint = "1702.03284",
    archivePrefix = "arXiv",
    primaryClass = "hep-ph",
    reportNumber = "TU-1039, IPMU17-0031",
    doi = "10.1088/1475-7516/2017/05/044",
    journal = "JCAP",
    volume = "05",
    pages = "044",
    year = "2017"
}

@article{Preskill:1982cy,
    author = "Preskill, John and Wise, Mark B. and Wilczek, Frank",
    editor = "Srednicki, M. A.",
    title = "{Cosmology of the Invisible Axion}",
    reportNumber = "HUTP-82-A048, NSF-ITP-82-103",
    doi = "10.1016/0370-2693(83)90637-8",
    journal = "Phys. Lett. B",
    volume = "120",
    pages = "127--132",
    year = "1983"
}

@article{Jain:2004gi,
    author = "Jain, Pankaj",
    title = "{Dark energy in an axion model with explicit Z(N) symmetry breaking}",
    eprint = "hep-ph/0411279",
    archivePrefix = "arXiv",
    doi = "10.1142/S0217732305016890",
    journal = "Mod. Phys. Lett. A",
    volume = "20",
    pages = "1763--1766",
    year = "2005"
}

@article{Kim:2009cp,
    author = "Kim, Jihn E. and Nilles, Hans Peter",
    title = "{Axionic dark energy and a composite QCD axion}",
    eprint = "0902.3610",
    archivePrefix = "arXiv",
    primaryClass = "hep-th",
    doi = "10.1088/1475-7516/2009/05/010",
    journal = "JCAP",
    volume = "05",
    pages = "010",
    year = "2009"
}

@article{Kim:2013jka,
    author = "Kim, Jihn E. and Nilles, Hans Peter",
    title = "{Dark energy from approximate U(1)$_{de}$ symmetry}",
    eprint = "1311.0012",
    archivePrefix = "arXiv",
    primaryClass = "hep-ph",
    doi = "10.1016/j.physletb.2014.01.031",
    journal = "Phys. Lett. B",
    volume = "730",
    pages = "53--58",
    year = "2014"
}

@article{Choi:2019jck,
    author = "Choi, Gongjun and Suzuki, Motoo and Yanagida, Tsutomu T.",
    title = "{Quintessence axion dark energy and a solution to the hubble tension}",
    eprint = "1910.00459",
    archivePrefix = "arXiv",
    primaryClass = "hep-ph",
    doi = "10.1016/j.physletb.2020.135408",
    journal = "Phys. Lett. B",
    volume = "805",
    pages = "135408",
    year = "2020"
}

@article{Daido:2015gqa,
    author = "Daido, Ryuji and Kitajima, Naoya and Takahashi, Fuminobu",
    title = "{Axion domain wall baryogenesis}",
    eprint = "1504.07917",
    archivePrefix = "arXiv",
    primaryClass = "hep-ph",
    reportNumber = "TU-993, IPMU15-0059",
    doi = "10.1088/1475-7516/2015/07/046",
    journal = "JCAP",
    volume = "07",
    pages = "046",
    year = "2015"
}

@article{DeSimone:2016bok,
    author = "De Simone, Andrea and Kobayashi, Takeshi and Liberati, Stefano",
    title = "{Geometric Baryogenesis from Shift Symmetry}",
    eprint = "1612.04824",
    archivePrefix = "arXiv",
    primaryClass = "hep-ph",
    reportNumber = "SISSA-64-2016-FISI",
    doi = "10.1103/PhysRevLett.118.131101",
    journal = "Phys. Rev. Lett.",
    volume = "118",
    number = "13",
    pages = "131101",
    year = "2017"
}

@article{Co:2019wyp,
    author = "Co, Raymond T. and Harigaya, Keisuke",
    title = "{Axiogenesis}",
    eprint = "1910.02080",
    archivePrefix = "arXiv",
    primaryClass = "hep-ph",
    reportNumber = "LCTP-19-27",
    doi = "10.1103/PhysRevLett.124.111602",
    journal = "Phys. Rev. Lett.",
    volume = "124",
    number = "11",
    pages = "111602",
    year = "2020"
}

@article{Im:2021xoy,
    author = "Im, Sang Hui and Jeong, Kwang Sik and Lee, Yeseong",
    title = "{Electroweak baryogenesis by axionlike dark matter}",
    eprint = "2111.01327",
    archivePrefix = "arXiv",
    primaryClass = "hep-ph",
    reportNumber = "CTPU-PTC-21-36, PNUTP-21-A15",
    doi = "10.1103/PhysRevD.105.035028",
    journal = "Phys. Rev. D",
    volume = "105",
    number = "3",
    pages = "035028",
    year = "2022"
}

@article{Jeong:2018jqe,
    author = "Jeong, Kwang Sik and Jung, Tae Hyun and Shin, Chang Sub",
    title = "{Adiabatic electroweak baryogenesis driven by an axionlike particle}",
    eprint = "1811.03294",
    archivePrefix = "arXiv",
    primaryClass = "hep-ph",
    reportNumber = "CTPU-18-34, PNUTP-18-A12",
    doi = "10.1103/PhysRevD.101.035009",
    journal = "Phys. Rev. D",
    volume = "101",
    number = "3",
    pages = "035009",
    year = "2020"
}

@article{Foster:2022ajl,
    author = "Foster, Joshua W. and Kumar, Soubhik and Safdi, Benjamin R. and Soreq, Yotam",
    title = "{Dark Grand Unification in the axiverse: decaying axion dark matter and spontaneous baryogenesis}",
    eprint = "2208.10504",
    archivePrefix = "arXiv",
    primaryClass = "hep-ph",
    reportNumber = "MIT-CTP/5458",
    doi = "10.1007/JHEP12(2022)119",
    journal = "JHEP",
    volume = "12",
    pages = "119",
    year = "2022"
}

@article{Brandenberger:2020gaz,
    author = {Brandenberger, Robert and Fr\"ohlich, J\"urg},
    title = "{Dark Energy, Dark Matter and Baryogenesis from a Model of a Complex Axion Field}",
    eprint = "2004.10025",
    archivePrefix = "arXiv",
    primaryClass = "hep-th",
    doi = "10.1088/1475-7516/2021/04/030",
    journal = "JCAP",
    volume = "04",
    pages = "030",
    year = "2021"
}

@article{Yao:2023qve,
    author = "Yao, Yan-Hong and Meng, Xin-He",
    title = "{Restoring cosmological concordance with axion-like early dark energy and dark matter characterized by a constant equation of state?}",
    eprint = "2312.04007",
    archivePrefix = "arXiv",
    primaryClass = "astro-ph.CO",
    month = "12",
    year = "2023"
}

@article{Chatziioannou:2024hju,
    author = {Chatziioannou, K. and Dent, T. and Fishbach, M. and Ohme, F. and P{\"u}rrer, M. and Raymond, V. and Veitch, J.},
    title = "{Compact binary coalescences: gravitational-wave astronomy with ground-based detectors}",
    eprint = "2409.02037",
    archivePrefix = "arXiv",
    primaryClass = "gr-qc",
    month = "9",
    year = "2024"
}

@article{Keitel:2025npi,
    author = "Keitel, David",
    title = "{LIGO{\textendash}Virgo{\textendash}KAGRA Results and Status of the Current Fourth Observing Run}",
    doi = "10.1007/978-3-031-88933-2_3",
    journal = "Springer Proc. Phys.",
    volume = "425",
    pages = "13--20",
    year = "2025"
}

@article{Olea-Romacho:2023rhh,
    author = "Olea-Romacho, Mar{\'\i}a Olalla",
    title = "{Primordial magnetogenesis in the two-Higgs-doublet model}",
    eprint = "2310.19948",
    archivePrefix = "arXiv",
    primaryClass = "hep-ph",
    doi = "10.1103/PhysRevD.109.015023",
    journal = "Phys. Rev. D",
    volume = "109",
    number = "1",
    pages = "015023",
    year = "2024"
}

@article{Balaji:2024rvo,
    author = "Balaji, Shyam and Fairbairn, Malcolm and Olea-Romacho, Maria Olalla",
    title = "{Magnetogenesis with gravitational waves and primordial black hole dark matter}",
    eprint = "2402.05179",
    archivePrefix = "arXiv",
    primaryClass = "hep-ph",
    doi = "10.1103/PhysRevD.109.075048",
    journal = "Phys. Rev. D",
    volume = "109",
    number = "7",
    pages = "075048",
    year = "2024"
}

@article{Balaji:2025tun,
    author = "Balaji, Shyam and Gon{\c{c}}alves, Jo{\~a}o and Marfatia, Danny and Morais, Ant{\'o}nio P. and Pasechnik, Roman",
    title = "{Primordial black holes and magnetic fields in conformal neutrino mass models}",
    eprint = "2505.08011",
    archivePrefix = "arXiv",
    primaryClass = "hep-ph",
    doi = "10.1088/1475-7516/2025/10/064",
    journal = "JCAP",
    volume = "10",
    pages = "064",
    year = "2025"
}

@article{ArteagaTupia:2025awh,
    author = "Arteaga Tupia, Martin and Ghoshal, Anish and Strumia, Alessandro",
    title = "{Primordial magnetogenesis from a supercooled dynamical electroweak phase transition}",
    eprint = "2506.16387",
    archivePrefix = "arXiv",
    primaryClass = "hep-ph",
    doi = "10.1007/JHEP10(2025)135",
    journal = "JHEP",
    volume = "10",
    pages = "135",
    year = "2025"
}

@article{Dolan:1973qd,
    author = "Dolan, L. and Jackiw, R.",
    title = "{Symmetry Behavior at Finite Temperature}",
    reportNumber = "MIT-CTP-406",
    doi = "10.1103/PhysRevD.9.3320",
    journal = "Phys. Rev. D",
    volume = "9",
    pages = "3320--3341",
    year = "1974"
}

@article{Anderson:1991zb,
    author = "Anderson, Greg W. and Hall, Lawrence J.",
    title = "{The Electroweak phase transition and baryogenesis}",
    reportNumber = "LBL-31169, UCB-PTH-91-41",
    doi = "10.1103/PhysRevD.45.2685",
    journal = "Phys. Rev. D",
    volume = "45",
    pages = "2685--2698",
    year = "1992"
}

@article{Linde:1981zj,
    author = "Linde, Andrei D.",
    title = "{Decay of the False Vacuum at Finite Temperature}",
    reportNumber = "LEBEDEV-81-265",
    doi = "10.1016/0550-3213(83)90072-X",
    journal = "Nucl. Phys. B",
    volume = "216",
    pages = "421",
    year = "1983",
    note = "[Erratum: Nucl.Phys.B 223, 544 (1983)]"
}

@article{Affleck:1980ac,
    author = "Affleck, Ian",
    title = "{Quantum Statistical Metastability}",
    reportNumber = "HUTP-80/A062",
    doi = "10.1103/PhysRevLett.46.388",
    journal = "Phys. Rev. Lett.",
    volume = "46",
    pages = "388",
    year = "1981"
}

@article{Linde:1980tt,
    author = "Linde, Andrei D.",
    title = "{Fate of the False Vacuum at Finite Temperature: Theory and Applications}",
    reportNumber = "LEBEDEV-80-92",
    doi = "10.1016/0370-2693(81)90281-1",
    journal = "Phys. Lett. B",
    volume = "100",
    pages = "37--40",
    year = "1981"
}

@article{Linde:1977mm,
    author = "Linde, Andrei D.",
    title = "{On the Vacuum Instability and the Higgs Meson Mass}",
    reportNumber = "LEBEDEV-77-112",
    doi = "10.1016/0370-2693(77)90664-5",
    journal = "Phys. Lett. B",
    volume = "70",
    pages = "306--308",
    year = "1977"
}

@article{Coleman:1977py,
    author = "Coleman, Sidney R.",
    title = "{The Fate of the False Vacuum. 1. Semiclassical Theory}",
    reportNumber = "HUTP-77-A004",
    doi = "10.1103/PhysRevD.16.1248",
    journal = "Phys. Rev. D",
    volume = "15",
    pages = "2929--2936",
    year = "1977",
    note = "[Erratum: Phys.Rev.D 16, 1248 (1977)]"
}

@article{Callan:1977pt,
    author = "Callan, Jr., Curtis G. and Coleman, Sidney R.",
    title = "{The Fate of the False Vacuum. 2. First Quantum Corrections}",
    reportNumber = "HUTP-77-A032",
    doi = "10.1103/PhysRevD.16.1762",
    journal = "Phys. Rev. D",
    volume = "16",
    pages = "1762--1768",
    year = "1977"
}

@article{Enqvist:1991xw,
    author = "Enqvist, K. and Ignatius, J. and Kajantie, K. and Rummukainen, K.",
    title = "{Nucleation and bubble growth in a first order cosmological electroweak phase transition}",
    reportNumber = "HU-TFT-91-35",
    doi = "10.1103/PhysRevD.45.3415",
    journal = "Phys. Rev. D",
    volume = "45",
    pages = "3415--3428",
    year = "1992"
}

@article{Grojean:2006bp,
    author = "Grojean, Christophe and Servant, Geraldine",
    title = "{Gravitational Waves from Phase Transitions at the Electroweak Scale and Beyond}",
    eprint = "hep-ph/0607107",
    archivePrefix = "arXiv",
    reportNumber = "CERN-PH-TH-2006-125",
    doi = "10.1103/PhysRevD.75.043507",
    journal = "Phys. Rev. D",
    volume = "75",
    pages = "043507",
    year = "2007"
}

@article{Sartore:2020gou,
    author = "Sartore, Lohan and Schienbein, Ingo",
    title = "{PyR@TE 3}",
    eprint = "2007.12700",
    archivePrefix = "arXiv",
    primaryClass = "hep-ph",
    doi = "10.1016/j.cpc.2020.107819",
    journal = "Comput. Phys. Commun.",
    volume = "261",
    pages = "107819",
    year = "2021"
}

@techreport{ATLAS:2013rir,
    collaboration = "ATLAS",
    title = "{Studies of the ATLAS potential for Higgs self-coupling measurements at a High Luminosity LHC}",
    number = "ATL-PHYS-PUB-2013-001",
    year = "2013"
}

@inproceedings{Yao:2013ika,
    author = "Yao, Weiming",
    title = "{Studies of measuring Higgs self-coupling with $HH\rightarrow b\bar b \gamma\gamma$ at the future hadron colliders}",
    booktitle = "{Snowmass 2013}: {Snowmass on the Mississippi}",
    eprint = "1308.6302",
    archivePrefix = "arXiv",
    primaryClass = "hep-ph",
    month = "8",
    year = "2013"
}

@article{CidVidal:2018eel,
    author = "Cid Vidal, Xabier and others",
    editor = "Dainese, Andrea and Mangano, Michelangelo and Meyer, Andreas B. and Nisati, Aleandro and Salam, Gavin and Vesterinen, Mika Anton",
    title = "{Report from Working Group 3}: {Beyond the Standard Model physics at the HL-LHC and HE-LHC}",
    eprint = "1812.07831",
    archivePrefix = "arXiv",
    primaryClass = "hep-ph",
    reportNumber = "CERN-LPCC-2018-05",
    doi = "10.23731/CYRM-2019-007.585",
    journal = "CERN Yellow Rep. Monogr.",
    volume = "7",
    pages = "585--865",
    year = "2019"
}

@inproceedings{Asner:2013psa,
    author = "Asner, D. M. and others",
    title = "{ILC Higgs White Paper}",
    booktitle = "{Snowmass 2013}: {Snowmass on the Mississippi}",
    eprint = "1310.0763",
    archivePrefix = "arXiv",
    primaryClass = "hep-ph",
    month = "10",
    year = "2013"
}

@article{Tian:2013yda,
    author = "Tian, Junping and Fujii, Keisuke",
    collaboration = "ILD",
    title = "{Measurement of Higgs couplings and self-coupling at the ILC}",
    eprint = "1311.6528",
    archivePrefix = "arXiv",
    primaryClass = "hep-ph",
    doi = "10.22323/1.180.0316",
    journal = "PoS",
    volume = "EPS-HEP2013",
    pages = "316",
    year = "2013"
}

@article{Abramowicz:2016zbo,
    author = "Abramowicz, H. and others",
    title = "{Higgs physics at the CLIC electron{\textendash}positron linear collider}",
    eprint = "1608.07538",
    archivePrefix = "arXiv",
    primaryClass = "hep-ex",
    reportNumber = "CLICDP-PUB-2016-001",
    doi = "10.1140/epjc/s10052-017-4968-5",
    journal = "Eur. Phys. J. C",
    volume = "77",
    number = "7",
    pages = "475",
    year = "2017"
}

@article{Barr:2014sga,
    author = "Barr, Alan J. and Dolan, Matthew J. and Englert, Christoph and Ferreira de Lima, Danilo Enoque and Spannowsky, Michael",
    title = "{Higgs Self-Coupling Measurements at a 100 TeV Hadron Collider}",
    eprint = "1412.7154",
    archivePrefix = "arXiv",
    primaryClass = "hep-ph",
    reportNumber = "IPPP-14-110, DCPT-14-220, SLAC-PUB-16194",
    doi = "10.1007/JHEP02(2015)016",
    journal = "JHEP",
    volume = "02",
    pages = "016",
    year = "2015"
}

@article{Chang:2018rso,
    author = "Chang, Jae Hyeok and Essig, Rouven and McDermott, Samuel D.",
    title = "{Supernova 1987A Constraints on Sub-GeV Dark Sectors, Millicharged Particles, the QCD Axion, and an Axion-like Particle}",
    eprint = "1803.00993",
    archivePrefix = "arXiv",
    primaryClass = "hep-ph",
    reportNumber = "YITP-SB-18-01, FERMILAB-PUB-17-432-T",
    doi = "10.1007/JHEP09(2018)051",
    journal = "JHEP",
    volume = "09",
    pages = "051",
    year = "2018"
}

@article{Dobrich:2015jyk,
    author = {D{\"o}brich, Babette and Jaeckel, Joerg and Kahlhoefer, Felix and Ringwald, Andreas and Schmidt-Hoberg, Kai},
    title = "{ALPtraum: ALP production in proton beam dump experiments}",
    eprint = "1512.03069",
    archivePrefix = "arXiv",
    primaryClass = "hep-ph",
    reportNumber = "CERN-PH-TH-2015-293, DESY-15-237",
    doi = "10.1007/JHEP02(2016)018",
    journal = "JHEP",
    volume = "02",
    pages = "018",
    year = "2016"
}

@article{Dolan:2017osp,
    author = "Dolan, Matthew J. and Ferber, Torben and Hearty, Christopher and Kahlhoefer, Felix and Schmidt-Hoberg, Kai",
    title = "{Revised constraints and Belle II sensitivity for visible and invisible axion-like particles}",
    eprint = "1709.00009",
    archivePrefix = "arXiv",
    primaryClass = "hep-ph",
    reportNumber = "DESY-17-127",
    doi = "10.1007/JHEP12(2017)094",
    journal = "JHEP",
    volume = "12",
    pages = "094",
    year = "2017",
    note = "[Erratum: JHEP 03, 190 (2021)]"
}

@article{NA64:2020qwq,
    author = "Banerjee, D. and others",
    collaboration = "NA64",
    title = "{Search for Axionlike and Scalar Particles with the NA64 Experiment}",
    eprint = "2005.02710",
    archivePrefix = "arXiv",
    primaryClass = "hep-ex",
    reportNumber = "CERN-EP-2020-068",
    doi = "10.1103/PhysRevLett.125.081801",
    journal = "Phys. Rev. Lett.",
    volume = "125",
    number = "8",
    pages = "081801",
    year = "2020"
}

@article{Mariotti:2017vtv,
    author = "Mariotti, Alberto and Redigolo, Diego and Sala, Filippo and Tobioka, Kohsaku",
    title = "{New LHC bound on low-mass diphoton resonances}",
    eprint = "1710.01743",
    archivePrefix = "arXiv",
    primaryClass = "hep-ph",
    reportNumber = "DESY-17-148",
    doi = "10.1016/j.physletb.2018.06.039",
    journal = "Phys. Lett. B",
    volume = "783",
    pages = "13--18",
    year = "2018"
}

@article{Aime:2022flm,
    author = "Aime, Chiara and others",
    title = "{Muon Collider Physics Summary}",
    eprint = "2203.07256",
    archivePrefix = "arXiv",
    primaryClass = "hep-ph",
    reportNumber = "FERMILAB-PUB-22-377-PPD",
    month = "3",
    year = "2022"
}

@article{Castelli:2025mqk,
    author = "Castelli, Luca",
    title = "{Higgs Physics at the Muon Collider}",
    doi = "10.3390/particles8010028",
    journal = "Particles",
    volume = "8",
    number = "1",
    pages = "28",
    year = "2025"
}

@article{CMS:2026nuu,
    author = "Aad, Georges and others",
    collaboration = "CMS, ATLAS",
    title = "{Combination of ATLAS and CMS searches for Higgs boson pair production at $\sqrt{s} = 13$ TeV}",
    eprint = "2602.23991",
    archivePrefix = "arXiv",
    primaryClass = "hep-ex",
    reportNumber = "CERN-EP-2026-011",
    month = "2",
    year = "2026"
}

@article{Schlickeiser:2013eca,
    author = "Schlickeiser, Reinhard and Krakau, Steffen and Supsar, Markus",
    title = "{Plasma Effects on Fast Pair Beams. II. Reactive versus Kinetic Instability of Parallel Electrostatic Waves}",
    eprint = "1308.4594",
    archivePrefix = "arXiv",
    primaryClass = "astro-ph.HE",
    doi = "10.1088/0004-637X/777/1/49",
    journal = "Astrophys. J.",
    volume = "777",
    pages = "49",
    year = "2013"
}

@article{Broderick:2011av,
    author = "Broderick, Avery E. and Chang, Philip and Pfrommer, Christoph",
    title = "{The Cosmological Impact of Luminous TeV Blazars I: Implications of Plasma Instabilities for the Intergalactic Magnetic Field and Extragalactic Gamma-Ray Background}",
    eprint = "1106.5494",
    archivePrefix = "arXiv",
    primaryClass = "astro-ph.CO",
    doi = "10.1088/0004-637X/752/1/22",
    journal = "Astrophys. J.",
    volume = "752",
    pages = "22",
    year = "2012"
}

@article{Perry:2021rgv,
    author = "Perry, Roy and Lyubarsky, Yuri",
    title = "{The role of resonant plasma instabilities in the evolution of blazar induced pair beams}",
    eprint = "2102.03190",
    archivePrefix = "arXiv",
    primaryClass = "astro-ph.HE",
    doi = "10.1093/mnras/stab324",
    journal = "Mon. Not. Roy. Astron. Soc.",
    volume = "503",
    number = "2",
    pages = "2215--2228",
    year = "2021"
}

@article{Rafighi:2017ise,
    author = "Rafighi, I. and Vafin, S. and Pohl, M. and Niemiec, J.",
    title = "{Plasma effects on relativistic pair beams from TeV blazars: PIC simulations and analytical predictions}",
    eprint = "1708.07797",
    archivePrefix = "arXiv",
    primaryClass = "astro-ph.HE",
    doi = "10.1051/0004-6361/201731127",
    journal = "Astron. Astrophys.",
    volume = "607",
    pages = "A112",
    year = "2017"
}

@article{Sironi:2013qfa,
    author = "Sironi, Lorenzo and Giannios, Dimitrios",
    title = "{Relativistic Pair Beams from TeV Blazars: A Source of Reprocessed GeV Emission rather than Intergalactic Heating}",
    eprint = "1312.4538",
    archivePrefix = "arXiv",
    primaryClass = "astro-ph.HE",
    doi = "10.1088/0004-637X/787/1/49",
    journal = "Astrophys. J.",
    volume = "787",
    pages = "49",
    year = "2014"
}

@article{Arrowsmith:2025apl,
    author = "Arrowsmith, Charles D. and others",
    title = "{Suppression of pair beam instabilities in a laboratory analogue of blazar pair cascades}",
    eprint = "2509.09040",
    archivePrefix = "arXiv",
    primaryClass = "astro-ph.HE",
    doi = "10.1073/pnas.2513365122",
    journal = "Proc. Nat. Acad. Sci.",
    volume = "122",
    number = "45",
    pages = "e2513365122",
    year = "2025"
}

@article{Kamada:2020bmb,
    author = "Kamada, Kohei and Uchida, Fumio and Yokoyama, Jun'ichi",
    title = "{Baryon isocurvature constraints on the primordial hypermagnetic fields}",
    eprint = "2012.14435",
    archivePrefix = "arXiv",
    primaryClass = "astro-ph.CO",
    reportNumber = "RESCEU-24/20",
    doi = "10.1088/1475-7516/2021/04/034",
    journal = "JCAP",
    volume = "04",
    pages = "034",
    year = "2021"
}

@article{Blunier:2025ddu,
    author = "Blunier, J. and Neronov, A. and Semikoz, D.",
    title = "{Revision of conservative lower bound on intergalactic magnetic field from Fermi and Cherenkov telescope observations of extreme blazars}",
    eprint = "2506.22285",
    archivePrefix = "arXiv",
    primaryClass = "astro-ph.HE",
    month = "6",
    year = "2025"
}

@article{CTA:2020hii,
    author = "Abdalla, H. and others",
    collaboration = "CTA",
    title = "{Sensitivity of the Cherenkov Telescope Array for probing cosmology and fundamental physics with gamma-ray propagation}",
    eprint = "2010.01349",
    archivePrefix = "arXiv",
    primaryClass = "astro-ph.HE",
    doi = "10.1088/1475-7516/2021/02/048",
    journal = "JCAP",
    volume = "02",
    pages = "048",
    year = "2021"
}

@article{Shakya:2023kjf,
    author = "Shakya, Bibhushan",
    title = "{Aspects of particle production from bubble dynamics at a first order phase transition}",
    eprint = "2308.16224",
    archivePrefix = "arXiv",
    primaryClass = "hep-ph",
    reportNumber = "DESY 23-125",
    doi = "10.1103/PhysRevD.111.023521",
    journal = "Phys. Rev. D",
    volume = "111",
    number = "2",
    pages = "023521",
    year = "2025"
}

@article{Mazumdar:2018dfl,
    author = "Mazumdar, Anupam and White, Graham",
    title = "{Review of cosmic phase transitions: their significance and experimental signatures}",
    eprint = "1811.01948",
    archivePrefix = "arXiv",
    primaryClass = "hep-ph",
    doi = "10.1088/1361-6633/ab1f55",
    journal = "Rept. Prog. Phys.",
    volume = "82",
    number = "7",
    pages = "076901",
    year = "2019"
}

@article{Vachaspati:2024vbw,
    author = "Vachaspati, Tanmay and Brandenburg, Axel",
    title = "{Spectra of magnetic fields from electroweak symmetry breaking}",
    eprint = "2412.00641",
    archivePrefix = "arXiv",
    primaryClass = "astro-ph.CO",
    reportNumber = "NORDITA-2024-047",
    doi = "10.1103/PhysRevD.111.043541",
    journal = "Phys. Rev. D",
    volume = "111",
    number = "4",
    pages = "043541",
    year = "2025"
}

@article{Gorghetto:2021fsn,
    author = "Gorghetto, Marco and Hardy, Edward and Nicolaescu, Horia",
    title = "{Observing invisible axions with gravitational waves}",
    eprint = "2101.11007",
    archivePrefix = "arXiv",
    primaryClass = "hep-ph",
    doi = "10.1088/1475-7516/2021/06/034",
    journal = "JCAP",
    volume = "06",
    pages = "034",
    year = "2021"
}

@article{Fu:2023nrn,
    author = "Fu, Bowen and Ghoshal, Anish and King, Stephen F.",
    title = "{Cosmic string gravitational waves from global U(1)$_{B-L}$ symmetry breaking as a probe of the type I seesaw scale}",
    eprint = "2306.07334",
    archivePrefix = "arXiv",
    primaryClass = "hep-ph",
    doi = "10.1007/JHEP11(2023)071",
    journal = "JHEP",
    volume = "11",
    pages = "071",
    year = "2023"
}

@article{Jinno:2016knw,
    author = "Jinno, Ryusuke and Takimoto, Masahiro",
    title = "{Probing a classically conformal B-L model with gravitational waves}",
    eprint = "1604.05035",
    archivePrefix = "arXiv",
    primaryClass = "hep-ph",
    reportNumber = "KEK-TH-1896",
    doi = "10.1103/PhysRevD.95.015020",
    journal = "Phys. Rev. D",
    volume = "95",
    number = "1",
    pages = "015020",
    year = "2017"
}

@article{Ellis:2020nnr,
    author = "Ellis, John and Lewicki, Marek and Vaskonen, Ville",
    title = "{Updated predictions for gravitational waves produced in a strongly supercooled phase transition}",
    eprint = "2007.15586",
    archivePrefix = "arXiv",
    primaryClass = "astro-ph.CO",
    reportNumber = "KCL-PH-TH/2020-40, CERN-TH-2020-129",
    doi = "10.1088/1475-7516/2020/11/020",
    journal = "JCAP",
    volume = "11",
    pages = "020",
    year = "2020"
}

@article{Ghoshal:2022hyc,
    author = "Ghoshal, Anish and Okada, Nobuchika and Paul, Arnab",
    title = "{Radiative plateau inflation with conformal invariance: Dynamical generation of electroweak and seesaw scales}",
    eprint = "2203.00677",
    archivePrefix = "arXiv",
    primaryClass = "hep-ph",
    doi = "10.1103/PhysRevD.106.055024",
    journal = "Phys. Rev. D",
    volume = "106",
    number = "5",
    pages = "055024",
    year = "2022"
}

@article{Ghoshal:2022qxk,
    author = "Ghoshal, Anish and Mukherjee, Debangshu and Rinaldi, Massimiliano",
    title = "{Inflation and primordial gravitational waves in scale-invariant quadratic gravity with Higgs}",
    eprint = "2205.06475",
    archivePrefix = "arXiv",
    primaryClass = "gr-qc",
    doi = "10.1007/JHEP05(2023)023",
    journal = "JHEP",
    volume = "05",
    pages = "023",
    year = "2023"
}

@article{Ghoshal:2024hfk,
    author = "Ghoshal, Anish and Okada, Nobuchika and Paul, Arnab and Raut, Digesh",
    title = "{Double inflation in classically conformal B-L model}",
    eprint = "2405.10537",
    archivePrefix = "arXiv",
    primaryClass = "astro-ph.CO",
    doi = "10.1088/1475-7516/2025/10/091",
    journal = "JCAP",
    volume = "10",
    pages = "091",
    year = "2025"
}

@article{Barman:2021lot,
    author = "Barman, Basabendu and Ghoshal, Anish",
    title = "{Scale invariant FIMP miracle}",
    eprint = "2109.03259",
    archivePrefix = "arXiv",
    primaryClass = "hep-ph",
    reportNumber = "PI/UAN-2021-698FT",
    doi = "10.1088/1475-7516/2022/03/003",
    journal = "JCAP",
    volume = "03",
    number = "03",
    pages = "003",
    year = "2022"
}

@article{Barman:2022njh,
    author = "Barman, Basabendu and Ghoshal, Anish",
    title = "{Probing pre-BBN era with scale invariant FIMP}",
    eprint = "2203.13269",
    archivePrefix = "arXiv",
    primaryClass = "hep-ph",
    reportNumber = "PI/UAN-2021-712FT",
    doi = "10.1088/1475-7516/2022/10/082",
    journal = "JCAP",
    volume = "10",
    pages = "082",
    year = "2022"
}

@inproceedings{Bardeen:1995kv,
    author = "Bardeen, William A.",
    title = "{On naturalness in the standard model}",
    booktitle = "{Ontake Summer Institute on Particle Physics}",
    reportNumber = "FERMILAB-CONF-95-391-T",
    month = "8",
    year = "1995"
}

@article{Hempfling:1996ht,
    author = "Hempfling, Ralf",
    title = "{The Next-to-minimal Coleman-Weinberg model}",
    eprint = "hep-ph/9604278",
    archivePrefix = "arXiv",
    reportNumber = "MPI-PHT-96-03",
    doi = "10.1016/0370-2693(96)00446-7",
    journal = "Phys. Lett. B",
    volume = "379",
    pages = "153--158",
    year = "1996"
}

@article{Espinosa:2007qk,
    author = "Espinosa, Jose Ramon and Quiros, Mariano",
    title = "{Novel Effects in Electroweak Breaking from a Hidden Sector}",
    eprint = "hep-ph/0701145",
    archivePrefix = "arXiv",
    reportNumber = "IFT-UAM-CSIC-07-01, UAB-FT-623",
    doi = "10.1103/PhysRevD.76.076004",
    journal = "Phys. Rev. D",
    volume = "76",
    pages = "076004",
    year = "2007"
}

@article{Chang:2007ki,
    author = "Chang, We-Fu and Ng, John N. and Wu, Jackson M. S.",
    title = "{Shadow Higgs from a scale-invariant hidden U(1)(s) model}",
    eprint = "hep-ph/0701254",
    archivePrefix = "arXiv",
    doi = "10.1103/PhysRevD.75.115016",
    journal = "Phys. Rev. D",
    volume = "75",
    pages = "115016",
    year = "2007"
}

@article{Foot:2007as,
    author = "Foot, Robert and Kobakhidze, Archil and Volkas, Raymond R.",
    title = "{Electroweak Higgs as a pseudo-Goldstone boson of broken scale invariance}",
    eprint = "0704.1165",
    archivePrefix = "arXiv",
    primaryClass = "hep-ph",
    doi = "10.1016/j.physletb.2007.06.084",
    journal = "Phys. Lett. B",
    volume = "655",
    pages = "156--161",
    year = "2007"
}

@article{Foot:2007ay,
    author = "Foot, Robert and Kobakhidze, Archil and McDonald, Kristian. L. and Volkas, Raymond. R.",
    title = "{Neutrino mass in radiatively-broken scale-invariant models}",
    eprint = "0706.1829",
    archivePrefix = "arXiv",
    primaryClass = "hep-ph",
    doi = "10.1103/PhysRevD.76.075014",
    journal = "Phys. Rev. D",
    volume = "76",
    pages = "075014",
    year = "2007"
}

@article{Foot:2007iy,
    author = "Foot, Robert and Kobakhidze, Archil and McDonald, Kristian L. and Volkas, Raymond R.",
    title = "{A Solution to the hierarchy problem from an almost decoupled hidden sector within a classically scale invariant theory}",
    eprint = "0709.2750",
    archivePrefix = "arXiv",
    primaryClass = "hep-ph",
    doi = "10.1103/PhysRevD.77.035006",
    journal = "Phys. Rev. D",
    volume = "77",
    pages = "035006",
    year = "2008"
}

@article{Meissner:2008gj,
    author = "Meissner, Krzysztof A. and Nicolai, Hermann",
    title = "{Neutrinos, Axions and Conformal Symmetry}",
    eprint = "0803.2814",
    archivePrefix = "arXiv",
    primaryClass = "hep-th",
    doi = "10.1140/epjc/s10052-008-0760-x",
    journal = "Eur. Phys. J. C",
    volume = "57",
    pages = "493--498",
    year = "2008"
}

@article{Iso:2009ss,
    author = "Iso, Satoshi and Okada, Nobuchika and Orikasa, Yuta",
    title = "{Classically conformal $B^-$ L extended Standard Model}",
    eprint = "0902.4050",
    archivePrefix = "arXiv",
    primaryClass = "hep-ph",
    reportNumber = "KEK-TH-1303",
    doi = "10.1016/j.physletb.2009.04.046",
    journal = "Phys. Lett. B",
    volume = "676",
    pages = "81--87",
    year = "2009"
}

@article{Iso:2009nw,
    author = "Iso, Satoshi and Okada, Nobuchika and Orikasa, Yuta",
    title = "{The minimal B-L model naturally realized at TeV scale}",
    eprint = "0909.0128",
    archivePrefix = "arXiv",
    primaryClass = "hep-ph",
    reportNumber = "KEK-TH-1327",
    doi = "10.1103/PhysRevD.80.115007",
    journal = "Phys. Rev. D",
    volume = "80",
    pages = "115007",
    year = "2009"
}

@article{Zhang:2019vsb,
    author = "Zhang, Yiyang and Vachaspati, Tanmay and Ferrer, Francesc",
    title = "{Magnetic field production at a first-order electroweak phase transition}",
    eprint = "1902.02751",
    archivePrefix = "arXiv",
    primaryClass = "hep-ph",
    doi = "10.1103/PhysRevD.100.083006",
    journal = "Phys. Rev. D",
    volume = "100",
    number = "8",
    pages = "083006",
    year = "2019"
}

@article{Inomata:2024rkt,
    author = "Inomata, Keisuke and Kamionkowski, Marc and Kasai, Kentaro and Shakya, Bibhushan",
    title = "{Gravitational waves from particles produced from bubble collisions in first-order phase transitions}",
    eprint = "2412.17912",
    archivePrefix = "arXiv",
    primaryClass = "astro-ph.CO",
    doi = "10.1103/k4s5-8zqy",
    journal = "Phys. Rev. D",
    volume = "112",
    number = "8",
    pages = "083523",
    year = "2025"
}

@article{ATLAS:2023tkt,
    author = "Aad, Georges and others",
    collaboration = "ATLAS",
    title = "{Combination of searches for invisible decays of the Higgs boson using 139 fb{\ensuremath{-}}1 of proton-proton collision data at s=13 TeV collected with the ATLAS experiment}",
    eprint = "2301.10731",
    archivePrefix = "arXiv",
    primaryClass = "hep-ex",
    reportNumber = "CERN-EP-2022-289",
    doi = "10.1016/j.physletb.2023.137963",
    journal = "Phys. Lett. B",
    volume = "842",
    pages = "137963",
    year = "2023"
}

@article{Jedamzik:2025cax,
    author = "Jedamzik, Karsten and Pogosian, Levon and Abel, Tom",
    title = "{Hints of primordial magnetic fields at recombination and implications for the Hubble tension}",
    eprint = "2503.09599",
    archivePrefix = "arXiv",
    primaryClass = "astro-ph.CO",
    reportNumber = "SCG-2025-01",
    doi = "10.1038/s41550-025-02737-x",
    journal = "Nature Astron.",
    volume = "10",
    number = "2",
    pages = "317--324",
    year = "2026"
}

@article{Dias:2014osa,
    author = "Dias, A. G. and Machado, A. C. B. and Nishi, C. C. and Ringwald, A. and Vaudrevange, P.",
    title = "{The Quest for an Intermediate-Scale Accidental Axion and Further ALPs}",
    eprint = "1403.5760",
    archivePrefix = "arXiv",
    primaryClass = "hep-ph",
    reportNumber = "DESY-14-020",
    doi = "10.1007/JHEP06(2014)037",
    journal = "JHEP",
    volume = "06",
    pages = "037",
    year = "2014"
}

@article{Jedamzik:2018itu,
    author = "Jedamzik, Karsten and Saveliev, Andrey",
    title = "{Stringent Limit on Primordial Magnetic Fields from the Cosmic Microwave Background Radiation}",
    eprint = "1804.06115",
    archivePrefix = "arXiv",
    primaryClass = "astro-ph.CO",
    doi = "10.1103/PhysRevLett.123.021301",
    journal = "Phys. Rev. Lett.",
    volume = "123",
    number = "2",
    pages = "021301",
    year = "2019"
}

@article{Jedamzik:2020krr,
    author = "Jedamzik, Karsten and Pogosian, Levon",
    title = "{Relieving the Hubble tension with primordial magnetic fields}",
    eprint = "2004.09487",
    archivePrefix = "arXiv",
    primaryClass = "astro-ph.CO",
    doi = "10.1103/PhysRevLett.125.181302",
    journal = "Phys. Rev. Lett.",
    volume = "125",
    number = "18",
    pages = "181302",
    year = "2020"
}

@article{Thiele:2021okz,
    author = "Thiele, Leander and Guan, Yilun and Hill, J. Colin and Kosowsky, Arthur and Spergel, David N.",
    title = "{Can small-scale baryon inhomogeneities resolve the Hubble tension? An investigation with ACT DR4}",
    eprint = "2105.03003",
    archivePrefix = "arXiv",
    primaryClass = "astro-ph.CO",
    doi = "10.1103/PhysRevD.104.063535",
    journal = "Phys. Rev. D",
    volume = "104",
    number = "6",
    pages = "063535",
    year = "2021"
}

@article{Takahashi:2016xsf,
    author = "Takahashi, Satoru and others",
    title = "{GRAINE 2015, a balloon-borne emulsion $\gamma$-ray telescope experiment in Australia}",
    doi = "10.1093/ptep/ptw089",
    journal = "PTEP",
    volume = "2016",
    number = "7",
    pages = "073F01",
    year = "2016"
}

@article{COrE:2011bfs,
    author = "Bouchet, F. R. and others",
    collaboration = "COrE",
    title = "{COrE (Cosmic Origins Explorer) A White Paper}",
    eprint = "1102.2181",
    archivePrefix = "arXiv",
    primaryClass = "astro-ph.CO",
    month = "2",
    year = "2011"
}

@article{EUCLID:2011zbd,
    author = "Laureijs, R. and others",
    collaboration = "EUCLID",
    title = "{Euclid Definition Study Report}",
    eprint = "1110.3193",
    archivePrefix = "arXiv",
    primaryClass = "astro-ph.CO",
    reportNumber = "ESA-SRE(2011)12",
    month = "10",
    year = "2011"
}

@article{NANOGrav:2023gor,
    author = "Agazie, Gabriella and others",
    collaboration = "NANOGrav",
    title = "{The NANOGrav 15 yr Data Set: Evidence for a Gravitational-wave Background}",
    eprint = "2306.16213",
    archivePrefix = "arXiv",
    primaryClass = "astro-ph.HE",
    doi = "10.3847/2041-8213/acdac6",
    journal = "Astrophys. J. Lett.",
    volume = "951",
    number = "1",
    pages = "L8",
    year = "2023"
}

@article{Bhattacharya:2025hme,
    author = "Bhattacharya, Subhaditya and Jahedi, Sahabub and Manna, Soumen Kumar and Sil, Arunansu",
    title = "{Probing ALP-portal Fermionic Dark Matter at the $e^+e^-$ Colliders}",
    eprint = "2505.00478",
    archivePrefix = "arXiv",
    primaryClass = "hep-ph",
    month = "5",
    year = "2025"
}

@article{Bisal:2025jwv,
    author = "Bisal, Subhadip",
    title = "{Constraining ALP-top interaction from the chromoelectric dipole moment of the top quark}",
    eprint = "2507.12570",
    archivePrefix = "arXiv",
    primaryClass = "hep-ph",
    doi = "10.1103/w6wh-knnp",
    journal = "Phys. Rev. D",
    volume = "112",
    number = "5",
    pages = "055027",
    year = "2025"
}

@article{Arteaga:2024vde,
    author = "Arteaga, Mart{\'\i}n and Ghoshal, Anish and Strumia, Alessandro",
    title = "{Gravitational waves and black holes from the phase transition in models of dynamical symmetry breaking}",
    eprint = "2409.04545",
    archivePrefix = "arXiv",
    primaryClass = "hep-ph",
    doi = "10.1088/1475-7516/2025/05/029",
    journal = "JCAP",
    volume = "05",
    pages = "029",
    year = "2025"
}

@article{DiLuzio:2020wdo,
    author = "Di Luzio, Luca and Giannotti, Maurizio and Nardi, Enrico and Visinelli, Luca",
    title = "{The landscape of QCD axion models}",
    eprint = "2003.01100",
    archivePrefix = "arXiv",
    primaryClass = "hep-ph",
    reportNumber = "DESY 20-036, DESY-20-036",
    doi = "10.1016/j.physrep.2020.06.002",
    journal = "Phys. Rept.",
    volume = "870",
    pages = "1--117",
    year = "2020"
}

@article{Kim:1979if,
    author = "Kim, Jihn E.",
    title = "{Weak Interaction Singlet and Strong CP Invariance}",
    reportNumber = "UPR-0120T",
    doi = "10.1103/PhysRevLett.43.103",
    journal = "Phys. Rev. Lett.",
    volume = "43",
    pages = "103",
    year = "1979"
}

@article{Shifman:1978by,
    author = "Shifman, Mikhail A. and Vainshtein, A. I. and Zakharov, Valentin I.",
    title = "{QCD and Resonance Physics: Applications}",
    reportNumber = "ITEP-94-1978, ITEP-81-1978",
    doi = "10.1016/0550-3213(79)90023-3",
    journal = "Nucl. Phys. B",
    volume = "147",
    pages = "448--518",
    year = "1979"
}

@article{Choi:1985cb,
    author = "Choi, Kiwoon and Kim, Jihn E.",
    title = "{DYNAMICAL AXION}",
    reportNumber = "SNUHE-84-05-REV, SNUHE-84-05",
    doi = "10.1103/PhysRevD.32.1828",
    journal = "Phys. Rev. D",
    volume = "32",
    pages = "1828",
    year = "1985"
}

@article{Dine:1981rt,
    author = "Dine, Michael and Fischler, Willy and Srednicki, Mark",
    title = "{A Simple Solution to the Strong CP Problem with a Harmless Axion}",
    reportNumber = "Print-81-0320 (IAS,PRINCETON)",
    doi = "10.1016/0370-2693(81)90590-6",
    journal = "Phys. Lett. B",
    volume = "104",
    pages = "199--202",
    year = "1981"
}

@article{Zhitnitsky:1980tq,
    author = "Zhitnitsky, A. R.",
    title = "{On Possible Suppression of the Axion Hadron Interactions. (In Russian)}",
    journal = "Sov. J. Nucl. Phys.",
    volume = "31",
    pages = "260",
    year = "1980"
}

@article{Diamond:2023cto,
    author = "Diamond, Melissa and Fiorillo, Damiano F. G. and Marques-Tavares, Gustavo and Tamborra, Irene and Vitagliano, Edoardo",
    title = "{Multimessenger Constraints on Radiatively Decaying Axions from GW170817}",
    eprint = "2305.10327",
    archivePrefix = "arXiv",
    primaryClass = "hep-ph",
    doi = "10.1103/PhysRevLett.132.101004",
    journal = "Phys. Rev. Lett.",
    volume = "132",
    number = "10",
    pages = "101004",
    year = "2024"
}

@article{Buckley:2024ldr,
    author = "Buckley, James H. and Dev, P. S. Bhupal and Ferrer, Francesc and Okawa, Takuya",
    title = "{Probing Heavy Axion-like Particles from Massive Stars with X-rays and Gamma Rays}",
    eprint = "2412.21163",
    archivePrefix = "arXiv",
    primaryClass = "hep-ph",
    reportNumber = "FERMILAB-PUB-24-0930-V",
    month = "12",
    year = "2024"
}

\end{document}